\documentclass[a4paper,12pt]{article}

\usepackage[]{color}
\usepackage{changes}
\usepackage{alltt}
\usepackage[left= 2.5cm,right = 2.5cm, bottom = 2.5cm,top=2cm]{geometry}
\usepackage[latin1]{inputenc} 
\usepackage{setspace}
\usepackage{amsmath}
\newcommand{\inv}{^{\raisebox{0.15ex}{$\scriptscriptstyle-1$}}}
\usepackage{amsfonts}
\usepackage{amssymb}
\usepackage{graphics}
\usepackage{graphicx}
\usepackage{epstopdf}
\usepackage[nottoc]{tocbibind}
\usepackage{array,multirow}
\usepackage{rotating} 
\usepackage{ragged2e}
\usepackage[super]{nth}
\usepackage{dsfont}

\usepackage{siunitx}                  
\usepackage{booktabs, threeparttable} 

\usepackage[referable]{threeparttablex}

\usepackage{caption}
\usepackage{colortbl}	
\usepackage{booktabs}				
\definecolor{MLgray}{gray}{0.9}
\definecolor{Lgray}{gray}{0.75}
\newcolumntype{g}{>{\columncolor{MLgray}}c}

\usepackage{abstract}

\usepackage{apacite} 
\newcommand{\GG}[1]{}
\newlength{\minuslength}
\usepackage{subfig}
\usepackage{float}  
\floatstyle{plaintop}
\restylefloat{table}

\usepackage{booktabs}
\usepackage{acronym} 
\usepackage{here} 
\allowdisplaybreaks
\usepackage{nccfoots}
\IfFileExists{upquote.sty}{\usepackage{upquote}}{}

\usepackage{titling}

\title{The Transmission of Monetary Policy via Common Cycles in the Euro Area\thanks{We thank seminar and conference participants at the Ruhr Graduate 
School in Economics in Essen, the \nth{7} International Conference on Applied Theory, Macro and Empirical Finance at the University of Macedonia in Thessaloniki, the \nth{24} IWH-CIREQ-GW-BOKERI Macroeconometric Workshop at the IWH Halle, the Workshop in Empirical Macroeconomics organized by the University of Innsbruck and the Liechtenstein Institute, the Workshop of the SGH Warsaw School of Economics, the \nth{28} International Conference on Macroeconomic Analysis and International Finance, the \nth{22} Conference of the European Economics and Finance Society, the 2024 European Seminar on Bayesian Econometrics, Philipp Ad\"ammer, Joscha Beckmann, Boris Blagov, Robert Czudaj, Sascha Keweloh, Martin Geiger and Mathias Klein for their valuable feedback. Jan Pr\"user gratefully
acknowledges the support of the German Research Foundation (DFG, 468814087).}}

\author{Lukas Berend$^{a}$ and Jan Pr\"user$^{b}$
\\[0.2cm]
$^{a}${\small FernUniversit\"at in Hagen\thanks{Corresponding Author: Fakult\"at f\"ur Wirtschaftswissenschaft, 58097 Hagen, Germany, e-mail: \texttt{lukas.berend@fernuni-hagen.de} }}
$^{b}${\small TU Dortmund\thanks{ Fakult\"at Statistik, 44221 Dortmund, Germany, e-mail: \texttt{prueser@statistik.tu-dortmund.de\,}} \hspace{0.1cm} 
} }

\begin{document}

\begin{titlepage}
    \maketitle
    \date{\today}
    \vspace{-0.5cm}
\begin{abstract}
	\begin{singlespace}
 \noindent
 We use a FAVAR model with proxy variables and sign restrictions to investigate the
role of the euro area's common output and inflation cycles in the transmission of monetary
policy shocks. Our findings indicate that common cycles explain most of the variation in output and inflation across member countries. However, Southern European economies exhibit a notable divergence from these cycles in the aftermath of the financial crisis. Building on this evidence, we demonstrate that monetary policy is homogeneously propagated to member countries via the common cycles. In contrast, country-specific transmission channels lead to heterogeneous country responses to monetary policy shocks. Consequently, our
empirical results suggest that the divergent effects of ECB monetary policy are attributable to heterogeneous country-specific exposures to financial markets, rather than to dis-synchronized economies within the euro area.
	\end{singlespace}
\end{abstract}
\bigskip


 \bigskip
\end{titlepage}


\thispagestyle{empty} 
\newpage

\pagenumbering{arabic}
\section{Introduction}

The transmission of monetary policy in the euro area to its member countries is marked by considerable heterogeneity. A number of empirical studies provide evidence of heterogeneous effects of monetary policy (see for example, \shortciteA{cmp13},  \shortciteA{bdp17}, \citeA{bg18}, \shortciteA{agkl22}, \shortciteA{msv22}, \citeA{g14}, and \citeA{ht20}). Differences in macroeconomic characteristics, such as labour markets, housing markets and financial markets, may prevent homogeneous outcomes of ECB monetary policy decisions. Also, it is motivated that the absence of common euro area cycles for output and inflation hampers an equal transmission to the member economies (\citeA{be92}, \citeA{fk15}, \citeA{b21}). However, there is a paucity of empirical evidence concerning the degree to which common cycles impede or facilitate monetary policy transmission.

In this paper, we employ an econometric framework to empirically identify common cycles for euro area output and inflation, the extent to which euro member countries are exposed to them and their relevance for monetary policy in the single currency area. Importantly, this framework enables us to untangle the propagation of monetary policy through area-wide co-movements in output and inflation from the country-specific transmission via financial variables. Our results reveal that euro member countries are largely linked to common euro cycles for output and inflation. Furthermore, the propagation of monetary policy via exposure to common co-movements is homogeneous, while country-specific exposures to financial variables drive heterogeneity in policy outcomes. As demonstrated by \shortciteA{agkl22}, we show that the heterogeneity induced via country-specific channels can be linked to structural characteristics of the member countries. Our results are highly robust across a range of different modeling choices.



In addition to the importance of similar macroeconomic environments, \citeA{m61} posits that optimal currency areas require a substantial degree of co-movement in output and inflation across member countries to facilitate effective monetary policy making. Otherwise, interventions by the central bank could have counterproductive effects in member countries that are not exposed to currency-wide dynamics. For example, a tightening would constitute a pro-cyclical policy in a contracting member country. In this scenario, the single monetary policy exacerbates fragmentation across the currency union.\footnote{\citeA{fk15} conclude that in light of the theory of \citeA{m61} asynchronous cycles may impair the effectiveness of common euro area monetary policy.} Consequently, the study of common cycles of output and inflation, and thus the degree of synchronization of the euro area economies is of significant relevance for a more profound understanding of the effectiveness of monetary policy.

In our analysis, we estimate distinct common cycles for output and inflation, which can be interpreted economically as they capture the co-movements of the euro area economies. In this manner, we identify the extent to which output and prices of ten member countries are exposed to these cycles over the period from January 2003 to December 2023. In accordance with \citeA{do07}, this approach allows us to investigate the impact of monetary policy on the aforementioned common cycles, as well as an examination of the extent to which the countries' exposures to these cycles facilitate the propagation of these effects to the member countries. From an intuitive standpoint, a higher degree of synchronization across member economies increases the central bank's ability to attain homogeneous policy outcomes across countries. 

We show that both output and inflation in the member countries are substantially exposed to common cycles. The co-movements in output and inflation account for up to 93\% of the variation in the countries indicating significant economic synchronization of business cycle fluctuations across the euro area economies. Nevertheless, Southern euro area economies exhibit diminished exposure to the output and inflation cycles in the wake of the financial crisis. In light of the strong co-movements in the economy, our findings indicate that common monetary policy is homogeneously propagated through the common cycles to the member countries. Hence, the observed heterogeneity in outcomes cannot be attributed to differential exposure to the cycles.

 In the second step of our empirical analysis, we allow for a direct country-specific transmission of monetary policy through a number of financial variables. This step allows us to untangle the propagation through the common cycles for output and inflation reflecting the synchronicity of the economies from country-specific transmission mechanisms. With respect to country-specific transmission, the financial crisis and the subsequent sovereign debt crisis in the euro area intensified fragmentation across member economies and impeded homogeneous stabilizing effects of accommodative monetary policy \cite{ecb17}.\footnote{In a report, the \citeA{ecb17} observes a decline in financial market convergence across euro area economies. \citeA{o19} and \citeA{aab22} find increasing financial fragmentation in the euro area.} To this end, our findings are consistent with the literature, as the effects of monetary policy become more heterogeneous once we allow for country-specific transmission via the financial variables (\citeA{h18}, \shortciteA{hmm18}, \citeA{bf15}). In the style of \shortciteA{agkl22} and \shortciteA{cdm22}, we undertake a correlation exercise that indicates that country channels may be driven by prevailing differences in macroeconomic structures across member countries. In contrast, the responses via common cycles appear to be detached from these. Consequently, it can be concluded that the divergent effects of the ECB's monetary policy are attributable to the presence of heterogeneous country-specific channels, rather than to the lack of synchronisation among the economies of the euro area.
 

To investigate the transmission of monetary policy via common cycles, we use a novel
Bayesian proxy FAVAR model with sign restrictions to jointly estimate the common cycles
for output and inflation, the exposure of countries to these cycles and to identify monetary policy shocks. By jointly estimating all cycles and coefficients we take all sources of estimation uncertainty into account. Following \shortciteA{kow03}, we estimate a dynamic factor model that allows for an economic interpretation of the factors as cycles for output and inflation. In accordance with \shortciteA{bbe05}, we incorporate the cycles into a VAR with macro-financial variables. We augment the VAR with a high-frequency instrument which is based on the dataset  provided by \shortciteA{abgmr19} and impose sign restrictions on the contemporaneous relationships between some of the endogenous variables in line with \citeA{Uhlig2005}. 

Recently, there has been a growing interest in combining external instruments with sign restrictions to identify monetary policy shocks (\citeA{jk20}, \shortciteA{arw21},  \shortciteA{cbs22},  \citeA{h24}, \citeA{bb23}, \citeA{f23} and \shortciteA{bbh23}). While sign restrictions alone set-identify structural shocks, the combination with a valid instrument allows for the point-identification of those.  Furthermore, sign restrictions
can serve as additional piece of information for shocks that are point-identified by instruments to sharpen inference.
Such information is especially valuable when the external variables are only weakly informative (\citeA{h24} and \citeA{bb23}). Therefore, our proposed model combines the benefits of this identification procedure with the advantages of a FAVAR model (\shortciteA{ggs23} and \shortciteA{cdm22}).\footnote{In a FAVAR setting, \shortciteA{ggs23} implement an instrument and order it first and perform a Cholesky decomposition to identify a monetary policy shock. \citeA{bruns21} follows the approach by \citeA{ch19} and adds a proxy equation to the matrix that maps the reduced form shocks to structural shocks. Our analysis integrates these recent approaches as we use \emph{pure} monetary policy shocks following \citeA{j22}, place a sign restriction on the common euro cycle for inflation to sharpen identification and extend the sampling approach by \citeA{h24} for a VAR to a FAVAR framework.} 

In this study, we follow the existing literature on the high-frequency identification of monetary policy shocks which employs interest rates changes around ECB policy announcements as external instruments.\footnote{Among others, this approach originates in the studies of \citeA{k01} and \shortciteA{gss05}. More recently, the \citeA{sw12}, \citeA{mr13} and \citeA{gk15} employ interest rate changes around FOMC announcements.} For the euro area, \citeA{jk20}, \citeA{ht20}, \citeA{j22} and \shortciteA{ggs23} utilize high-frequency data provided by \shortciteA{abgmr19} to estimate the impact of monetary policy shocks.\footnote{\shortciteA{cdm22} use changes in the 1-year EONIA swap rate in a longer time span than \shortciteA{abgmr19}.} As ECB announcements may contain information about the economic outlook, the above studies use changes in asset prices to isolate the \emph{pure} monetary policy shock as proposed by \citeA{j22}.\footnote{In their studies, \citeA{bs23a} and \citeA{bs23b} question the existence of information advantages of the FED and argue for a "FED response to news channel".}\\



\noindent
\textbf{Literature} Our study relates to several strands of the literature. It is linked to the literature on FAVAR models that study monetary policy. Unlike our study, the estimated factors in these model have no pre-specified economic meaning (\citeA{bbe05}, \citeA{st05}, \shortciteA{fglr09}, \citeA{fg10}). As central banks are required to consider a vast array of information sets, FAVAR models assist in reducing the dimension of the model. With regard to the euro area, \shortciteA{bgm08}, \shortciteA{bcl14} find substantial heterogeneous effects of monetary policy across member countries.\footnote{\citeA{lr19}, \citeA{jk20}, \citeA{ht20}, \citeA{ep20}, \shortciteA{hhs22} and others study the effect of monetary policy on euro area aggregates.} More recently, \shortciteA{cdm22} and \shortciteA{ggs23} provide further evidence for a heterogeneous transmission of ECB policy shocks. While our analysis focuses on the common cycles as transmitters of monetary policy, \shortciteA{cdm22} examine the importance of country-specific structural characteristics and their importance for heterogeneous policy outcomes. In our analysis, we indicate that these characteristics cannot be related to the responses propagated through the common cycles.  

More generally, heterogeneous effects of monetary policy in the euro area are found in studies using different empirical frameworks. \shortciteA{cmp13} employ a panel-VAR model, \shortciteA{bdp17} a near-VAR, and \citeA{bg18} global VAR respectively. Moreover, \shortciteA{agkl22} utilize local projections, while \shortciteA{msv22} employ a large-scale Bayesian VAR model.\footnote{\citeA{g14} and \shortciteA{msv22} motivate heterogeneous effects across countries with, amongst other, differences in the industry mix and the labour market.} \citeA{ht20} analyze euro aggregate and country-specific effects of monetary policy in separate regression settings, positing heterogeneous financial markets as underlying drivers. 

 Finally, in the literature on common business cycles, Bayesian dynamic factor models are a widely used method for modeling joint dynamics across countries (\shortciteA{jkoo16}, \citeA{b21}). In their seminal work, \citeA{ow98} and \citeA{kn98} employ single factor models to examine the co-movements across countries. \shortciteA{kow03} and \shortciteA{kow08} extend these studies to multi-factor models, providing evidence for strong co-movements of macroeconomic aggregates.\footnote{\citeA{ms12} and \citeA{mm21} analyze global and regional co-movements in inflation. \citeA{kl13} model co-movements of sectoral production across industry sectors of major developed economies. In a multi-level factor model, \citeA{fk15} model co-movements for major euro economies and find synchronous cycles for output. \citeA{b21} investigates co-movements of output growth for EU countries and observes heightened convergence of economic activity up to the beginning of the financial crises.} One advantages of Bayesian dynamic factor models is that the factors can be interpreted economically, as zero restrictions can be imposed on some factor loadings, allowing factors to only load on pre-specified groups of countries. Methodologically, our work is related to that of \citeA{do08}, \shortciteA{joz18} and \citeA{mr20}, who estimate economically interpretable factors that are subsequently stacked into VAR models.\footnote{\citeA{do07} further motivate that the implementation of common factors in VARs can be more suitable than aggregates as they possibly mix opposite and disproportionate movements of the member countries.} However, our approach differs from these studies in that we distinguish between between cycles for output and inflation and differentiate between the two propagation mechanisms outlined above.

The paper proceeds as follows. Section 2 lays out our econometric framework. Section 3 introduces our dataset and outlines our identification strategy. Section 4 presents our empirical results. Section 5 concludes. Additional material is relegated to the online appendix.

\section{A Sign-Restricted Proxy FAVAR Model}

In this section, we present and discuss our Bayesian sign-restricted proxy FAVAR model. We use the model to estimate euro area common factors for output and inflation. The factors have a clear economic interpretation as common cycles. Additionally, the model allows us to examine how each country is influenced by these cycles and how monetary policy shocks are propagated via these cycles.
Our model consists of two equations and departs from the conventional FAVAR  model: 
\vspace{-10pt}
\begin{eqnarray}
    \boldsymbol{x}_{t} &=& \boldsymbol{\Lambda} \boldsymbol{f}_{t} + \boldsymbol{\Lambda}^{\boldsymbol{z}}\boldsymbol{z}_{t}+ \boldsymbol{e}_{t}  \\  \label{eqn_1} 
    \begin{bmatrix}
        \boldsymbol{f}_{t} \\ \boldsymbol{z}_{t}
    \end{bmatrix} &=& \boldsymbol{A}
    \begin{bmatrix}
        \boldsymbol{f}_{t-1} \\
        \boldsymbol{z}_{t-1}
    \end{bmatrix} + \ldots + \boldsymbol{B} \boldsymbol{\epsilon}_{t}, \label{eqn_2} 
\end{eqnarray}

Equation (1) represents the factor equation, while Equation (2) illustrates the SVAR equation. The factor equation serves to model the manner in which the country-specific variables $\boldsymbol{x}_{t}$ depend on the common factors $\boldsymbol{f}_{t}$, which are shared by all countries. The SVAR equation captures the joint dynamics between the common factors and other macro-financial variables, denoted by $\boldsymbol{z}_{t}$. The monetary policy shock is identified by augmenting the SVAR in Equation (2) with a proxy equation for the external instrument, and sign restrictions are employed following the approaches of \citeA{h24}, \citeA{bb23} and \shortciteA{bbh23} (see section 3.2). In our baseline specification, we set $\boldsymbol{\Lambda}^{z}=\boldsymbol{0}$. In in section 4.4, however, we allow for a direct effect of the macro-financial variables $\boldsymbol{z}_{t}$ via $\boldsymbol{\Lambda}^{z}$.

\subsection{Factor Equation}
We commence with an introduction to the factor equation. In the baseline model, country-specific output growth is explained by a single common factor $f_{t}^{OUT}$, and country-specific inflation is explained by a single common factor $f_{t}^{INF}$ (compare with e.g. \shortciteA{kow03}
\shortciteA{bbe05}):\footnote{In their analysis, \shortciteA{joz18} estimate a area-wide common factor that loads on both countries' output and inflation. However, our aim is to explicitly distinguish between the two economic indicators in order to allow for differing degrees of synchronization of these.}
\vspace{-10pt}
\begin{eqnarray}
	\begin{bmatrix} 
		x_{EA19,t}^{OUT} \\ x_{1,t}^{OUT} \\ x_{2,t}^{OUT} \\ \vdots \\ x_{EA19,t}^{INF} \\ x_{1,t}^{INF} \\ x_{2,t}^{INF} \\ \vdots 
	\end{bmatrix} = 
	\begin{bmatrix}
		1 & 0 \\ \lambda_{1}^{OUT} & 0 \\
		\lambda_{2}^{OUT} & 0 \\ \vdots & \vdots \\ 0 & 1 \\ 0 & \lambda_{1}^{INF}  \\  0& \lambda_{2}^{INF} \\ \vdots & \vdots
	\end{bmatrix}
	\begin{bmatrix}
		f_{t}^{OUT}  \\ f_{t}^{INF} 
	\end{bmatrix}+
	\begin{bmatrix}
		e_{EA19,t}^{OUT} \\ e_{1,t}^{OUT} \\ e_{2,t}^{OUT} \\ \vdots \\ e_{EA19,t}^{INF} \\ e_{1,t}^{INF} \\ e_{2,t}^{INF} \\ \vdots
	\end{bmatrix}, 
 \label{eqn_3}
\end{eqnarray}

where $x_{i,t}^{OUT}$ and $x_{i,t}^{INF}$ represent country-specific output
and inflation for $N$ countries and $T$ periods, $i=1,\dots,N$ and $t=1,\dots,T$. The factor loadings $\lambda_{i}^{OUT}$ 
and $\lambda_{i}^{INF}$ determine how each country-specific variables depend on the common cycle, in a manner analogous to \shortciteA{kow03} and \shortciteA{jkoo16}. The error terms $e_{i,t}^{OUT}$ and $e_{i,t}^{INF}$ measure country-specific time-varying idiosyncratic 
components in the member countries' output growth and inflation and are distributed $e_{i,t}^{OUT} \sim N(0,\sigma_{i,t}^{OUT})$ 
and $e_{i,t}^{INF} \sim N(0,\sigma_{i,t}^{INF})$. Consequently, the factor equation differentiates the movement of each variable into a common component shared by all countries and an idiosyncratic country-specific part. Due to the structure of zeros incorporated into the matrix of factor loadings in Equation (\ref{eqn_3}), the output factor is only allowed to explain the variation of the country-specific output growth, while the inflation factor is only allowed to explain the variation of the country-specific inflation. In accordance with \cite{do07}, it is crucial to note that this enables us to economically interpret $f_{t}^{OUT}$ and $f_{t}^{INF}$ as common cycles for output and inflation. In order to achieve statistical identification, we include output and inflation for the euro area aggregates, denoted as $x^{OUT}_{EA19,t}$ and $x^{INF}_{EA19,t}$ above the country-specific observations for $x^{OUT}_{i,t}$ and $x^{INF}_{i,t}$, and normalize the corresponding factor loadings to 1. In this manner, we follow the approach set forth by \citeA{kl13} and \citeA{b21} which permits us to interpret the loadings of the countries relative to the aggregate euro area.\footnote{Since we only include ten Euro area countries, the Euro area aggregate of the 19 member countries does not represent the sum of the countries under investigation.} In the robustness analysis, we allow for lags in the factor equation to accommodate disproportional country responses and demonstrate that the results remain unchanged.\footnote{In the online appendix, we follow the argument by \shortciteA{adp21} and motivate the model specifications with higher lag orders.}


The extent to which common cycles account for fluctuations in country output and inflation may undergo changes over time. Accordingly, we allow for time variation in the idiosyncratic country-specific component. In particular, $\sigma_{i,t}^{OUT}$ and $\sigma_{i,t}^{INF}$ are allowed to vary over time according to geometric random walks:
\vspace{-10pt}
\begin{eqnarray}
	h_{i,t}^j = h_{i,t-1}^j + \xi_{i,t}^j, \quad \xi_{i,t} \sim N(0,V_{h_{i,t}}^j), \label{eqn_4} 
\end{eqnarray}
where $\sigma_{i,t}^{j} = exp(h_{i,t}^{j})$ for $j = \text{OUT, INF}$. The variance $V_{h_{i,t}}^j$ determines the degree of time variation. We estimate $V_{h_{i,t}}^j$ using a hierarchical horseshoe prior which comprises a global and a local shrinkage component to allow for smooth as well as sudden changes in 
the idiosyncratic part of the model \cite{p21}. This allows us to study if any disconnection from one country from the common cycle occurs abruptly or more gradually. The prior specifications for the factor equation are presented in appendix C.

\subsection{SVAR Equation}
Now we direct our attention to the second equation of our model, the structural proxy sign-restricted VAR model following the implementation by \citeA{h24} and which draws upon the concept of \citeA{bb23}.
The common factors $f_{t}^{OUT}$ and $f_{t}^{INF}$, macro-financial variables $\boldsymbol{z}_{t}$ (see section 3.1) and the external instrument $m_{t}$ (see section 3.2 for the structural identification) are stacked into $\boldsymbol{y}_t=(f_{t}^{OUT},f_{t}^{INF}, \boldsymbol{z}_{t}', m_{t})'$. The  $n \times 1$ vector $\boldsymbol{y}_{t}$ is modeled by a structural proxy VAR model, where $n=r+k$. In this context, $r$ denotes the sum of the factors and the variables in $\boldsymbol{z}_{t}$, while $k$ denotes the number of external instruments:
\vspace{-10pt}
\begin{eqnarray}
	\boldsymbol{y}_t &=& \boldsymbol{c} + \boldsymbol{A}_{1} \boldsymbol{y}_{t-1} + \cdots + \boldsymbol{A}_{L} \boldsymbol{y}_{t-L} + \boldsymbol{B}\boldsymbol{\epsilon}_{t},\quad \boldsymbol{\epsilon}_{t} \sim N(\boldsymbol{0},\boldsymbol{I}_{n}), \label{eqn_7} 
\end{eqnarray}
where $\boldsymbol{A}_{1}, \ldots, \boldsymbol{A}_{L}$ are $n \times n$ 
and contain the autoregressive VAR coefficient matrices and $\boldsymbol{B}$ denotes the (invertible) contemporaneous impact matrix of the structural shocks $\boldsymbol{\epsilon}_{t}$. In order to maintain a modest level of parsimony, we set $L=6$ in the baseline model. The online appendix presents the results for lag lengths of $L=3$ and $L=12$.

The shocks $\boldsymbol{\epsilon}_{t}$ comprise the structural shocks of the endogenous variables $\boldsymbol{\epsilon}_{t}^{r'}$ and the shock of the instrumental variable $\boldsymbol{\epsilon}_{t}^{k'}$. Since the shock of the instrumental variable lacks information for the endogenous variables $f_{t}^{OUT}$, $f_{t}^{INF}$, and $\boldsymbol{z}_{t}'$, the coefficient matrices in \eqref{eqn_7} are as follows:
\vspace{-10pt}
\begin{eqnarray}
    \boldsymbol{A}_{l} = 
    \begin{bmatrix}
      \boldsymbol{\Gamma}_{l} & \boldsymbol{0}_{r \times k} \\
      \boldsymbol{\Phi}_{l,1} & \Phi_{l,2} 
    \end{bmatrix}, \text{    }
    \boldsymbol{B} = 
    \begin{bmatrix}
      \boldsymbol{\Gamma}_{0} & \boldsymbol{0}_{r \times k} \\
      \boldsymbol{\Phi}_{0,1} & \Phi_{0,2}
    \end{bmatrix},  \text{    } l = 1, \ldots,L,  \label{eqn_8} 
\end{eqnarray}
where $\boldsymbol{\Gamma}_{l}$ is $r \times r$, $\boldsymbol{\Phi}_{l,1}$ is $k \times r$ and $\Phi_{l,2}$ is $k \times k$ for $0 \leq l \leq L$. In $\boldsymbol{A}_{l}$, $\boldsymbol{\Phi}_{l,1}$ and $\Phi_{l,2}$ are set to $0$, thereby ensuring that the instrument variable $m_{t}$ does not exert a lagged effect on the endogenous variables. The autoregressive parameters in $\boldsymbol{\Gamma}_{l}$ are subject to adaptive asymmetric Minnesota-type shrinkage priors.
In $\boldsymbol{B}$, $\boldsymbol{\Phi}_{0,1}$ is composed of a row of zeros, with the exception of  the coefficient that describes the relationship between the instrument variable $m_{t}$ and the structural monetary policy shock.\footnote{\citeA{h24} discusses the structural proxy VAR in more depth.} In the remainder of the paper, we use $\Phi_{0,1}$ and only refer to this single non-zero element in $\boldsymbol{\Phi}_{0,1}$. As in \citeA{h24}, the non-zero elements of $\boldsymbol{B}$ are assumed to follow independent (sign-restricted) Gaussian priors. A comprehensive discussion of the prior specifications can be found in appendix C. The sign restrictions in the contemporaneous impact matrix $\boldsymbol{B}$ are jointly presented with the external instrument in section 3.2. The autoregressive coefficients in $\boldsymbol{A}_{l}$ are estimated following \shortciteA{ccm19} and \shortciteA{cccm22}. The estimation of the impact matrix $\boldsymbol{B}$ is implemented as in \citeA{h24}, employing a parameter transformation scheme that avoids the computationally inefficient Metropolis Hastings algorithm.

\subsection{The Gibbs Sampler}
We employ a fully Bayesian framework to estimate the FAVAR model with proxy variables and sign restrictions using the Gibbs sampler. In this section, we provide a brief overview of our proposed sampler, which incorporates the sampling algorithms of \shortciteA{bbe05}, \citeA{p21}, and \citeA{h24}. In contrast to a principal component analysis, our novel Gibbs sampler takes the uncertainty of the dynamic factor model into account, resulting in posterior distributions for both the factors and their loadings. Furthermore, we address the issue of over-parameterization and estimate the scaling parameters for the stochastic volatilities in the factor equation and the VAR parameters within the sampler in a data-driven approach. In addition, our proposed sampler incorporates sign restrictions on the contemporaneous impact matrix $\boldsymbol{B}$ as well as external instruments for the purpose of identifying structural shocks to the VAR. This is achieved in an efficient manner that minimizes the computational costs substantially in comparison to other more recent attempts to combine sign restrictions and instruments \cite{h24}. We simulate 18,000 iterations and reject the first 3,000 as burn-in and retain every fifth draw. In appendix D, we describe the Gibbs sampler in detail.
\vspace{-10pt}
\begin{itemize}
     \itemsep-10pt 
    \item[\bf{Step 1}] Draw factor loadings from the conditional posterior of $\lambda_{i}^{j}$.
    \item[\bf{Step 2}] Draw stochastic volatilities $h_{i,t}^{j}$ using the auxiliary mixture sampler of \shortciteA{ksc98} in combination with the precision sampler of \citeA{cj09}.
    \item[\bf{Step 3}] Draw the initial conditions for $h_{i,0}^{j}$ using the precision sampler of \citeA{cj09}.
    \item[\bf{Step 4}] Draw the local and global shrinkage components $\lambda_{h_{i,t}^{j}}$ and $\tau_{h_{i}^{j}}$  of the stochastic volatilities $h_{i,t}^{j}$ using the sampler for half-Cauchy distributions of \citeA{ms16}. 
    \item[\bf{Step 5}] Draw the VAR coefficients $\boldsymbol{A} = (\boldsymbol{c}, \boldsymbol{A}_{1},\ldots,\boldsymbol{A}_{L})'$ using the equation-by-equation approach of \shortciteA{ccm19}.
    \item[\bf{Step 6}] Draw the contemporaneous impact matrix $\boldsymbol{B}$ with zero and sign restrictions using the parameter transformation scheme of \citeA{h24}.
    \item[\bf{Step 7}] Draw the shrinkage components $\kappa_1$ and $\kappa_2$ of the VAR coefficients $\boldsymbol{A} = (\boldsymbol{c}, \boldsymbol{A}_{1},\ldots,\boldsymbol{A}_{L})'$ from truncated inverse Gamma distributions.
    \item[\bf{Step 8}] Draw the common factors $f_{t}^{j}$ using the algorithm of \citeA{ck94} and following \shortciteA{bbe05}.
\end{itemize}

\section{Data and Structural Identification}
\subsection{Data}

In our analysis, we use monthly time series data from January 2003 to December 2023.\footnote{We include data on loans to non-financial institutions which are only available starting in January 2003. Further, the high-frequency data on the monetary policy surprises shortly after the introduction of the Euro is noisy \cite{ggs23}.} Consequently, the sample commences shortly after the introduction of the single currency and encompasses the global pandemic of 2020 and the recent surge in inflation. The examination of the large fluctuations in gross domestic product (GDP) and inflation over the past four years facilitates the comprehension of the causal relationships between structural shocks and our target variables. 

In the factor Equation \eqref{eqn_3}, we use annual growth rates of the real gross domestic products and the harmonized index of consumer prices (HICP) of the ten largest euro area members to estimate the common cycles for GDP growth and inflation.\footnote{We follow \citeA{pb22} and omit Ireland, since it adjusted its accounting standards for the GDP in 2015 which portrays a structural break in the time series and include Austria (AT), Belgium (BE), Germany (DE), Greece (GR), Spain (ES), Finland (FI), France (FR), Italy (IT), the Netherlands (NL) and Portugal (PT).} In both cases, we employ the seasonally adjusted data series. In order to obtain monthly time series for GDP growth, the quarterly series are interpolated using industrial production and unemployment following \citeA{jk20} and \citeA{cl71}. Although our models effectively capture the volatilities associated with the global pandemic, we remove outliers from the GDP and industrial production time series following \citeA{cl93}. In this regard, we adhere to the methodology proposed by \citeA{eurostat20} and control for additive outliers in the country time series. As the outbreak of the pandemic marked an unprecedented event with several structural shocks occurring simultaneously, the removal of outliers ensures that the slump and recovery in GDP are not disproportionately attributed to monetary policy.\footnote{In the online appendix, we present the results using the original time series and show that the effects of a monetary policy surprise appear unreasonably large indicating that the pandemic drives the results.}
In our SVAR, we employ the German one-year government bond yield as it represents the safest euro area interest rate. In order to account for the financial channels, we include the logarithmic monthly average of the Eurostoxx50 and the BBB bond spread which is consistent with \citeA{jk20}. Furthermore, we refer to \citeA{ht20} and include the German ten-year government bond yield, the effective real exchange rate, and loans to non-financial institutions. The German ten-year government bond yield is employed as
benchmark for euro area long-term financial expectations. The real effective exchange rate drives
capital flows, while loans to non-financial institutions reflect the liquidity conditions in the
financial market. This allows the financial market a greater scope to drive heterogeneity in the country-specific responses. These variables are stacked into $\boldsymbol{z}_{t}$.\footnote{We standardize the variables in $\boldsymbol{z}_{t}$ to $N(0,1)$ and de-standardize to portray the impulse responses. Country GDP and inflation are not standardized since they are included as annual growth rates.} As instrument variable $m_{t}$, we use the first principal component of high-frequency changes in overnight index swap rates which we discuss in detail in the next section. 

A detailed description of the data is outlined in the Data section in the online appendix.

\subsection{Structural Identification}
We use the Euro Area Monetary Policy Event-Study Database of \shortciteA{abgmr19} which encompassed changes in swap rates and asset prices in the vicinity of ECB announcements. The identification of monetary policy surprises through the use of high-frequency data is based on the reasoning that the shock in question is unlikely to be confounded with other structural shocks \shortcite{gss05}. Nevertheless, statements by central banks on monetary policy may be influenced by their outlook on the broader economic situation, potentially obscuring the clarity of a \emph{pure} policy shock.\footnote{In their analysis, \citeA{bs23a} and \citeA{bs23b} introduce an alternative to the information shock and motivate a \emph{reaction to news shock}.} \citeA{jk20} and \citeA{j22} employ changes in the Eurostoxx50 around ECB announcements to purge \emph{pure} policy shock from these information shocks.\footnote{\citeA{jk20} argue that when asset prices react positively to a policy rate hike, markets interpret the tightening as a positive assessment of the economic situation by the ECB. On the contrast, a negative reaction of the Eurostoxx50 is in line with the textbook \emph{pure} monetary policy shock, as a tightening reduces economic activity.} For our analysis, we construct the proxy variable for the policy shock $m_{t}$ following the approach by \citeA{j22} which we outline in the online appendix.\footnote{\shortciteA{ggs23} proceed analogously in their study heterogeneous effects across the euro area. The approach by \citeA{ht20} is very similar as they use changes in German government bond yields and asset prices.} \\ 
For the proxy variable $m_{t}$ to be a valid instrument, it must satisfy the relevance and the exogeneity conditions. The relevance condition requires it to be correlated with the structural monetary policy shock. The exogeneity condition posits that the instrument is uncorrelated with other structural shocks. In particular, the structural shocks $\boldsymbol{\epsilon}_{t}^{r'}$ can be split into $\boldsymbol{\epsilon}_{1,t}^{r'}$ and $\boldsymbol{\epsilon}_{2,t}^{r'}$. The former represents the monetary policy shock, while the latter encompasses all other shocks. In accordance with \citeA{h24}, the relevance and exogeneity assumptions can be summarized as follows:
\vspace{-10pt}
\begin{eqnarray}
    \mathbb{E}(m_{t} \boldsymbol{\epsilon}_{t}^{r'}) = 
    \begin{pmatrix} \mathbb{E}(m_{t} \epsilon_{1,t}^{r'}) & \mathbb{E}(m_{t} \boldsymbol{\epsilon}_{2,t}^{r'}) & \end{pmatrix} = 
    \begin{pmatrix} \mathbb{E}(m_{t} \epsilon_{1,t}^{r'}) & 0 \end{pmatrix}. \label{eqn_9} 
\end{eqnarray}
Following \eqref{eqn_9}, we implement the relationship between the structural monetary policy shock and the instrument $m_{t}$ via the proxy equation as in \citeA{ch19}. From \eqref{eqn_7} and \eqref{eqn_8} in section 2.2, we obtain:
\vspace{-10pt}
\begin{eqnarray}
    m_{t} &=& \Phi_{0,1} \epsilon_{1,t}^{r'} + \Phi_{0,2} \epsilon_{t}^{k'}, \quad \epsilon_{t}^{k'} \sim N(0,1) \text{ and } \epsilon_{1,t}^{r'} \perp \epsilon_{t}^{k'}. \label{eqn_18}
\end{eqnarray}
In order to incorporate our prior beliefs regarding the relevance of the proxy, we shrink the relevance of the measurement error $\epsilon_{t}^{k'}$ towards 0 (see appendix C). Following \citeA{mr13}, we provide a quantitative understanding of the relevance of the instrument and estimate the reliability indicator:
\vspace{-10pt}
\begin{eqnarray}
    \rho &=& \text{CORR}(m_{t} \epsilon^{r'}_{1,t})^{2} = \frac{\Phi_{0,1}^{2}}{\Phi_{0,1}^{2} + \Phi_{0,2}^{2}}. \label{eqn_19}
\end{eqnarray}
In our baseline model, the median value of $\rho$ is $0.11$ with 68\% posterior percentiles ranging from $0.07$ to $0.16$. Notwithstanding the justifiable assumption of exogeneity and the prior shrinkage of the measurement error, this indicates a relatively weak relevance of the instrument in comparison with \citeA{ch19}. Accordingly, we adopt the approach by \shortciteA{bbh23} to enhance the identification strategy and integrate sign restrictions into the contemporaneous impact matrix $\boldsymbol{B}$.\footnote{Specifically, \shortciteA{bbh23} assume that a expansionary policy shock lowers the poor man's proxy by \citeA{jk20} and the shadow rate by \citeA{k13}.} We follow \citeA{f98} and \citeA{Uhlig2005} and assume that a contractionary policy shock raises the German one-year government bond yield and lowers the inflation factor $f_{t}^{INF}$.\footnote{Among others, \citeA{p05}, \citeA{fp11}, \citeA{bb13}, \citeA{bp21}, \citeA{w22} and \citeA{ks24} also place sign restrictions on inflation. Next to this, they assume that output responds negatively to a monetary tightening.} No sign restrictions are imposed on the country-specific factor loadings $\lambda_{i}^{OUT}$ and $\lambda_{i}^{INF}$ in \eqref{eqn_3}. \\
From the perspective of proxy identification, our sign restrictions add information to the identified shock, as Equation \eqref{eqn_19} suggests a certain degree of weak informativeness of our external instrument \cite{bb23}. In particular, our instrument may be subject to noise as market participants may interpret information within monetary policy announcements differently than the ECB intends to communicate \shortcite{abgmr19}. From the perspective of sign restriction identification, the instrument pins down the set of admissible impulse responses \cite{bb23}. \\
For our unified model, we argue that the combined identification strategy provides a suitable degree of agnosticism while ensuring informative impulse responses in a heavily parameterized estimation procedure.

\section{Empirical Results}

This section presents and discusses the empirical results. First, we present the factor loadings $\lambda_{i}^{OUT}$ and $\lambda_{i}^{INF}$ which demonstrate the countries' exposure to the common cycles. We find that euro cycles account for the majority of the variation in country-specific output and inflation. Building on this, we examine how expansionary monetary policy shocks are propagated via these common co-movements and find considerable homogeneous responses across all countries. This picture alters when we allow the macro-financial variables $\boldsymbol{z}_{t}'$ to directly affect the countries via an extension of the factor model in section 4.4. Nevertheless, we provide evidence indicating that asymmetric responses of the countries to monetary policy shocks are driven by heterogeneous country-specific exposure to the financial variables and not by dis-synchronized common euro cycles. We corroborate these findings by estimating the coefficients of variation proposed by \shortciteA{cdm22}. Furthermore, we link the heterogeneous responses induced by the country-specific channels to structural characteristics of the economies, in accordance with \shortciteA{cdm22} and \shortciteA{agkl22}.

\subsection{Common Euro Area Cycles}
The following subsection examines the importance of the common cycles for output and inflation for the euro area countries. As previously stated, the euro area cycles for output and inflation are represented by the common factors $f_{t}^{OUT}$ and $f_{t}^{INF}$. The country-specific factor loadings $\lambda_{i}^{OUT}$ and $\lambda_{i}^{INF}$ quantify the degree to which the common cycles influence country $i$ \cite{do07}.
\begin{table}[H]
	\centering
  \begin{threeparttable}
	\begin{tabular}{cccc}
		\toprule
	\textbf{	Country name} &\textbf{ Median} &\textbf{ Lower Bound}  &\textbf{ Upper Bound} \\
		\midrule
        EA19 & 1 & - & - \vspace{-5pt} \\
        AT & 1.0055 & 0.9835  & 1.0299 \vspace{-5pt} \\
        BE & 0.9002 & 0.8628 & 0.9385  \vspace{-5pt} \\
		DE & 0.9954 & 0.9528 & 1.0317 \vspace{-5pt} \\
		GR & 0.6823 & 0.6633 & 0.7049  \vspace{-5pt} \\
		ES & 1.1735 & 1.1515 & 1.1964 \vspace{-5pt} \\
		FI & 1.2320 & 1.2052 & 1.2611 \vspace{-5pt}  \\
		FR & 0.7946 &  0.7744 & 0.8140 \vspace{-5pt} \\
		IT & 0.6495 & 0.6262 & 0.6749 \vspace{-5pt} \\
		NL & 1.0967 & 1.0713 & 1.1231 \vspace{-5pt}\\
		PT & 0.8504 &  0.8041 & 0.8989 \vspace{-5pt}\\
		\bottomrule
	\end{tabular}
 \begin{tablenotes}[flushleft]\footnotesize
    \linespread{.5}\small
    \item\hspace*{-\fontdimen2\font}\note The first column lists the country codes. The second column reports the medians of the posterior distribution of the factor loadings. The third and fourth columns report the lower and upper bounds of the 68\% credible bands. The factor loadings for EA19 is set to 1.
    \end{tablenotes}
     \end{threeparttable}
	\caption{Country Factor Loadings with respect to Common Cycle for Output, $\lambda_{i}^{OUT}$.}
	\label{tab_1}
\end{table}
\vspace{-10pt}
Table \ref{tab_1} reports the country factor loadings with respect to the output cycle. For all countries under consideration, the posterior densities obtained from the draws are concentrated around the median. The estimated factor loadings are lowest in IT and GR and highest in FI and ES. As they deviate from the pre-set factor loading of the euro area aggregate, it can be reasoned that the corresponding countries exhibit a minor influence on the euro area cycle for output. In contrast, the estimated factor loadings of AT, DE, and NL are close to unity and thus align with the benchmarked euro area aggregate. Therefore, it can be concluded that it is the Northern economies that drive the common cycle for output. Nevertheless, countries with factor loadings above unity demonstrate a greater sensitivity to changes in the cycle.

\begin{table}[H]
	\centering
 \begin{threeparttable}
	\begin{tabular}{cccc}
		\toprule
	\textbf{	Country name} &\textbf{ Median} &\textbf{ Lower Bound}  &\textbf{ Upper Bound} \\
		\midrule
        EA19 & 1 & - & - \vspace{-5pt} \\
        AT & 1.0034 &  0.9905 &  1.0177 \vspace{-5pt} \\
        BE & 1.1384 & 1.1083 & 1.1700 \vspace{-5pt} \\
		DE & 0.9032 & 0.8908 & 0.9159 \vspace{-5pt} \\
		GR & 1.2801 & 1.2573 & 1.3063 \vspace{-5pt} \\
		ES & 1.2815 & 1.2592 & 1.3013 \vspace{-5pt} \\
		FI & 0.8241 &  0.7992 & 0.8448 \vspace{-5pt}  \\
		FR & 0.8800 & 0.8666 & 0.8933 \vspace{-5pt} \\
		IT & 0.9649 & 0.9531 &  0.9768 \vspace{-5pt} \\
		NL & 0.7810 & 0.7620 &  0.8039\vspace{-5pt}\\
		PT & 1.0036 & 0.9852 & 1.0183 \vspace{-5pt}\\
		\bottomrule
	\end{tabular}
 \begin{tablenotes}[flushleft]\footnotesize
    \linespread{.5}\small
    \item\hspace*{-\fontdimen2\font}\note The first column lists the country codes. The second column reports the medians of the posterior distribution of the factor loadings. The third and fourth columns report the lower and upper bounds of the 68\% credible bands. The factor loadings for EA19 is set to 1.
    \end{tablenotes}
     \end{threeparttable}
	\captionof{table}{Country Factor Loadings with respect to Common Cycle for Inflation, $\lambda_{i}^{INF}$.}
	\label{tab_2}
\end{table}
\vspace{-10pt}

Table \ref{tab_2} presents the factor loadings of the countries with respect to inflation. As with the factor loadings for output, the posterior densities for inflation are also closely centered around the medians. In this instance, the factor loadings for NL, FI and FR are the lowest. The countries with the highest factor loadings are ES and GR. The factor loadings for AT, PT and IT are in close proximity to the euro area aggregate benchmark. In contrast to the factor loadings for output, the inflation factor loadings are more dispersed, indicating that there is no discernible pattern within the countries. Rather, the inflation cycle appears to be the results of an averaging out of the countries under investigation. Additionally, inflation rates in the euro area countries may be more susceptible to domestic economic dynamics (see stochastic volatilities in Figure \ref{fig_A2}). As with the output factor loadings, countries with factor loadings exceeding 1, such as ES and GR, exhibit a more pronounced response to fluctuations in the euro inflation cycle. \\ 
Subsequently, we fit the country exposure to the common cycle for output and inflation to the actual time series for output and inflation of the countries. In this regard, the country-specific exposure is specified as the product of the factor loadings $\lambda_{i}^{OUT}$ and $\lambda_{i}^{INF}$ and the corresponding common cycles $f_{t}^{OUT}$ and $f_{t}^{INF}$ (see Equation \eqref{eqn_3}).\footnote{\citeA{do07} proceed analogously in their analysis on the co-movement of housing prices across the US.}
This approach allows us to investigate the extent to which the fluctuations of output and inflation in the countries are explained by the common euro area cycles. In this regard, significant or more prolonged deviations of country output and inflation from the predicted exposure can be regarded as periods of under- or overperformance during which the country-specific business cycles are dominated by idiosyncratic forces.  \\ 

\begin{center}
    \begin{figure}[H]
        \includegraphics[width=0.99\textwidth]{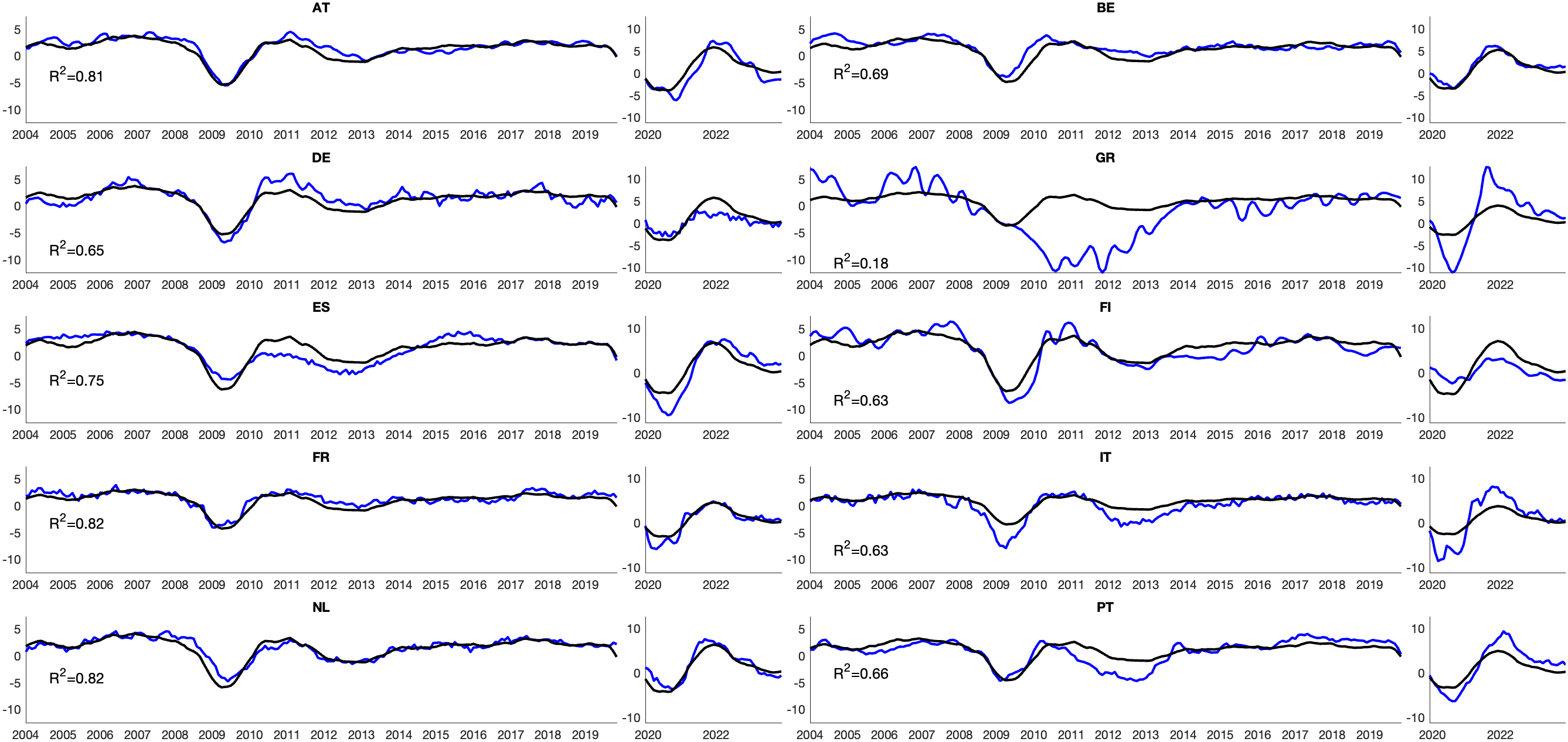}
		\caption{Countries' exposure to common cycle for output growth. The solid blue line depicts actual interpolated output growth and the solid black line portrays the common euro cycle for output growth multiplied by the corresponding country-specific factor loadings. $R^{2}$ denotes the R-squared.} \label{fig_1}
    \end{figure}
\end{center}
\vspace{-40pt}

Figure \ref{fig_1} illustrates the actual country-level time series for output growth rates ($x_{i,t}^{OUT}$) and the corresponding exposures to the common cycle. With the exception of GR, output cycle of the euro area accounts for at least 63\% of the variation in output growth rates across countries. In the case of AT, FR, and NL, the euro area cycle accounts for more than 80\% of the overall variation. Therefore, our parsimonious model specification with one common factor effectively captures the majority of the country-specific variation in output which suggests a high degree of synchronization across the euro area. It is noteworthy that DE and BE exhibited a more robust recovery from the financial crisis than the other euro area countries, outperforming the euro area cycle. Economic activity during the financial crisis in FI and IT contracted to a greater extent than would have been predicted based on their exposure to the common cycle. Most notably, the Southern economies of GR, ES, PT and IT faced severe economic challenges during the sovereign debt crisis and exhibited prolonged below-average performance to the euro area cycle.\footnote{In the joint business cycle literature, \citeA{fk15} and \citeA{b21} find similar results.} In GR, country output demonstrated above-average growth prior to the financial crisis but subsequently remained below the expected exposure to the common cycle until 2014. In comparison to the other countries, the underperformance of GR represents the most persistent deviation and underlines the peculiarity of GR during the euro debt crisis. The diminished exposure of GR to the common cycle coincides with a pronounced surge in volatility in 2009 which suggests the preponderance of the idiosyncratic component (see Figure A.1). To a lesser extent, a similar pattern can be observed in ES and PT. The unprecedented economic downturn that commenced in 2020, subsequent to the shock of the pandemic, proceeded in a synchronized manner throughout the euro area, with repeatedly higher degrees of deviation from the cycle in GR, ES, IT and PT. The diminished exposure to the common cycle in the Southern economies is partly reflected in stronger increases in idiosyncratic volatility. With regard to this, dissimilarities in the components of output, such as the service sector, might explain the heightened influence of the idiosyncratic component.

\begin{center}
    \begin{figure}[H]
        \includegraphics[width=0.99\textwidth]{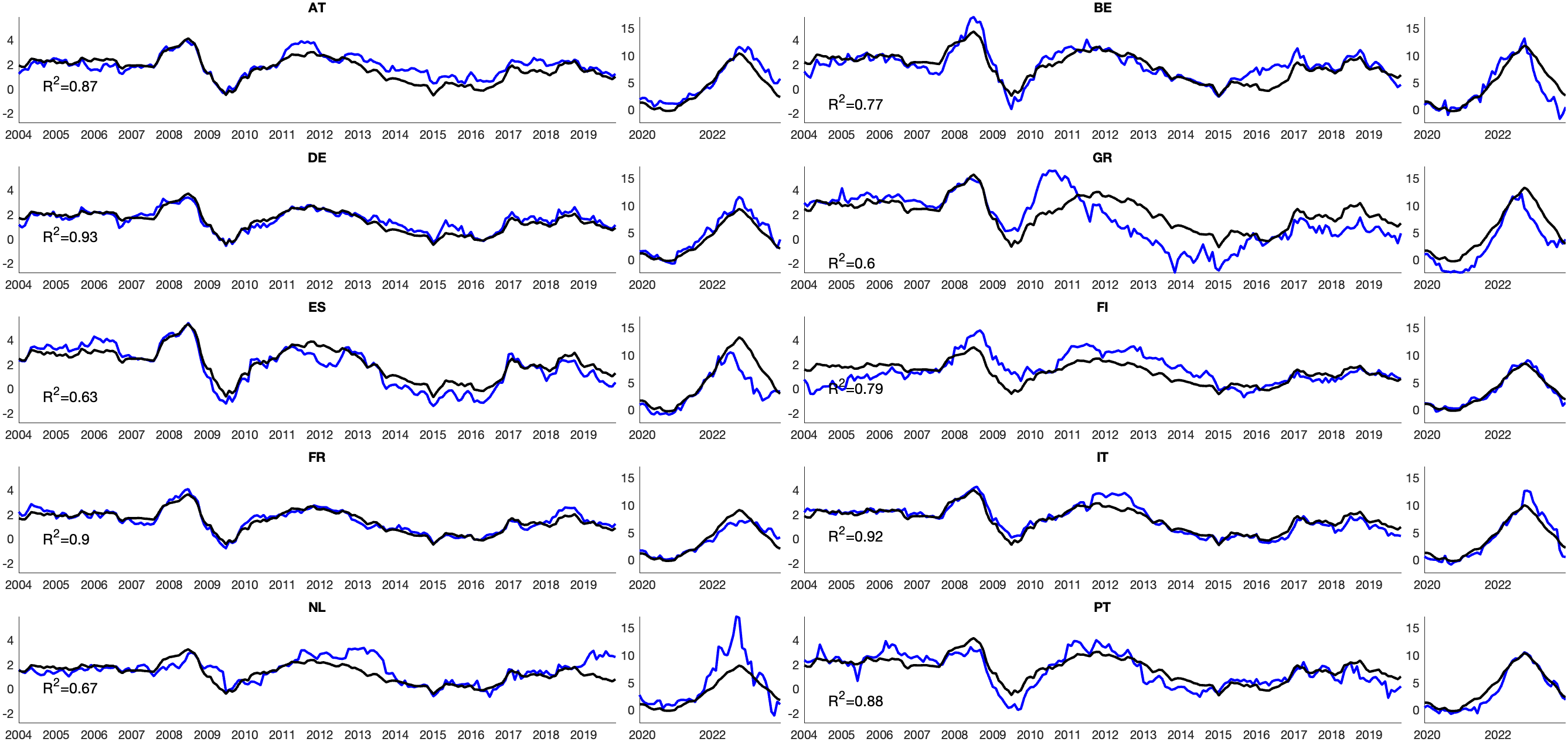}
		\caption{Countries' exposure to common cycle for inflation. The solid blue line depicts actual inflation and the solid black line portrays the common euro cycle for inflation multiplied by the corresponding country-specific factor loadings. $R^{2}$ denotes the R-squared.} \label{fig_2}
    \end{figure}
\end{center}
\vspace{-40pt}

Figure \ref{fig_2} depicts the actual country-level time series for inflation ($x_{i,t}^{INF}$) and the exposure to the euro cycle for prices, as estimated in 
\eqref{eqn_3}. In all countries, the euro area cycle for inflation accounts for at least 60\% of the variation. In AT, DE, FR, IT and PT the inflation cycle accounts for more than 87\% of the variation, with the $R^{2}$ in DE reaching 93\%. In Southern Europe, GR experienced heightened inflationary pressures in the wake of the financial crisis. However, from 2012 until 2023, there was a consistent reduction in exposure to and downward deviation from the euro area cycle. In the remaining Southern economics, this pattern is only marginally evident in ES, whereas IT and PT at times exhibited a positive deviation from the cycle. In FI and AT, realized inflation remained consistently above the fitted exposure to the euro cycle. With the exception of NL, the recent surge in inflation rates is captured to a considerable extent by the common cycle. In contrast to the common cycle for output, the deviations from the inflation cycle cannot be attributed to a group of countries, such as the Southern economies. Indeed, temporary upward or downward deviations appear to be related to the idiosyncratic component, which may be driven by domestic drivers. This is reflected, in part, in the changes in volatility (see Figure A.2). \\
In conclusion, both Figure \ref{fig_1} and \ref{fig_2} present evidence that business cycles for output and inflation in the euro area are largely synchronized business cycles. Notwithstanding the transitory economic underperformance in the Southern economies and minor country-specific divergences from the common inflation cycle, our parsimonious factor model elucidates a substantial proportion of fluctuations in the euro area member countries. In the appendix, Figures \ref{fig_A1} and \ref{fig_A2} illustrate the idiosyncratic residual volatilities $\sigma_{i,t}^{OUT}$ and $\sigma_{i,t}^{INF}$. The volatilities exhibited the largest upswings after the outbreak of the pandemic in 2020 for both country output and inflation. Additionally, the financial crisis increased the importance of the idiosyncratic components in some countries, particularly in GR. As detailed in section 2.1, the implementation of time-varying idiosyncratic components enables the model to identify periods in which domestic forces exert a dominant influence on exposure to the common cycles.

\subsection{Effects on Common Cycles for Output and Inflation}

As previously stated in the introduction, the common cycles can be interpreted economically, as the factors are restricted to only load on the corresponding country time series for output and inflation. In accordance with \citeA{do08} and \citeA{mr20}, the SVAR equation enables the estimation of impact of monetary policy on the common euro cycles for output and inflation.

\begin{center}
    \begin{figure}[H]
        \includegraphics[height=10\baselineskip,width=0.99\textwidth]{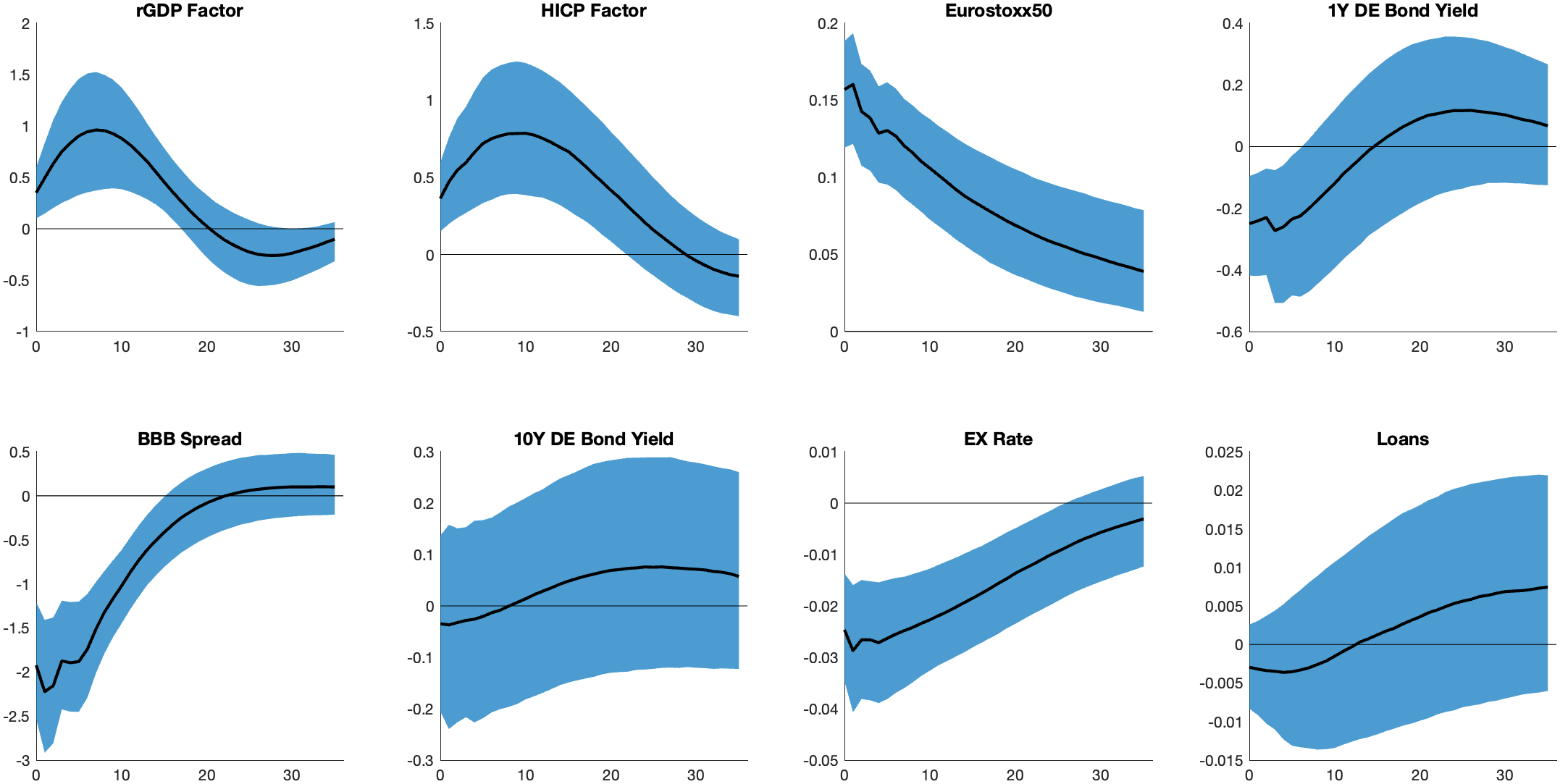}
		\caption{Median impulse responses to an expansionary monetary policy shock. The solid black line depicts the median and the shaded blue area the  68\% credible bands.} \label{fig_5}
    \end{figure}
\end{center}
\vspace{-50pt}

Figure \ref{fig_5} illustrates the impulse responses of variables in the SVAR to an expansionary monetary policy shock. The initial responses are normalized such that the German one-year government bond yield is lowered by 0.25 in $h=0$. Both the common co-movements for output and inflation exhibit a positive reaction for a sustained period of time to a monetary easing. Additionally, both the output and inflation factors follow a hump-shaped pattern, responding the greatest nine months after the shock. The inflation factor exhibits a positive median response for a horizon of $h=30$, while the output factor exhibits a positive median response for a horizon of $h=20$. The monetary expansion increases the Eurostoxx50 which is consistent with the identification of a \emph{pure} monetary policy shock that is untangled from any information shock (see section 3.2 and \citeA{jk20}). The expansionary shock results in a reduction of the BBB spread, thereby facilitating more favourable borrowing conditions in the financial markets. The depreciation of the euro represents another transmission channel. The German ten-year government bond yield and loans to non-financials do not react significantly.\footnote{Restricting the sample to the pre-COVID period heightens the importance of the loans.} 

\subsection{Transmission of Monetary Policy via Common Cycles}

The following section examines the transmission of monetary policy shocks via the common euro cycles to the euro area member countries. We present the country-level impulse responses of output growth and inflation to an expansionary ECB policy shock instrumented by the \emph{pure} policy surprises following \citeA{j22} and accompanied by conventional sign restrictions.\footnote{In the online appendix, we portray the joint impulse responses instead of the point-wise responses following the novel approach of \citeA{ik22}.} Using the impulse responses of the output and inflation cycles, which were estimated via the SVAR and depicted in the previous section, we recover the country responses through the factor equations. This is done by multiplying the responses of the cycles by the corresponding factor loadings. \\

\begin{center}
    \begin{figure}[H]
        \includegraphics[height=12\baselineskip,width=0.99\textwidth]{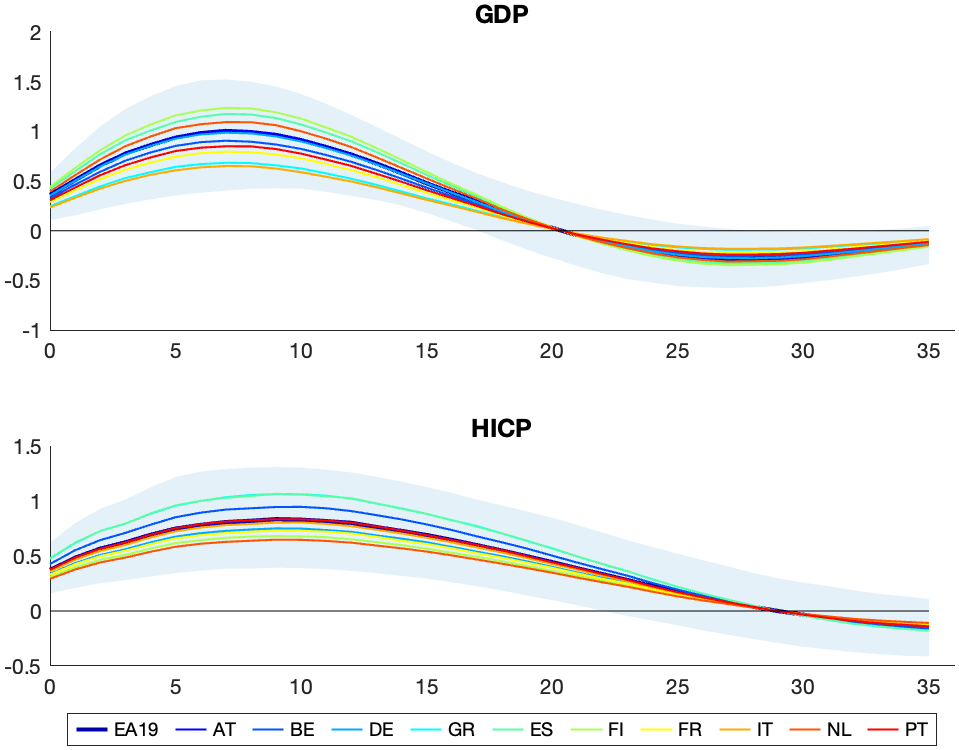}
		\caption{Country-level median impulse responses of country-specific output growth and inflation to an expansionary monetary policy shock. The solid blue line depicts the median responses of the euro area aggregate (EA19) and the shaded light blue areas the 68\% credible bands. The other coloured lines portray the median responses of the euro area member countries.} \label{fig_6}
    \end{figure}
\end{center}
\vspace{-50pt}

Figure \ref{fig_6} presents the country-level responses to an expansionary monetary policy shock identified in the aforementioned SVAR. The solid dark blue lines represent the median responses of the euro area aggregate (EA19), while the shaded light blue areas indicate the corresponding \nth{16} and \nth{84} percentiles of the posterior distribution of the impulse response function of the EA19. The remaining colored solid lines illustrate the median responses of the ten euro area member countries.\footnote{The credible bands of country-specific impulse responses can be found in the online appendix.} Previously stated, the responses are recovered from impulse responses of the common cycles for output growth and inflation, as well as the country-specific factor loadings (see Equation \eqref{eqn_3}). \\
An expansionary monetary policy surprise is observed to enhance output for the EA19 aggregate and for all ten euro countries under consideration. By construction, the country responses reach their maximum at the same time as the common cycle, as illustrated in Figure \ref{fig_5}. In general, the countries exhibit a very comparable exposure to the common output cycle, as indicated by their factor loadings. Nevertheless, it is evident that IT, GR, FR, and PT demonstrate a comparatively weaker growth trajectory in relation to the euro area aggregate. Conversely, FI, ES and NL demonstrate a more pronounced growth trajectory than the aggregate. In terms of prices, the monetary easing has the effect of increasing inflation, with a similar magnitude as output growth. Additionally, inflation rates across countries exhibit a notable degree of homogeneity. It is noteworthy that GR (and AT) exhibit a more pronounced increase in prices than the aggregate. Prices in NL and FI increase by less than their EA19 counterpart.

\subsection{Common Cycles vs. Country-Specific Channels}

In this section, we extend the baseline model by allowing for a direct country-specific exposure to financial variables in addition to the propagation mechanism of the common cycles. The results of the previous section indicate that monetary policy is propagated homogeneously across euro area member countries via the cycles. As previously stated in the introduction, the propagation of monetary policy via the cycles reflects the degree to which monetary policy impacts country output and inflation due to synchronized economies. However, in the factor equations of our baseline model in \eqref{eqn_3}, differences could only arise from a differing exposure to the common cycles, while a direct country-specific impact of the macro-financial variables $\boldsymbol{z}_{t}$ remained muted. Therefore, any heterogeneity in country responses could merely be explained by ECB's inability to steer aggregate demand and prices to the same extent across countries. As the literature indicates that the malfunctioning of country-specific financial markets impedes the homogeneous transmission of monetary policy, it is essential to consider these potential drivers of heterogeneity \cite{ecb17}. \\
Accordingly, we extend our model in \eqref{eqn_3} to permit the variables in $\boldsymbol{z}_{t}$ to have a direct impact on the country's output and inflation via the factor equations. In other words, monetary policy shocks can now affect country-level output and inflation through the common cycles and via country-specific exposures to the financial variables.\footnote{Note that we set $\boldsymbol{\lambda}_{EA19}^{OUT,\boldsymbol{z}}$ and $\boldsymbol{\lambda}_{EA19}^{INF,\boldsymbol{z}}$ to 0 to ensure statistical identification of $f_{t}^{OUT}$ and $f_{t}^{INF}$ (see section 2.1).} 
\vspace{-20pt}
\begin{eqnarray}
    x_{i,t}^{OUT} &=& \lambda_{i}^{OUT} f_{t}^{OUT} + \boldsymbol{\lambda}_{i}^{OUT,\boldsymbol{z}} \boldsymbol{z}_{t} + e_{i,t}^{OUT}, \label{eqn_20} \\
    x_{i,t}^{INF} &=& \lambda_{i}^{INF} f_{t}^{INF} + \boldsymbol{\lambda}_{i}^{INF,\boldsymbol{z}} \boldsymbol{z}_{t} + e_{i,t}^{INF}, \label{eqn_21}
\end{eqnarray}
where $\boldsymbol{\lambda}_{i}^{OUT,\boldsymbol{z}}$ and $\boldsymbol{\lambda}_{i}^{INF,\boldsymbol{z}}$ represent the direct impact of the macro-financial variables on the member countries. Accordingly, we allow the common policy shock to engender heterogeneity in the responses through both the common co-movements, e.g. the common cycles, and the country-specific exposure to financial markets. We then recover the country-specific responses by multiplying the impulse responses of the SVAR with the corresponding factor loadings.

\begin{center}
    \begin{figure}[H]
        \includegraphics[height=12\baselineskip,width=0.99\textwidth]{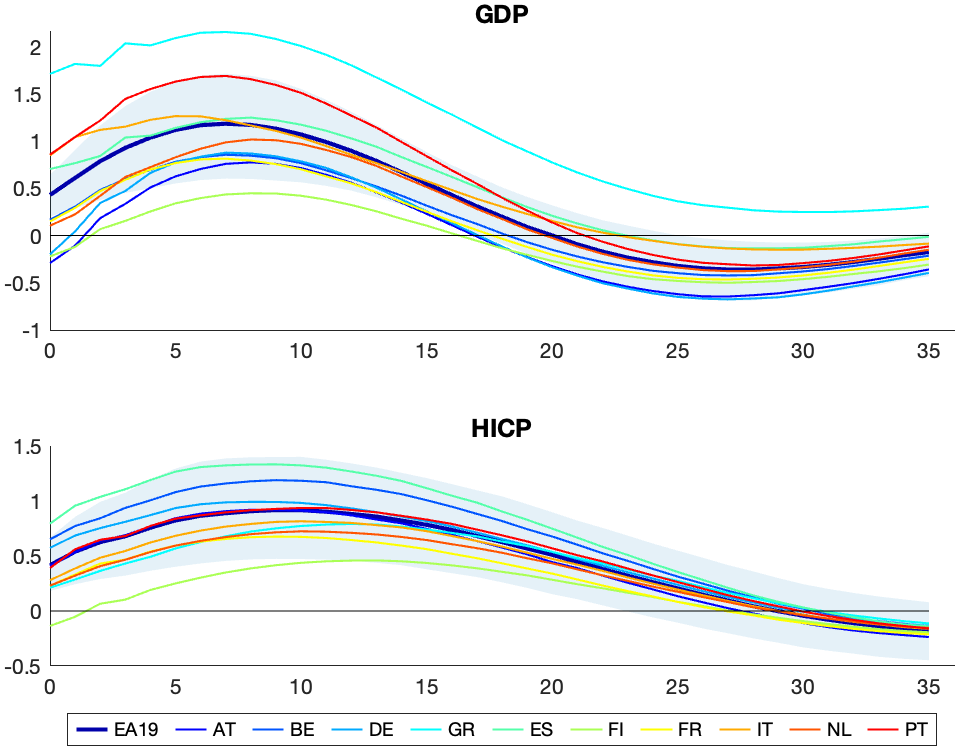}
		\caption{Country-level median impulse responses of country-specific output growth and inflation to an expansionary monetary policy shock. The solid blue line depicts the median responses of the euro area aggregate (EA19) and the shaded light blue areas the 68\% credible bands. The other coloured lines portray the median responses of the euro area member countries.} \label{fig_7}
    \end{figure}
\end{center}
\vspace{-50pt}

Figure \ref{fig_7} illustrates the country-level responses to an expansionary monetary policy shock identified in the aforementioned SVAR. As in the preceding figure, monetary easing leads to an increase in output for the euro aggregate and all member countries under consideration. In comparison to the baseline model with a muted direct country-specific impact via financial variables, the responses exhibit a considerably greater degree of heterogeneity across countries. Once more, GR outperforms the euro aggregate. PT also demonstrates consistent growth in excess of the aggregate. FI and other Northern economies grow by less than their Southern counterparts. In contrast with the results presented in Figure \ref{fig_6}, the country responses require a considerably longer period of time to converge back to the same growth rate in output. \\
As in the previous baseline model, a monetary policy shock raises inflation rates across all member countries. However, the responses are less heterogeneous than those observed for output. Nevertheless, the patterns of the responses are more heterogeneous once we allow for the direct impact of financial variables. In this case, FI, GR, NL and IT consistently demonstrate lower price increases than other member countries, while ES and BE exhibit inflation rates above the euro aggregate. \\
Therefore, the inclusion of the country-specific exposure to financial variables introduces heterogeneity to country-specific responses of output and inflation. The following figure presents a decomposition of the total effects of monetary policy, distinguishing between the propagation via the common cycles and the country-specific channels. This allows for the visualization of the heterogeneity in the extended baseline model, which can be attributed to the country-specific channels. In other words, the homogeneous transmission via common cycles that reflect the synchronicity of the economies persists across member countries when the direct effects of the macro-financial variables are taken into account.

\begin{center}
    \begin{figure}[H]
    \begin{minipage}{.5\textwidth}
      	\includegraphics[width=0.99\textwidth]{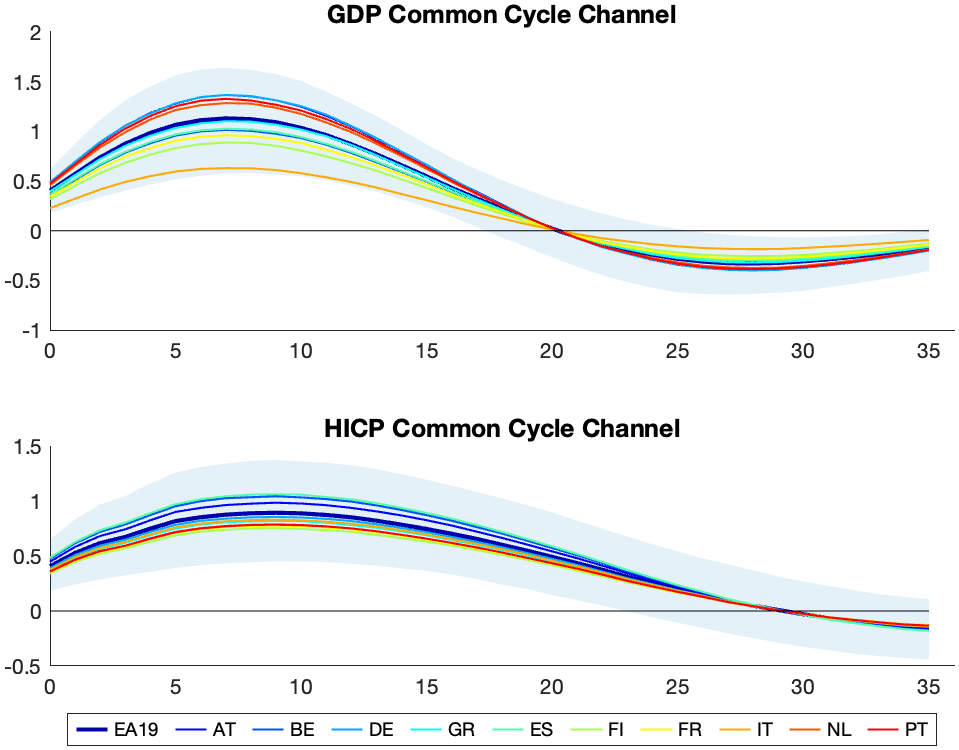}
       \end{minipage}
        \begin{minipage}{.5\textwidth}
      	\includegraphics[width=0.99\textwidth]{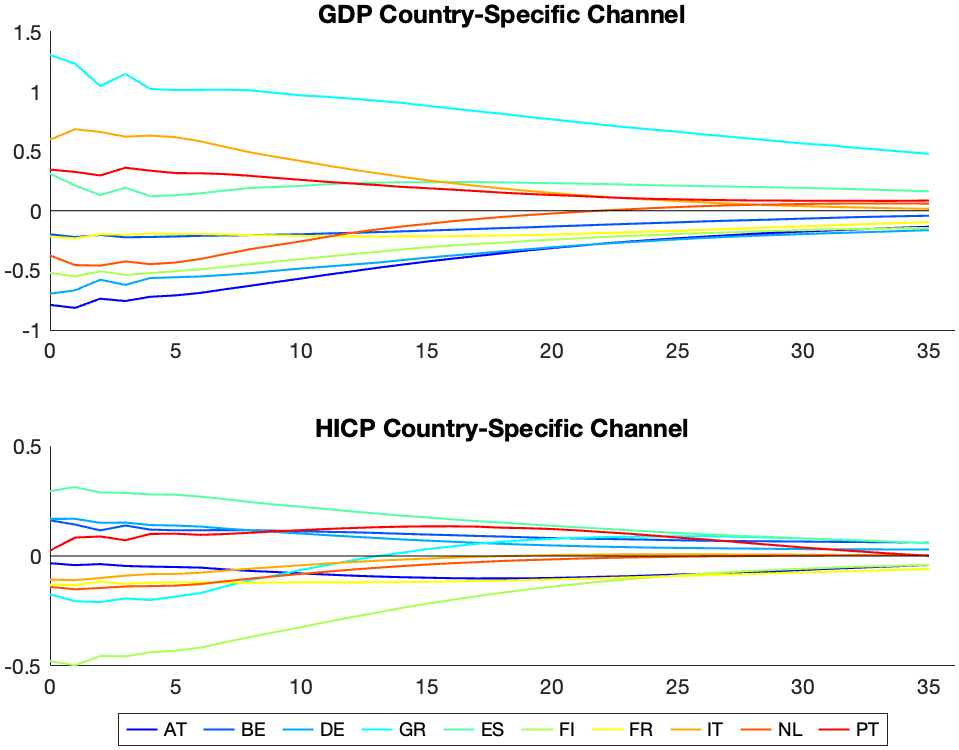}
       \end{minipage}
    
		  \caption{Country-level median impulse responses of country-specific output growth and inflation to an expansionary monetary policy shock. The left-hand side plots show the propagation via the common cycles and the right-hand side plots show the direct impact via the country-specific channels. The solid blue lines depicts the median responses of the euro area aggregate (EA19) and the shaded light blue areas the 68\% credible bands.} \label{fig_8}
    \end{figure}
\end{center}
\vspace{-50pt}

Figure \ref{fig_8} depicts the country responses via the common cycles on the left-hand side and the effects via the country exposure to the financial variables on the right-hand side. The total responses, therefore, are the sums of the two propagation mechanisms as illustrated in the previous figure. \\
With regard to output, it is evident that the overall heterogeneity is attributable to country-specific exposure to the financial variables. The responses via the common cycles are analogous to the results from our baseline model with muted direct effects, albeit of a larger magnitude (see Figure \ref{fig_6}). Once more, the expansion in output in IT (and FI) is less pronounced than in the other countries. The countries with the greatest exposure to the common cycle are DE and PT. The transmission via country-specific channels is characterised by considerable heterogeneity. It is noteworthy that the impact of the policy shock is amplified via the country-specific channels in the Southern economies of GR, IT, ES and PT. In the remaining countries, the efficacy of monetary policy is constrained by the country-specific exposure to financial markets. With regard to inflation, the policy shock is once more transmitted in a homogeneous manner via common cycle (see Figure \ref{fig_6}). Again, the direct channels create a discrepancy between the total country responses. In FI and GR, the transmission for inflation is impeded. In ES, PT, DE and BE, the country-specific exposures to financial markets facilitate the overall transmission. \\
As previously stated, the results demonstrate that the ECB is capable of effectively increasing aggregate demand and inflation via the common cycles that reflect the high degree of integration of euro member economies. Moreover, the findings suggest that the relationship between output and inflation exhibits comparable patterns across countries, indicating a degree of a comparability in the Phillips curves. In light of the effects induced via country-specific exposure to financial variables, it is evident that domestic environments that facilitate similar monetary policy outcomes cannot be argued for. Indeed, the findings encourage the formulation of policies that mitigate the countries' disparate exposures to the dynamics of the financial markets. While the real economy displays considerable synchronicity, the exposure to financial markets appears to be marked by prevailing non-convergence. 

In the preceding figures, we made reference to the median responses in order to identify and examine the heterogeneities that exist across country-specific responses, the common cycles, and the country-specific exposure to financial markets. To provide a further measure for the degree of heterogeneous responses, the coefficients of variation as proposed by \shortciteA{cdm22} are estimated. The coefficient of variation is a measure of the standard deviation of responses across countries with respect to the average of country responses at a given horizon. To facilitate comparison across variables, the coefficients are scaled such that the average responses are set to 1.\footnote{We differ from \shortciteA{cdm22} and \shortciteA{ggs23} and use the average of country responses as benchmark as we do not estimate responses via the financial variables for the euro area aggregate. In appendix B, we present the coefficients of variation using the euro area aggregate as benchmark and confirm our results.} It is important to note that our model allows for a comparison of the degree of heterogeneity for the total responses and for the responses via the common cycles and country-specific channels.
\begin{table}[H]
	\centering
  \begin{threeparttable}
	\begin{tabular}{lcccc}
		\toprule
	\textbf{Response} & $\mathbf{h=0}$ & $\mathbf{h=6}$  & $\mathbf{h=12}$ & $\mathbf{h=24}$ \\
		\midrule
        Aggregate & 0.11 & 0.1 & 0.08 & 0.06 \vspace{-5pt} \\
         & (0.09, 0.14) & (0.07,  0.13)  & (0.06, 0.11) & (0.04,0.09) \vspace{-5pt} \\ \midrule
        Common Cycle & 0.02 & 0.04 &  0.03 & 0.01 \\
		    & (0.01, 0.02) & (0.02, 0.06) & (0.02, 0.05) & (0.00, 0.02) \vspace{-5pt} \\ \midrule
		Country Channels  & 0.12 & 0.1 & 0.08 &  0.06 \vspace{-5pt} \\
		 & (0.09, 0.14) & (0.08, 0.13) & (0.06, 0.11) & (0.03,0.09) \vspace{-5pt} \\
		\bottomrule
	\end{tabular}
 \begin{tablenotes}[flushleft]\footnotesize
    \linespread{.5}\small
    \item\hspace*{-\fontdimen2\font}\note The Table depicts the coefficients of variation for the responses of country output at $h=0$, $h=6$, $h=12$ and $h=24$. For each retained draw of the Gibbs sampler, we compute the coefficients of variation of the responses  and calculate the medians and the 68\% credible bands. The coefficients are rounded to two decimal digits. Note that we use the mean response of the countries as benchmark in order to enable the estimation of the coefficients of variation for the country-specific channels.
    \end{tablenotes}
     \end{threeparttable}
	\caption{Coefficients of Variation - Output}
	\label{tab_3}
\end{table}
\vspace{-10pt}
Table \ref{tab_3} presents the coefficients of variation for output at horizons $h=0$, $h=6$, $h=12$, and $h=24$. The degree of heterogeneity is largest on impact and then decreases for both the aggregate response and the sub-aggregate responses via the country-specific exposure to financial markets. The dispersion of the responses via the common cycle is greatest at $h=6$. Most notably, the degree of heterogeneity is markedly diminished for the common cycle across all horizons, thereby corroborating the findings presented earlier.
\begin{table}[H]
	\centering
  \begin{threeparttable}
	\begin{tabular}{lcccc}
		\toprule
	\textbf{Response} & $\mathbf{h=0}$ & $\mathbf{h=6}$  & $\mathbf{h=12}$  & $\mathbf{h=24}$ \\
		\midrule
        Aggregate &  0.05 &     0.05 & 0.04 &   0.02 \vspace{-5pt} \\
         & (0.04,  0.06) & (0.04, 0.06)  & (0.03, 0.05) & (0.01, 0.03) \vspace{-5pt} \\ \midrule
        Common cycle & 0.01 & 0.02 &  0.02 & 0.01 \\
		    & (0.00, 0.01) & (0.01, 0.03) & (0.01, 0.03) & (0.00, 0.01) \vspace{-5pt} \\ \midrule
		Country channel  &  0.04 & 0.04 & 0.03 &  0.02 \vspace{-5pt} \\
		 & (0.03, 0.05) & (0.03, 0.05) & (0.02, 0.04) & (0.01, 0.03) \vspace{-5pt} \\
		\bottomrule
	\end{tabular}
 \begin{tablenotes}[flushleft]\footnotesize
    \linespread{.5}\small
    \item\hspace*{-\fontdimen2\font}\note The Table depicts the coefficients of variation for the responses of country inflation at $h=0$, $h=6$, $h=12$ and $h=24$. For each retained draw of the Gibbs sampler, we compute the coefficients of variation of the responses  and calculate the medians and the 68\% credible bands. The coefficients are rounded to two decimal digits. Note that we use the mean response of the countries as benchmark in order to enable the estimation of the coefficients of variation for the country-specific channels. 
    \end{tablenotes}
     \end{threeparttable}
	\caption{Coefficients of Variation - Inflation}
	\label{tab_4}
\end{table}
\vspace{-10pt}
Table \ref{tab_4} presents the coefficients of variation for inflation at horizons $h=0$, $h=6$, $h=12$, and $h=24$. The coefficients indicate a reduction in the degree of heterogeneity over the horizons for both the aggregate responses and the responses via the  country-specific exposure to financial markets. Once more, the greatest degree of heterogeneity via the common cycle is observed at $h=6$. Moreover, the responses via the common cycle demonstrate a diminished degree of heterogeneity which serves to reinforce the preceding results. \\
As outlined in the introduction, the heterogeneous outcomes of monetary policy are frequently observed to correlate with differences in macroeconomic characteristics across the euro area (\citeA{g15}, \citeA{bg18} \shortciteA{msv22}, \shortciteA{agkl22}, \shortciteA{cdm22}). In the following exercise, we seek to establish a correlation between a number of structural characteristics and the peak responses for output that are induced via both the country-specific channel and the common cycle channel. This analysis allows us to scrutinize whether differences in the macroeconomic environment can be linked to the two propagation mechanisms. In this manner, we consider the following characteristics. GDP per capita is employed as a proxy for general economic development, the public debt to GDP ratio is used as a proxy for the fiscal constraints of the member countries, and the unemployment rate is used as a proxy for labor market conditions. Furthermore, the ease of doing business index, which represents the regulatory environment of the economies in question, is considered. In examining the characteristics of the housing market, four indicators are considered. First, we employ the share of flexible mortgage rate contracts as monthly payments adjust with monetary policy shocks (\shortciteA{cms13}, \shortciteA{fksv20}). Second, we correlate the peak responses with the home ownership rate, the home ownership rates with mortgages and without mortgages. Among other factors, monetary policy shocks may transmit through the collateral channel for homeowners \shortcite{chik19}.\footnote{We resort to annualized data within our baseline sample ranging from 2003 to 2023, if available, and take the average over the years. In the online appendix, we provide a detailed description on the data and its transformation.}
\begin{table}[H]
	\centering
 \small
  \begin{threeparttable}
	\begin{tabular}{lccc}
		\toprule
	        & Country Channel  & Common Channel & Country$\vert$Common  \\
		\midrule
        
        GDP per capita & $-0.78^{***}$ & $0.01$ &  $-0.81^{***}$ \vspace{-5pt} \\ 
        Public Debt to GDP ratio &  $0.9^{***}$ & $-0.38$ & $0.89^{***}$  \vspace{-5pt} \\
        Unemployment Rate & $0.72^{**}$ &  $-0.32$ & $0.70^{**}$ \vspace{-5pt} \\
        Ease of Doing Business &  $0.89^{***}$ & $-0.35$ & $0.88^{***}$ \vspace{-5pt} \\
        Flexible Mortgage Rate & $0.38$ & $-0.14$ & $0.36$ \vspace{-5pt}\\
        Home Ownership &  $0.56^{*}$ & $-0.51$ & $0.51$ \vspace{-5pt}\\
        Home Ownership w. Mortgage & $-0.57^{*}$ & $0.26$ & $-0.53$ \vspace{-5pt}\\
        Home Ownership w/o Mortgage &  $0.80^{***}$ & $-0.51$ & $0.81^{***}$ \vspace{-5pt}\\
		\bottomrule
	\end{tabular}
 \begin{tablenotes}[flushleft]\footnotesize
    \linespread{.5}\small
    \item\hspace*{-\fontdimen2\font}\note The columns \emph{Country Channel} and \emph{Common Channel} show the correlation coefficients between the peak responses of output induced via the country-specific channels and via the common cycle channel respectively, and the structural variables. The column \emph{Country$\vert$Common} shows the semi-partial correlation coefficients between peak responses of output induced via the country-specific channels and the structural variables which controls for the correlation coefficients of the second column ($^{*} p < 0.1$, $^{**} p < 0.05$, $^{***} p < 0.01$).  Calculations of \emph{GDP per capita}, \emph{Public Debt to GDP}, \emph{Ease of Doing Business}, \emph{Unemployment Rate} are based on data from the World Bank, \emph{Flexible Mortgage Rate} are retrieved from Eurostat and \emph{Home Ownership}, \emph{Home Ownership w. Mortgage} and \emph{Home Ownership w/o Mortgage} are downloaded from the ECB database. 
    \end{tablenotes}
     \end{threeparttable}
	\caption{Correlations between Channel Responses and Structural Characteristics}
	\label{tab_5}
\end{table}
The results presented in Table \ref{tab_5} reveal two key findings. Firstly, the signs (and magnitudes) of correlation with respect to the country-specific channels in the second column are in accordance with the findings of the existing literature. Secondly, and of particular significance, the structural characteristics significantly correlate with the country-specific channel, whereas the correlation coefficients with the common cycle channel in the third column are insignificant. The fourth column corroborates this finding, as the semi-partial correlation of the responses via the country-specific channel with the characteristics is estimated while controlling for the common channel. Therefore, the differences in macroeconomic environments across countries may be attributed to the heterogeneity induced via country-specific channels, rather than to the minor differences in exposure to the common cycle. This supports our finding that country heterogeneity in response to monetary policy shocks is driven by country-specific features, which at the same time do not appear to impede the integration of business cycles across the euro area. \\
In conclusion, the allowance for a direct transmission to the euro area member countries via the financial variables results in a greater degree of heterogeneity in the responses to a monetary policy shock. Nevertheless, our findings indicate that euro area cycles persist as homogeneous propagation mechanisms. In consideration of the theory on currency unions, the potential decoupling of business cycles among member countries does not represent a major driver contributing to fragmentation within the euro area. In addition, the member countries are predominantly affected by monetary policy shocks in a homogeneous manner through the common cycles. However, differences in the country-specific channels result in heterogeneity in the overall policy outcomes. Furthermore, the heterogeneity in the country  channels can be linked to differences in macroeconomic characteristics across member countries. However, deviations from the common cycle do not appear to be driven by these characteristics. From a policy perspective, the distinction between the two propagation mechanisms is of great importance in identifying the sources of heterogeneity and uncovering the corresponding underlying structural characteristics.

\subsection{Robustness Analysis}

Further robustness tests are performed to assess the sensitivity of the results to different model specifications. These alternative approaches are presented in the online appendix and are briefly discussed in this section.\footnote{In the online appendix, we present a selection of the results obtained in the robustness analysis. Further results can be obtained from the authors on request.} \\
First, we re-estimate the extended model from section 4.4 and follow a more agnostic identification of the monetary policy shock, allowing the response of the cycle for inflation to remain unrestricted. The main results are unaffected as heterogeneity in the transmission is characterized by the country-specific channels. Subsequently, the external instrument is replaced with the poor man's proxy proposed by \citeA{jk20}.\footnote{\citeA{jk20} set changes on announcement days to $0$ if the signs of changes in the three-month OIS swap rate and the Eurostoxx50 are identical.} Additionally, the results are presented using only the three-month OIS swap rate and the first principal component of the OIS swap rates with maturities ranging from one month to one year. We then replace the German one-year government bond yield with the euro area average of the one-year government bond yields, and the respective sign restrictions are imposed. The heterogeneities within and across the propagation mechanisms are similar to the above results, but the effects are considerably larger in size. As the euro area average yield does not constitute a risk-free policy variable, the transmission via the BBB bond spread is slightly confounded. Furthermore, we set a less informative prior on the measurement error $\epsilon_{t}^{k'}$ and obtain the same results. \\
Next, the interpolated monthly output growth rate is replaced with the growth rate in industrial production. It should be noted that the use of industrial production for the purpose of interpolating output is not without its disadvantages, given that the importance of the industrial production for output varies considerably between countries. Nevertheless, the same caveat applies to the interpretation of the results for industrial production as proxy for economic activity. The results demonstrate that the euro area cycle for industrial production aligns with the country-specific production series to a considerable extent. The monetary policy shock has a more pronounced impact on the common cycle for industrial production than on output. Furthermore, the propagation mechanism of the euro cycle for industrial production displays a higher degree of heterogeneity than that of the common output cycle. This evidenced by the markedly smaller effects observed in GR, PT and ES. Nevertheless, this finding is consistent with existing literature (\citeA{g14} and \shortciteA{msv22}). Furthermore, additional results for industrial production are presented, employing smoothing and other outlier detection algorithms on the original time series. \\
In light of the unprecedented circumstances of the pandemic and the subsequent years, we stop the sample period in December 2019 and re-estimate the model. The results are similar to those of the baseline model, although the cycles for output and inflation exhibit a diminished response. Furthermore, the transmission is driven more by loans to non-financial corporations than by the BBB bond spread and the stock market. The transmission to the member countries via the output euro area cycle is less homogeneous (see GR). For some countries, output growth is considerably dampened by their exposures to financial markets, especially for GR. As in the baseline model, inflation is homogeneously propagated via the euro area cycle, and the country-specific exposure to financial variables adds heterogeneity to the total responses. Nevertheless, ending the sample in December 2019 indicates that the aftermath of the GFC was marked by a disintegration of the member countries. \\
Furthermore, we adhere to the approach by \shortciteA{adp21}, which allows for more disproportionate and richer dynamics within the country responses. To this end, we set the lag orders in the factor equation to one, two and three respectively. Despite, euro member countries may either lead or follow the common cycles. The results reported in the online appendix indicate a more heterogeneous pattern when the lagged effect for both propagation mechanisms is considered. Nevertheless, the homogeneous responses through the euro area cycles persist. A more detailed description of the model extension can be found in the online appendix. \\
Additionally the online appendix presents the results for further model specifications. The lag order in the SVAR equation is altered to $L=3$ and $L=12$. Subsequently, the annual growth is replaced with the log annual growth rates and the log growth rates proposed by \citeA{bh23}. Also, we re-estimate the model with time series for interpolated output and industrial production that do not control for outliers. Here, we demonstrate that employing the original time series results in an over-attribution of the economic downturn in second quarter of 2020 to the monetary policy shock, despite the fact that multiple shocks occurred simultaneously. Finally, we present impulse responses in accordance with the approach by \citeA{ik22}. In lieu of estimating pointwise posterior medians and other percentiles which may be absent from the set of impulse responses, the authors compute a joint credible set of responses that accounts for the joint uncertainty in the interval bands. Herein, we depict the 68\% joint impulse responses color-sorted by the maximum sum of the responses of EA19. Collectively, the primary results are robust to alternative model specifications.

\section{Conclusion}

We employ a novel Bayesian proxy FAVAR with sign restrictions to provide new insights into how ECB's monetary policy shocks are transmitted to individual euro area member countries and the underlying reasons for the observed asymmetry in these responses across countries.
We establish a dynamic factor model with stochastic volatility to estimate economically interpretable cycles for output and inflation while allowing for country-specific idiosyncratic fluctuations. In the SVAR, we model the joint dynamics of the cycles, a set of macro-financial variables, and an external instrument. To address the issue of overfitting, we impose data-driven adaptive asymmetric Minnesota-type shrinkage priors, as proposed by \citeA{c21} and \citeA{h24}. Furthermore, we identify a \emph{pure} monetary policy shock that is purged from an information shock in the style of \citeA{j22} using high-frequency changes in swap rates and asset prices. To sharpen and strengthen inference, we impose sign restrictions on the contemporaneous relationships of the endogenous variables, as proposed by \citeA{Uhlig2005} and \shortciteA{bbh23}. Our proxy and sign-restricted FAVAR model combines several methodological advantages. It simultaneously shrinks a high-dimensional dataset to two factors and point-identifies a structural shock using both an external instrument and sign restrictions. Furthermore, the posterior distributions are sampled cost-efficiently, as in \citeA{h24}. \\
The present study offers empirical evidence indicating that both output growth and inflation in the euro area member countries exhibit each a common cycle. Country-specific output growth and inflation are explained by up to 93\% of the cycles, thereby capturing significant fluctuations such as the financial crisis and the global pandemic. However, a reduction in exposure to these cycles is observed in Southern European economics in the aftermath of the crisis. It is of pivotal importance to our analysis that our baseline model finds that monetary policy shocks are propagated homogeneously to the member countries via these common cycles. Once direct channels of macro-financial variables on countries' output growth and inflation rates are allowed for in the extended model, the responses exhibit a more heterogeneous pattern. As we separate the aggregate responses to identify the effects via the common cycles and the country-specific exposures to the financial variables, we find that the common cycles continue to serve as homogeneous propagators of monetary policy. In contrast, the country-specific channels either amplify or attenuate the impacts on output and inflation. The coefficients of variation for both the aggregate responses and responses via the common cycles and the country-specific exposure to financial markets variables corroborate our results. The heterogeneity induced via country-specific channels can be linked to structural characteristics of the member countries, whereas the responses via common cycles appear to be detached from these. Our results are highly robust across a range of different modeling choices. \\
The transmission via the common cycles portrays the ECB as being capable of steering both aggregate demand and inflation in the member countries to a comparable degree. Therefore, the homogeneous responses via the cycles indicate the presence of highly synchronized economies and comparable Phillips curves in the member countries. Moreover, in accordance with the theory proposed by \cite{m61}, the integrated cycles for output and inflation in the euro area provide the ECB with a robust foundation for policy making. With regard to the heterogeneous transmission via the country-specific channels, it is evident that a greater degree of structural convergence across the euro area is necessary to further facilitate the implementation of a common monetary policy by the ECB.
 
\clearpage

\small{\setstretch{1.25}  
\addcontentsline{toc}{section}{References}
\bibliographystyle{apacite}
\bibliography{CountryFAVAR.bib}} 
\appendix

\renewcommand{\thesection}{Appendix \Alph{section}}
\setcounter{table}{0}
\renewcommand{\thetable}{\Alph{section}.\arabic{table}}
\setcounter{figure}{0}
\renewcommand{\thefigure}{\Alph{section}.\arabic{figure}}

\newpage

\section{Additional Figures}

\begin{center}
    \begin{figure}[H]
        \includegraphics[height=7\baselineskip,width=0.99\textwidth]{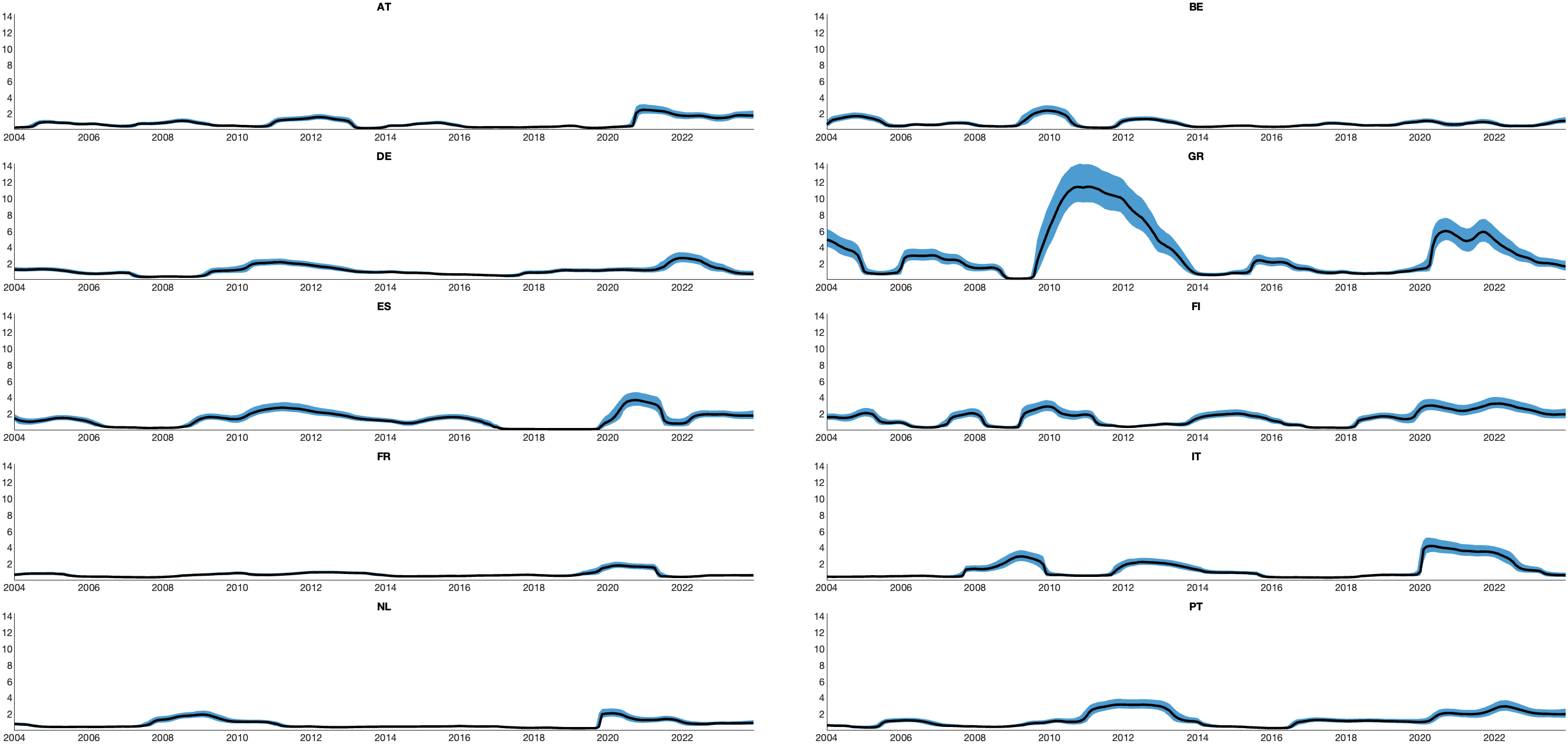}
		\caption{Residual stochastic volatilities of output growth. The solid black line depicts the median and the shaded blue area the 68\% credible bands.} \label{fig_A1}
    \end{figure}
\end{center}
\vspace{-50pt}

\begin{center}
    \begin{figure}[H]
        \includegraphics[height=6\baselineskip,width=0.99\textwidth]{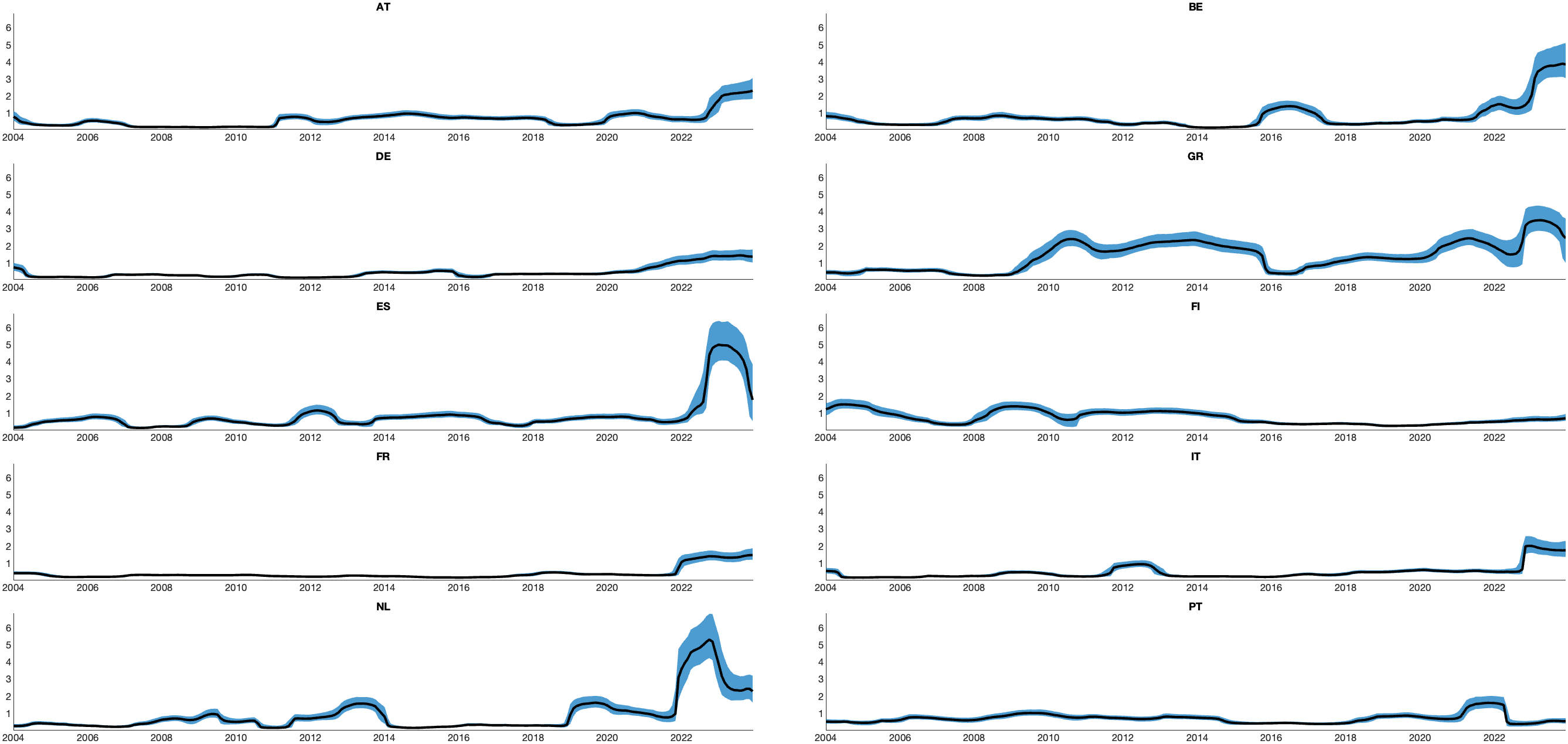}
		\caption{Residual stochastic volatilities of inflation. The solid black line depicts the median and the shaded blue area the 68\% credible bands.} \label{fig_A2}
    \end{figure}
\end{center}
\vspace{-50pt}

\newpage
\section{Additional Tables}

\begin{table}[H]
	\centering
  \begin{threeparttable}
	\begin{tabular}{lcccc}
		\toprule
	\textbf{Response} & $\mathbf{h=0}$ & $\mathbf{h=6}$  & $\mathbf{h=12}$ & $\mathbf{h=24}$  \\
		\midrule
        Aggregate & 0.12 & 0.1 & 0.08 & 0.06\vspace{-5pt} \\
         & (0.09, 0.14) & (0.07, 0.13)  & (0.06,  0.11) & (0.04, 0.09) \vspace{-5pt} \\ \midrule
        Common cycle & 0.02 & 0.04 & 0.03 & 0.01 \\
		    & (0.01, 0.02) & (0.02, 0.06) & (0.02, 0.05) & (0.00, 0.02) \vspace{-5pt} \\
		\bottomrule
	\end{tabular}
 \begin{tablenotes}[flushleft]\footnotesize
    \linespread{.5}\small
    \item\hspace*{-\fontdimen2\font}\note The Table depicts the coefficients of variation for the responses of country output at $h=0$, $h=6$ and $h=12$. For each retained draw of the Gibbs sampler, we compute the coefficients of variation of the responses  and calculate the medians and the 68\% credible bands. The coefficients are rounded to two decimal digits. Note that we use the response of the EA19 as benchmark and therefore, cannot compute the the coefficients of variation for the financial cycle due to the statistical identification procedure (see section 2.1 and 4.4).
    \end{tablenotes}
     \end{threeparttable}
	\caption{Coefficients of Variation - Output}
	\label{tab_app1}
\end{table}
\vspace{-10pt}

\begin{table}[H]
	\centering
  \begin{threeparttable}
	\begin{tabular}{lcccc}
		\toprule
	\textbf{Response} & $\mathbf{h=0}$ & $\mathbf{h=6}$  & $\mathbf{h=12}$ & $\mathbf{h=24}$ \\
		\midrule
        Aggregate & 0.05 & 0.05 & 0.04 & 0.02 \vspace{-5pt} \\
         & (0.04, 0.06) & (0.04, 0.06)  & (0.03, 0.05) & (0.01, 0.03) \vspace{-5pt} \\ \midrule
        Common cycle & 0.01 & 0.02 & 0.02 & 0.01 \\
		    & (0.00, 0.01) & (0.01, 0.03) & (0.01, 0.03) & (0.00, 0.01) \vspace{-5pt} \\
		\bottomrule
	\end{tabular}
 \begin{tablenotes}[flushleft]\footnotesize
    \linespread{.5}\small
    \item\hspace*{-\fontdimen2\font}\note The Table depicts the coefficients of variation for the responses of country inflation at $h=0$, $h=6$ and $h=12$. For each retained draw of the Gibbs sampler, we compute the coefficients of variation of the responses  and calculate the medians and the 68\% credible bands. The coefficients are rounded to two decimal digits. Note that we use the response of the EA19 as benchmark and therefore, cannot compute the the coefficients of variation for the financial cycle due to the statistical identification procedure (see section 2.1 and 4.4).
    \end{tablenotes}
     \end{threeparttable}
	\caption{Coefficients of Variation - Inflation}
	\label{tab_app2}
\end{table}
\vspace{-10pt}

\newpage
\section{Prior Specification}
In this section, we discuss the specification of the prior distributions. First, we present the priors for the factor equation in \eqref{eqn_3}. The priors of the factor loadings $\lambda_{i,p}^{j}$ with $i=1,\ldots$,$N, p=,\ldots,P$ and $j=(OUT,INF)$ for both the common factors and macro-financial variables are kept loose and follow:
\vspace{-10pt}
\begin{eqnarray}
    \lambda_{i,p}^{j} &\sim& N(\mu_{\lambda_{i,p}^{p}},\Sigma_{\lambda_{i,p}^{j}}), \quad \forall i = 1,\ldots,N,\quad \forall p = 0,\ldots,P, \label{eqn_10}  \\
    \lambda_{i,p}^{j,\boldsymbol{z}_{t}} &\sim& N(\mu_{\lambda_{i,p}^{j,\boldsymbol{z}_{t}}},\Sigma_{\lambda_{i,p}^{j,\boldsymbol{z}_{t}}}), \quad \forall i = 1,\ldots,N,\quad \forall p = 0,\ldots,P, \label{eqn_11} 
\end{eqnarray}
with $\mu_{\lambda_{i,p}^{j}}$ and $\mu_{\lambda_{i,p}^{j,\boldsymbol{z}_{t}}} = 0$ and $\Sigma_{\lambda_{i,p}^{j}}$ and $\Sigma_{\lambda_{i,p}^{j,\boldsymbol{z}_{t}}} = 10$. \\
The prior variances of the stochastic volatilities $V_{h_{i,t}^{j}}$ in \eqref{eqn_4} read as:
\vspace{-10pt}
\begin{eqnarray}
    V_{h_{i,t}^{j}} = \tau_{h_{i}^{j}} \lambda_{h_{i,t}^{j}}, \label{eqn_12} 
\end{eqnarray}
for $j=(OUT,INF)$ and where $\tau_{h_{i}^{j}}$ and $\lambda_{h_{i,t}^{j}}$ depict the global and the local shrinkage components. We follow \shortciteA{cps10} and assume a half-Cauchy prior for the components:
\vspace{-10pt}
\begin{eqnarray}
	\sqrt{\tau_{h_{i}^j}} \sim C^+(0,1), \quad \sqrt{\lambda_{h_{i,t}^j}}\sim C^+(0,1). \label{eqn_13} 
\end{eqnarray}
These hierarchical priors define the horseshoe prior and allow for both gradual changes and larger breaks in $V_{h_{i,t}^{j}}$ and therefore, half-Cauchy distributions can depict a more suitable prior than inverse Gamma distributions (\citeA{g06}, \citeA{p21}). \\
Further, we avoid overfitting in the VAR model by implementing an adaptive asymmetric Minnesota-type shrinkage prior for the parameters $\boldsymbol{\Gamma} = (\boldsymbol{\Gamma}_{1},\ldots,\boldsymbol{\Gamma}_{L})$ in \eqref{eqn_7} and \eqref{eqn_8} (\shortciteA{chp20}, \citeA{c21} and \citeA{h24}). This prior combines the strengths of both the commonly employed Minnesota prior (see \citeA{l86} and \shortciteA{bgr10}) and more recent hierarchical priors (see \citeA{c21}). Briefly, the Minnesota prior places more weight to coefficient of own recent lags, while the other coefficients are shrunk to zero. The hierarchical priors which control the degree of shrinkage are obtained in a data-driven fashion instead of being selected a priori. \\
More precisely, let $\boldsymbol{\Gamma}_{ij,l}$ be the (\emph{i,j})-th entry $\boldsymbol{\Gamma}_{l}$ which follows an independent normal prior:
\vspace{-10pt}
\begin{eqnarray}
    \boldsymbol{\Gamma}_{ij,l} &\sim& N(\gamma_{ij,l}, V_{ij,l}), \quad l = 1,\ldots,L, \label{eqn_14} 
\end{eqnarray}
where
\vspace{-10pt}
\begin{eqnarray}
    \gamma_{ij,l} = \begin{cases}
      1, & l = 1, i=j \text{ and } i, \neq 1,2, \\
      0, & \text{otherwise}, \\
    \end{cases} \text{      } V_{ij,l} = 
    \begin{cases}
        \frac{\kappa_1}{l^{2}}, & i = j, \\
        \frac{\kappa_1 \kappa_2 \sigma_{i}^{2}}{l^{2}\sigma_{j}^{2}}, & i \neq j, \nonumber
    \end{cases}
\end{eqnarray}
and where $\sigma_{i}^{2}$ is a scale parameter and equal to the residual variance of an AR(L) model for variable 
$i = 1,\ldots,r$. Note that for $i=1, 2$ we assume $\gamma_{11,1}$ and $\gamma_{22,1}$ to be $0$ since the factors $f_{t}^{OUT}$ and $f_{t}^{INF}$ depict growth rates \shortcite{cky24}. The remaining variables in $\boldsymbol{z}_{t}$ are assumed to follow a random walk. The constants $\boldsymbol{c}$ follow a normal prior with $N(0,100^{2})$. The prior for shrinkage parameters $\kappa_1$ and $\kappa_2$ follow an uninformative prior:
\vspace{-10pt}
\begin{eqnarray}
    p(\kappa_1) \propto c_{1}, \quad p(\kappa_2) \propto c_{2}. \label{eqn_15} 
\end{eqnarray}
As noted in the previous subsection, the non-zero elements in $\boldsymbol{B}$ follow independent (sign-restricted) normal priors. Let $b_{i,j}$ be the (\emph{i,j})-th element $\boldsymbol{\Gamma}_{0}$ in $\boldsymbol{B}$ such that
\vspace{-10pt}
\begin{eqnarray}
    b_{i,j} = \begin{cases}
        N(\beta_{i,j},v_{i,j}), & \text{if } b_{i,j} \text{ is unrestricted}, \\
        N(\beta_{i,j},v_{i,j})\mathds{1}(b_{i,j} > 0), & \text{if } b_{i,j} > 0 \\
        N(\beta_{i,j},v_{i,j})\mathds{1}(b_{i,j} < 0), & \text{if } b_{i,j} < 0 \\
        \delta_0(b_{i,j}), & \text{if } b_{i,j} = 0, \label{eqn_16} 
    \end{cases}
\end{eqnarray}
where $\mathds{1}$ denotes a indicator function and $\delta_{0}(\cdot)$ the Dirac delta function at zero. We set $\beta_{i,j} = 0$ and $\nu_{i,j} = 1$. \\
The (non-zero) parameters in the proxy variable equation ($\Phi_{0,1}$, $\Phi_{0,2}$) are assumed to be independently normal distributed:
\vspace{-10pt}
\begin{eqnarray}
    \Phi_{0,1} \sim N(\phi_{0,1},V_{0,1}), \quad \Phi_{0,2} \sim N(\phi_{0,2},V_{0,2}). \label{eqn_17} 
\end{eqnarray}
We set $\phi_{0,1} = 0, \phi_{0,2} = 0, V_{0,1} = 10$ and $V_{0,2} = 0.01^2$. Doing so, we follow \citeA{ch19}, \shortciteA{arw21} and \shortciteA{bgs22} and impose our prior belief that our proxy variable depicts a relevant instrument. More specifically, with $\phi_{0,2} = 0$ and $V_{0,2} = 0.01^2$ we assume that movements in our proxy $m_{t}$ cannot be attributed to the measurement error to a large degree. Instead, by setting a very loose variance on $\phi_{0,1}$, we allow the structural monetary policy shock $\epsilon_{1,t}^{r'}$ to largely explain the variation in $m_{t}$. We discuss the proxy equation more thoroughly in section 3.2. In the online appendix, we present the results with other assumptions on the informativeness of the proxy variable.

\newpage
\section{The Gibbs Sampler for the FAVAR Model}

Here, we describe the Markov Chain Monte Carlo (MCMC) algorithm which allows to sample from the joint posterior distributions of all parameters and hyperparameters. The algorithm depicts an integration of sampling procedures implemented by \citeA{bbe05}, \shortciteA{cky24}, \citeA{p21} and \citeA{h24}. Further, we borrow from \citeA{ck94}, \shortciteA{ksc98}, \citeA{cj09}, \citeA{v09}, \citeA{ms16}, \citeA{b17}, \shortciteA{ccm19} and \shortciteA{cccm22}. \\
As usual for a Gibbs sampler, initializing values for chosen parameters are required in order to start sampling from the conditional 
posterior of the first parameters of interest. In our case, we set the following initial values.
For the common factors $f_{t}^{OUT}$ and $f_{t}^{INF}$, we set the initial values by the principal components analysis 
and their variance to $10$. The initial values of stochastic volatilities $h_{i,t}^{j}$ are computed by the logarithm of the variance 
of the countries' realizations of $x_{i,t}^{j}$ for output growth and inflation. We add a constant equal to 1 to ensure strictly positive values. The variance $V_{h_{i,t}}^j$ is the product of the initial 
values of $\tau_{h_{i}^j}$ and $\lambda_{h_{i,t}^j}$ which are both set to $1$. The initial values for $\boldsymbol{B}$ is set to an identity matrix $\boldsymbol{I}_{n}$ and the autoregressive matrices $\boldsymbol{A}_{l}$ are set to zero matrices.\footnote{Note that the Markov chain property of the Gibbs sampler ensures that 
the draws are drawn from large parameter space and hence, the initial values do not influence the outcome of the sampled posterior 
distribution in a decisive manner.} \\
To sequentially obtain draws from the joint posterior distributions, we implement the following sampling algorithm:
	\begin{itemize}
		\item[\bf{Step 1}] Draw the factor loadings $\lambda_{i,p}^{j}$ from the posterior distribution in a standard linar Bayesian regression:
        \vspace{-10pt}
        \begin{eqnarray}
            \lambda_{i,p}^{j} \mid x_{i,t}^{j},z_{t},f_t^{j},h_{i,t}^{j} \sim N(\lambda_{i,p}^{j\ast},
            \Sigma^{\ast}_{\lambda_{i,p}^{j}}), \nonumber
        \end{eqnarray}
        with
        \vspace{-10pt}
        \begin{eqnarray}
            \lambda_{i,p}^{j\ast} = \Big( \Sigma^{-1}_{\lambda_{i,p}^{j}} + \frac{1}{h_{i,t}^{j}}Z_{t}'Z_{t} \Big)\inv \Big( \Sigma^{-1}_{\lambda_{i,p}^{j}}\lambda_{i,p}^{j0} + \frac{1}{h_{i,t}^{j}}Z_{t}'x_{i,t}^{j}    \Big) \nonumber
        \end{eqnarray}
        and
        \begin{eqnarray}
            \Sigma^{\ast}_{\lambda_{i,p}^{j}}= \Big( \Sigma^{-1}_{\lambda_{i,p}^{j}} + \frac{1}{h_{i,t}^{j}}Z_{t}'Z_{t}\Big)\inv, \nonumber
        \end{eqnarray}
        where $Z_t = \{ x_{i,t}^{j},\boldsymbol{z}_{t},\ldots, x_{i,t-p}^{j},\boldsymbol{z}_{t-p} \}$ and $j = \text{OUT, INF}$. Note that $Z_t$ varies with $P$ and the inclusion or exclusion of the macro-financial variables of the VAR equation into the factor model. Further, $\lambda_{EA19,0}^{j}$ are set to 1 to ensure identification and $\lambda_{EA19,p}^{j}$ for $p \neq 0$ is set to 0. $\lambda_{i,p}^{j0}$ and $\Sigma^{-1}_{\lambda_{i,p}^{j}}$ denote the priors. See \shortciteA{bbe05} for further details.    
        \item[\bf{Step 2}] Draw the stochastic volatilities $h_{i,t}^{j}$ using the auxiliary mixture sampler of \shortciteA{ksc98} in combination with the precision sampler of \citeA{cj09}:
        
            \item[] In \eqref{eqn_1}, $e_{i,t}^{j}$ is normally distributed with N(0,$\sigma_{i,t}^{j,2}$) and $\sigma_{i,t}^{j,2} = \exp{(h_{i,t}^{j})} \epsilon_{h_{i,t}^{j}}$, where
            \vspace{-10pt}
            \begin{eqnarray}
                h_{i,t}^{j} = h_{i,t-1}^{j} + \xi_{i,t}^j, \quad \xi_{i,t} \sim N(0,V_{h_{i,t}}^j) \nonumber
            \end{eqnarray}
            and $\epsilon_{h_{i,t}^{j}}$ is standard normal white noise distributed and $j = \text{OUT, INF}$. To set up the auxiliary mixture sampler, we transform $\sigma_{i,t}^{j,2} = \exp{(h_{i,t}^{j})} \epsilon_{h_{i,t}^{j}}$ into a linear model and take the logarithm of the squares of the observed $e_{i,t}^{j}$ such that $\log (e_{i,t}^{j,2}) = h_{i,t}^{j} + \log(\epsilon_{h_{i,t}^{j}})$. \\
            This linear approximation can be represented by an offset mixture time series model:
            \vspace{-10pt}
            \begin{eqnarray}
                e_{i,t}^{j,\ast} = h_{i,t}^{j} + z_{t}^{\ast} \nonumber
            \end{eqnarray}
            with $e_{i,t}^{j,\ast} = \log (e_{i,t}^{j,2}+c)$ and $c=.0001$. Thus, the density of $\log(\epsilon_{h_{i,t}^{j}})$ is approximated by the density of $z_{t}^{\ast}$ which \shortciteA{ksc98} set up using a mixture of 7 normal densities. Further, the authors represent the mixture density with help of a component indicator variable $s_{t}$ such that:
            \vspace{-10pt}
            \begin{eqnarray}
                z_{t}^{\ast} \mid s_{t} \sim N(m_{k}-1.2704,\nu_{k}^{2}), \nonumber
            \end{eqnarray}
            with $k=1,\ldots,7$ and $Pr(s_t = k) = q_k$ (see \shortciteA{ksc98} for the selection of $m_k$, $q_k$ and $\nu_{k}^{2}$). \\
            Once we sampled $s_{t}$, we use the corresponding $m_k$ and $\nu_{k}^{2}$ to sample $h_{i,t}^{j}$ using the precision sampler of \citeA{cj09} as follows: \\
            We re-write \eqref{eqn_2} in more compact form as
            \vspace{-10pt}
            \begin{eqnarray}
                \boldsymbol{H}_{h_{i}^{j}} \boldsymbol{h}_{i}^{j} = \tilde{\boldsymbol{\alpha}}_{h_{i}^{j}} + \boldsymbol{\xi}_{i}^j, \quad \boldsymbol{\xi}_{i}^j \sim N(\boldsymbol{0},\boldsymbol{V}_{h_{i}}^j), \nonumber
            \end{eqnarray}
            where $\tilde{\boldsymbol{\alpha}}_{h_{i}^{j}} = (\boldsymbol{h}_{i,0}^{j},\boldsymbol{0,\ldots,\boldsymbol{0}})$ and
            \vspace{-10pt}
            \begin{eqnarray}
                \boldsymbol{H}_{h_{i}^{j}} = \begin{bmatrix} 
            1 & 0 & \ldots & 0 \\
            -1 & 1 & \ddots & 0 \\
            \vdots & \ddots & \ddots & \vdots \\
            0 & \ldots & -1 & 1
            \end{bmatrix}. \nonumber
            \end{eqnarray}
            It follows that
            \vspace{-10pt}
            \begin{eqnarray}
                (\boldsymbol{h}_{i^j} \mid \boldsymbol{V}_{h_i ^ j}, \boldsymbol{h}_{i,0}^{j}) \sim N(\boldsymbol{\alpha}_{h _ i ^ j}, \boldsymbol{H}_{h _ i ^ j}' \boldsymbol{V}_{h _ i}^{j \mathbf{-1}} \boldsymbol{H}_{h_i^j}^{\mathbf{-1}}), \nonumber
            \end{eqnarray}
            where $\boldsymbol{\alpha}_{h_{i}^{j}} = \boldsymbol{H}_{h_{i}^{j}}^{\mathbf{-1}} \tilde{\boldsymbol{\alpha}}_{h_{i}^{j}}$. \\
            The conditional posterior of $\boldsymbol{h}_{i}^{j}$ then is
            \vspace{-10pt}
            \begin{eqnarray}
                (\boldsymbol{h}_{i}^{j} \mid \boldsymbol{e}_{i}^{j,\ast}, \boldsymbol{h}_{i,0}^{j}, \boldsymbol{V}_{h_{i}}^{j}, m_{k}, \nu_{k}^{2}) \sim N(\boldsymbol{\hat{h}}_{i}^{j},\boldsymbol{K}_{h_{i}}^{j \mathbf{-1}}), \nonumber
            \end{eqnarray}
            where $\boldsymbol{\hat{h}}_{i}^{j} = \boldsymbol{K}_{h_{i}}^{j \mathbf{-1}} \boldsymbol{d}_{h_{i}}^{j}$ with
            \vspace{-10pt}
            \begin{eqnarray}
                \boldsymbol{K}_{h_{i}}^{j} = \boldsymbol{H}_{h_{i}^{j'}} \boldsymbol{V}_{h_{i}}^{j \mathbf{-1}} \boldsymbol{H}_{h_{i}^{j'}} + \boldsymbol{\nu}_{k}^{2 \mathbf{-1}}, \boldsymbol{d}_{h_{i}}^{j} = \boldsymbol{H}_{h_{i}}^{j'} \boldsymbol{V}_{h_{i}}^{j \mathbf{-1}} \boldsymbol{H}_{h_{i}}^{j} \boldsymbol{\alpha}_{h_{i}^{j}} + \boldsymbol{\nu}_{k}^{2 \mathbf{-1}} (\boldsymbol{e}_{i}^{j,\ast} - m_{k}). \nonumber
            \end{eqnarray}
            As the precision matrix $\boldsymbol{K}_{h_{i}}^{j}$ is a band matrix that only contains non-zero bands close to the main diagonal, we can apply the precision sampler of \citeA{cj09} and draw $\boldsymbol{\hat{h}}_{i}^{j}=(h_{i1}^{j},\ldots,h_{iT}^{j})'$ sequentially for $i=1,\ldots,n$.
		\item[\bf{Step 3}] Draw the initial conditions for $h_{i,0}^{j}$ using the precision sampler of \citeA{cj09} from  the following posterior:
            \vspace{-10pt}
            \begin{eqnarray}
                (\boldsymbol{h_{0}^{j}} \mid \boldsymbol{e}_{i}^{j,\ast}, \boldsymbol{h}_{i}^{j}, \boldsymbol{V}_{h_{i}}^{j}) \sim N(\boldsymbol{\hat{h}}_{i,0}^{j},\boldsymbol{K}_{h_{i,0}}^{j \mathbf{-1}}), \nonumber
            \end{eqnarray}
            where $\boldsymbol{K}_{h_{i,0}}^{j} = diag(V_{h_{i1}},\ldots,V_{h_{in}})^{\mathbf{-1}} + \frac{1}{10} \boldsymbol{I}_{n}$ and $\boldsymbol{\hat{h}}_{0}^{j} = \boldsymbol{K}_{h_{i,0}}^{j \mathbf{-1}} diag(V_{h_{i1}},\ldots,V_{h_{in}})\inv \boldsymbol{h}_{1}^{j}$.
		\item[\bf{Step 4}] Draw the local shrinkage component $\lambda_{h_{i,t}^j}$ and the global shrinkage component $\tau_{h_{i}^{j}}$ using the scalar mixture representation of the half-Cauchy distribution as done by \citeA{ms16}.
            The conditional posterior distributions of the hyperparameters from the scale mixture representation of the half-Cauchy distribution are conditionally independent and are inverse Gamma distributed:
            \vspace{-10pt}
            \begin{eqnarray}
                (\nu_{\lambda_{h_{i,t}^{j}}} \mid \lambda_{h_{}i,t}^{j}) &\sim& IG(1,1+\frac{1}{ \lambda_{h_{}i,t}^{j}}), \quad i = 1,\ldots,n, \nonumber \\
                (\lambda_{h_{i,t}^j} \mid \boldsymbol{h}_{i}^{j}, \boldsymbol{h}_{0}^{j}, \nu_{\lambda_{h_{i}^j}}, \tau_{h_{i}^{j}}) &\sim& IG(1,\frac{1}{\nu_{\lambda_{h_{i}^{j}}}}+\frac{1}{2} \frac{(h_{i,t}^{j}-h_{i,t-1}^{j})^{2}}{\tau_{h_{i}^{j}}}), \quad i = 1,\ldots,n, \nonumber \\
                (\nu_{\tau_{h_{i,t}^{j}}} \mid \tau_{h_{i}^{j}}) &\sim& IG(1,1+\frac{1}{\tau_{h_{i}^{j}}}), \quad i = 1,\ldots,n, \nonumber \\
                (\tau_{h_{i}^{j}} \mid \boldsymbol{h}_{i}^{j}, \nu_{\tau_{h_{i,t}^{j}}}) &\sim& IG(\frac{T+1}{2},\frac{1}{\nu_{\tau_{h_{i,t}^{j}}}}+\frac{1}{2} \sum_{t=1}^{T} \frac{(h_{i,t}^{j}-h_{i,t-1}^{j})^{2}}{\lambda_{h_{i}^{j}}}), \quad i = 1,\ldots,n \nonumber
            \end{eqnarray}
            Using the posterior draws, we compute $V_{h_{i,t}}^j = \tau_{h_{i}^j} \lambda_{h_{i,t}^j}$.
		\item[\bf{Step 5}] Draw the VAR coefficients $\boldsymbol{A} = (\boldsymbol{c},\boldsymbol{A}_1,\ldots,\boldsymbol{A}_L)'$ using the equation-by-equation approach of \shortciteA{ccm19} and \shortciteA{cccm22}. We stack \eqref{eqn_7} over $t = 1,\ldots,T$ and obtain:
        \vspace{-10pt}
        \begin{eqnarray}
            \boldsymbol{Y} = \boldsymbol{X}\boldsymbol{A} + \boldsymbol{E}\boldsymbol{B}', \quad vec(\boldsymbol{E}) \sim N(\boldsymbol{0},\boldsymbol{I}_{T_n}) \nonumber,
        \end{eqnarray}
        where $\boldsymbol{Y} = (\boldsymbol{y}_{1},\ldots,\boldsymbol{y}_{T})', \boldsymbol{X} = (\boldsymbol{x}_{1},\ldots,\boldsymbol{x}_{T})'$ with $\boldsymbol{x_t} = (1,\boldsymbol{y}_{t-1}^{'},\ldots,\boldsymbol{y}_{t-p}^{'})$ and $\boldsymbol{E} = (\boldsymbol{\epsilon}_1,\ldots,\boldsymbol{\epsilon}_{T})'$. Following \citeA{h24}, we define $\boldsymbol{y}_{a,i} = vec((\boldsymbol{Y}-\boldsymbol{X}\boldsymbol{A}_{i=0})\boldsymbol{B}^{'-1})$ and $\boldsymbol{X}_{a,i} = (\boldsymbol{B}^{-1})_{i} \bigotimes \boldsymbol{X}$, where $\boldsymbol{A}_{i=0}$ is the $\boldsymbol{A}$ matrix with its \emph{i}th column replaced by zeros and $(\boldsymbol{B}^{-1})_{i}$ is the \emph{i}th column of $\boldsymbol{B}^{-1}$. Thus, we can re-write
        \vspace{-10pt}
        \begin{eqnarray}
            \boldsymbol{y}_{a,i} = \boldsymbol{X}_{a,i} \boldsymbol{a}_{i} + vec(\boldsymbol{E}), \quad vec(\boldsymbol{E}) \sim N(\boldsymbol{0},\boldsymbol{I}_{T_{n}}) \nonumber,
        \end{eqnarray}
        where $\boldsymbol{a}_{i}$ is the \emph{i}th column of $\boldsymbol{A}$. \\
        Let $\boldsymbol{S}_{\gamma,i}$ be the selection matrix such that $\boldsymbol{a}_{i} = \boldsymbol{S}_{\gamma,i} \boldsymbol{\gamma}_{i}$ with $\boldsymbol{\gamma}$ being the \emph{i}th row of $\Gamma$ with the prior $\boldsymbol{\gamma}_{i} \sim N(\boldsymbol{\gamma}_{0,i},\boldsymbol{V}_{\gamma,i})$. The prior mean $\boldsymbol{\gamma}_{0,i}$ and the covariance matrix $\boldsymbol{V}_{\gamma,i}$ is obtained from \eqref{eqn_14}. Thus, the conditional prior of $\boldsymbol{a}_{i}$ is
        \vspace{-10pt}
        \begin{eqnarray}
            (\boldsymbol{\gamma}_{i} \mid \boldsymbol{A}_{i=0},\boldsymbol{B},\boldsymbol{y}) \sim N(\hat{\boldsymbol{\gamma}}_{i},\boldsymbol{K}_{\gamma,i}^{-1}), \nonumber
        \end{eqnarray}
        with $\boldsymbol{K}_{\gamma,i} = \boldsymbol{S}_{\gamma,i}^{'} \boldsymbol{X}_{a,i}^{'} \boldsymbol{X}_{a,i} \boldsymbol{S}_{\gamma,i} + \boldsymbol{V}_{\gamma,i}^{-1}$ and $\hat{\boldsymbol{\gamma}}_{i} = \boldsymbol{K}_{\gamma,i}^{-1} (\boldsymbol{S}_{\gamma,i}^{'} \boldsymbol{X}_{a,i}^{'} \boldsymbol{y}_{a,i} + \boldsymbol{V}_{\gamma,i}^{-1} \boldsymbol{\gamma}_{0,i}$.
		\item[\bf{Step 6}] Draw the contemporaneous impact matrix $\boldsymbol{B}$ with zero and sign restrictions following \citeA{h24}. We sample the entries of $\boldsymbol{B}$ column-by-column from the density $p(\boldsymbol{b}_{i} \mid \boldsymbol{B}_{-i}, \boldsymbol{y})$ for $i = 1,\ldots,n$ and $i$ denoting the \emph{i}th column of $\boldsymbol{B}$. To sample $\boldsymbol{b}_{i}$ from $\boldsymbol{B}$, we implement a parameter transformation which we explain for \nth{1} column in the following as the ordering of the column can be set arbitrarily.
        \vspace{-10pt}
        \begin{eqnarray}
            \boldsymbol{B} = \begin{bmatrix}
                b_{1,1} & \boldsymbol{b}_{12}^{'} \\
                \boldsymbol{b}_{-1} & \boldsymbol{B}_{22} \nonumber
            \end{bmatrix},
        \end{eqnarray}
        where $\boldsymbol{b}_{-1} = (b_{2,1},\ldots,b_{n,1}'), \boldsymbol{b}_{12}$ is $(n-1) \times 1$ and $\boldsymbol{B}_{22}$ is $(n-1) \times (n-1)$.\footnote{It is assumed that $b_{1,1} \neq 0$ and $\boldsymbol{B}_{22}$ is invertible \cite{h24}.} \\
        The parameter transformation maps the \nth{1} column of $\boldsymbol{B}$ which is $\boldsymbol{b}_{1} = (b_{1,1},\boldsymbol{b}_{-1}^{'})'$  to $(u, \boldsymbol{w})$:
        \vspace{-10pt}
        \begin{eqnarray}
            \boldsymbol{w} = \boldsymbol{b}_{-1}, \quad u = b_{1,1} - \boldsymbol{b}_{12}^{'} \boldsymbol{B}_{22}^{-1} \boldsymbol{b}_{-1}. \nonumber
        \end{eqnarray}
        Since our sampling is conditioned on zero and sign restrictions, we establish the index set $G$ that the indices $j$ such that $b_{j,1} \neq 0$ for $j = 2,\ldots,n$ which reads as $G = \{j \in \{2,\ldots,n\} : b_{j,1} \neq 0\}$. Thus, we can re-write the priors of the non-zero parameters in $\boldsymbol{b}_{1}$:
        \vspace{-10pt}
        \begin{eqnarray}
            b_{1,1} \sim N(\beta_{1,1},\nu_{1,1}) \mathds{1}(R_1), \quad b_{j,1} \sim N(\beta_{j,1},\nu_{j,1}) \mathds{1}(R_j), \quad j \in G, \nonumber
        \end{eqnarray}
        where $R_j$ is given by \\
        \vspace{-10pt}
        \begin{eqnarray}
        R_{j} = \begin{cases}
            \mathbb{R}, & \text{if } b_{j,1} \text{ is unrestricted}, \\
            \{b_{j,1} > 0 \}, & \text{if  } b_{j,1} >0, \quad \quad  \quad \quad \quad \quad \text{for  } j \in \{1\} \cup G. \\
            \{b_{j,1} < 0 \}, & \text{if  } b_{j,1} <0,
        \end{cases} \nonumber
        \end{eqnarray}
        We define the support of $(u,w)$ as $\tilde{R}_{j}$ implied by $R_{j}$ for $j \in \{1\} \cup G$ and write:
        \vspace{-10pt}
        \begin{eqnarray}
            \tilde{R}_{1} &=& \begin{cases}
                \mathbb{R}, & \text{if } R_{1} = \mathbb{R}, \\
                \{(u,\boldsymbol{w}) : u + \boldsymbol{b}_{12}^{'} \boldsymbol{B}_{22}^{-1} \boldsymbol{w} > 0 \}, & \text{if } R_{1} = b_{1,1} > 0, \\
                \{(u,\boldsymbol{w}) : u + \boldsymbol{b}_{12}^{'} \boldsymbol{B}_{22}^{-1} \boldsymbol{w} < 0 \}, & \text{if } R_{1} = b_{1,1} < 0,
            \end{cases}  \nonumber \\
             \tilde{R}_{j} &=& \begin{cases}
                \mathbb{R}, & \text{if } R_{j} = \mathbb{R}, \\
                \{w_{j} > 0 \} & \text{if } R_{j} = \{b_{j,1} > 0 \}, \text{  },\quad \quad  \quad \quad \quad \quad \text{for  } j \in G. \\
                \{w_{j} < 0 \} & \text{if } R_{j} = \{b_{j,1} < 0 \},
            \end{cases} \nonumber
         \end{eqnarray} \nonumber
         The $\tilde{k}$ elements in $G$ that are associated with the non-zero parameters in $\boldsymbol{w}$. Further, we denote the vector $\tilde{\boldsymbol{w}}$ of dimension $\tilde{k} \times 1$ that gathers these non-zero parameters. The selection matrix $\boldsymbol{S}$ of dimension $(n-1) \times \tilde{k}$ then guarantees that $\boldsymbol{w} = \boldsymbol{S} \tilde{\boldsymbol{w}}$. \\
         Given the parameter transformation to $(u,\boldsymbol{w})$ and $\boldsymbol{w} = \boldsymbol{S} \tilde{\boldsymbol{w}}$, we write the prior density function as
         \vspace{-10pt}
         \begin{eqnarray}
             p(u,\tilde{\boldsymbol{w}} \propto e^{(-\frac{1}{2} (\tilde{\boldsymbol{w}}-\tilde{\boldsymbol{\beta}}{-1})' \tilde{\boldsymbol{V}}_{-1}^{-1}(\tilde{\boldsymbol{w}}-\tilde{\beta}_{-1}))} \times e^{(-\frac{1}{2\nu_{1,1}}(u + \boldsymbol{b}_{12}^{'} \boldsymbol{B}_{22}^{-1}\boldsymbol{S}\tilde{\boldsymbol{w}}-\beta_{1,1})^{2})} \times \mathds{1}(\bigcap_{j \in G} \tilde{R}_{i}) \times \mathds{1}(\tilde{R}_{1}), \nonumber
         \end{eqnarray}
         where $\tilde{\beta}_{-1}$ is a $\tilde{k} \times 1$ vector and $\tilde{\boldsymbol{V}}_{-1}$ is a $\tilde{k} \times \tilde{k}$ diagonal matrix and denote the priors.\footnote{The priors for the non-zero parameters $\tilde{\boldsymbol{w}}$ are derived from the unrestricted prior for $p(\boldsymbol{b}_{}1) = p(b_{1,1},\boldsymbol{b}_{-1}) \propto e^{-\frac{1}{2}(\boldsymbol{b}_{-1}-\boldsymbol{\beta}_{-1})'\boldsymbol{V}_{-1}^{-1}(\boldsymbol{b}_{-1}-\boldsymbol{\beta}_{-1})} \times e^{-\frac{1}{2\nu_{1,1}}(b_{1,1}-\beta{1,1})^{2}}$. where $\boldsymbol{\beta}_{-1} = (\beta_{2,1},\ldots,\beta_{n,1})'$ and $\boldsymbol{V}_{-1} = diag(\nu_{2,1},\ldots,\nu_{n,1}).$} \\
         It can be shown that the induced conditional likelihood of $\tilde{\boldsymbol{w}}$ follows:
         \vspace{-10pt}
         \begin{eqnarray}
             p(\boldsymbol{y} \mid \tilde{\boldsymbol{w}}, u, \tilde{\boldsymbol{w}}\boldsymbol{B}_{-1}) &\propto& e^{-\frac{1}{2}(\boldsymbol{q}-\boldsymbol{Z}\boldsymbol{w})'(\boldsymbol{q}-\boldsymbol{Z}\boldsymbol{w})} \nonumber \\
             &\propto& e^{-\frac{1}{2}(\boldsymbol{q}-\tilde{\boldsymbol{Z}}\tilde{\boldsymbol{w}})'(\boldsymbol{q}-\tilde{\boldsymbol{Z}}\tilde{\boldsymbol{w}})} \nonumber,
         \end{eqnarray}
         where $\tilde{\boldsymbol{Z}} = \boldsymbol{Z}\boldsymbol{S}$, $\boldsymbol{q} = (q_{1},\ldots,q_{T})'$, $\boldsymbol{Z}=(\boldsymbol{Z}_{1}^{'},\ldots,\boldsymbol{Z}_{T}^{'})'$ and
         \vspace{-10pt}
         \begin{eqnarray}
             q_{t} = \begin{bmatrix}
                 u^{-1}y_{1,t} - u^{-1}\boldsymbol{b}_{12}^{'} \boldsymbol{B}_{22}^{-1}\boldsymbol{y}_{-1,t} \\
                 \boldsymbol{B}_{22}^{-1} \boldsymbol{y}_{-1,t}
             \end{bmatrix}, \quad \quad \boldsymbol{Z}_{t} = \begin{bmatrix}
                 \boldsymbol{0} \\
                 u^{-1} \boldsymbol{B}_{22}^{-1} y_{1,t}-u^{-1} ((\boldsymbol{y}_{-1,t}^{'}\boldsymbol{B}_{22}^{'-1}\boldsymbol{b}_{12}) \bigotimes \boldsymbol{B}_{22}^{-1})
             \end{bmatrix}. \nonumber
         \end{eqnarray}
         Then, the conditional posterior of $\tilde{\boldsymbol{w}}$ reads:
         \vspace{-10pt}
         \begin{eqnarray}
             (\tilde{\boldsymbol{w}} \mid u, \boldsymbol{B}_{-1}, \boldsymbol{y}) \sim N(\hat{\tilde{\boldsymbol{w}}},\boldsymbol{D}_{\tilde{w}})\mathds{1}(\bigcap_{j \in G} \tilde{R}_{j}) \times \mathds{1} (\tilde{R}_{1}),  \nonumber
         \end{eqnarray}
         where
         \vspace{-10pt}
         \begin{eqnarray}
             \hat{\tilde{\boldsymbol{w}}} &=& \boldsymbol{D}_{\tilde{w}}(\tilde{\boldsymbol{Z}}^{'}\boldsymbol{q} + \tilde{\boldsymbol{V}}_{-1}^{-1}\tilde{\boldsymbol{\beta}}_{-1}-\nu_{1,1}^{-1}(u-\tilde{\beta}_{1,1})\boldsymbol{B}_{22}^{'-1}\boldsymbol{b}_{12}), \nonumber \\ 
             \boldsymbol{D}_{\tilde{w}}^{-1} &=& \tilde{\boldsymbol{Z}}^{'}\tilde{\boldsymbol{Z}} + \tilde{\boldsymbol{V}}_{-1}^{-1} + \nu_{1,1}^{-1} \boldsymbol{B}_{22}^{'-1} \boldsymbol{b}_{12} \boldsymbol{b_12}^{'} \boldsymbol{B}_{22}^{-1}. \nonumber
         \end{eqnarray}
         We use the algorithm of \citeA{b17} to sample a proposal draw $\tilde{\boldsymbol{w}}^{\ast}$ from the truncated normal distribution $N(\hat{\tilde{\boldsymbol{w}}},\boldsymbol{D}_{\tilde{w}})\mathds{1}(\bigcap_{j \in G} \tilde{R}_{j}))$ and set $\tilde{\boldsymbol{w}} = \tilde{\boldsymbol{w}}^{\ast}$ if $(u,\tilde{\boldsymbol{w}}^{\ast}) \in \tilde{R}_{1}$. \\
         The conditional posterior of u is sampled from using the normal mixture approximation of \citeA{v09}:
         \vspace{-10pt}
         \begin{eqnarray}
            p(u \mid \tilde{\boldsymbol{w}}, \boldsymbol{B}_{-1}, \boldsymbol{y}) \propto |u|^{-T} e^{-\frac{1}{2}(u^{-1}\gamma_{u,1}+u^{-2}\gamma_{u,2})} \times e^{-\frac{1}{2\nu_{1,1}}(u+\boldsymbol{b}_{12}^{'} \boldsymbol{B}_{22}^{-1} \tilde{\boldsymbol{w}}-\beta_{1,1})^{2}} \times \mathds{1}(\tilde{R}_{1}). \nonumber
         \end{eqnarray}
         We return $u=u^{\ast}$ with probability
         \vspace{-10pt}
         \begin{eqnarray}
             min \left\{\!\begin{aligned}
                 1, & \frac{e^{-\frac{1}{2\nu_1}(u^{\ast}+\boldsymbol{b}_{12}^{'}\boldsymbol{B}_{22}^{-1}\tilde{\boldsymbol{w}}-\beta_{1,1})^{2}}}{e^{-\frac{1}{2\nu_1}(u+\boldsymbol{b}_{12}^{'}\boldsymbol{B}_{22}^{-1}\tilde{\boldsymbol{w}}-\beta_{1,1})^{2}}}
            \end{aligned}\right\} \times \mathds{1}(\tilde{R}_{1}) \nonumber.
         \end{eqnarray}
         The draws for $(u,\boldsymbol{w}$ are then mapped such that $b_{1,1} = u + \boldsymbol{b}_{12}^{'}\boldsymbol{B}_{22}^{-1}\boldsymbol{w}$ and $\boldsymbol{b}_{1} = \boldsymbol{w}$. It follows the sampling of the remaining columns $i = 2,\ldots,n$ and the computation of the variance $\boldsymbol{B}\boldsymbol{B}^{'}$.
		\item[\bf{Step 7}] Draw the shrinkage parameters $\kappa_1$ and $\kappa_2$ for the autoregressive parameters in $\boldsymbol{A}$. To do so, we establish:
        \vspace{-10pt}
        \begin{eqnarray}
            \tilde{\boldsymbol{\Gamma}}_{ij,l} = \begin{cases}
                \frac{\sigma_{j}^{2}}{\sigma_{i}^{2}}l^{2}(\Gamma_{ij,l}-\gamma_{ij,l})^{2}, & \text{for } i=j, \\
                \frac{\sigma_{j}^{2}}{\kappa_{2} \sigma_{i}^{2}}l^{2}(\Gamma_{ij,l}-\gamma_{ij,l})^{2}, & \text{for } i\neq j,
            \end{cases} 
            \quad \tilde{\tilde{\boldsymbol{\Gamma}}}_{ij,l} = 
            \begin{cases}
                0, & \text{for } i=j, \\
                \frac{\sigma_{j}^{2}}{\kappa_{1} \sigma_{i}^{2}}l^{2}(\boldsymbol{\Gamma}_{ij,l}-\gamma_{ij,l})^{2}, & \text{for } i\neq j,
            \end{cases} \nonumber
        \end{eqnarray} \nonumber
        Using this, the conditional posteriors read as follows:
        \vspace{-10pt}
        \begin{eqnarray}
            p(\kappa_1 \mid \kappa_2, \boldsymbol{\Gamma}) &\propto& (\kappa_1)^{-\frac{r^{2}L}{2}e{-\frac{1}{\kappa_1}\sum_{i=1}^{r}\sum_{l=1}^{L}\tilde{\boldsymbol{\Gamma}}_{ij,l}}}, \nonumber \\
            p(\kappa_2 \mid \kappa_1, \boldsymbol{\Gamma}) &\propto& (\kappa_2)^{-\frac{r(r-1)L}{2}e{-\frac{1}{\kappa_2}\sum_{i=1}^{r}\sum_{l=1}^{L}\tilde{\tilde{\boldsymbol{\Gamma}}}_{ij,l}}}, \nonumber
        \end{eqnarray}
        and can be drawn from truncated inverse Gamma distributions:
        \vspace{-10pt}
        \begin{eqnarray}
            (\kappa_{1} \mid \kappa_{2}, \boldsymbol{\Gamma}) &\sim& IG(\frac{r^{2}l}{2}-1, \frac{1}{2} \boldsymbol{\sum}_{i=1}^{r} \boldsymbol{\sum}_{l=1}^{L} \tilde{\boldsymbol{\Gamma}}_{ij,l}), \nonumber \\
            (\kappa_{2} \mid \kappa_{1}, \boldsymbol{\Gamma}) &\sim& IG(\frac{r(r-1)l}{2}-1, \frac{1}{2} \boldsymbol{\sum}_{i=1}^{r} \boldsymbol{\sum}_{l=1}^{L} \tilde{\tilde{\boldsymbol{\Gamma}}}_{ij,l}). \nonumber
        \end{eqnarray}
        \item[\bf{Step 8}] Draw the common factors $f_t^{j}$ using the algorithm of \citeA{ck94}. In order to implement the algorithm which employs the Kalman (forward) filter and the backward smoother, we transform the FAVAR model to its state-space representation.\\
        The observation equation
        The observation equation for the model specification in \eqref{eqn_1}, \eqref{eqn_3} and \eqref{eqn_3} reads as follows:
        \vspace{-10pt}
        \begin{eqnarray}
            \tilde{x}_{t} = H \beta_{t} + r_{t}, \nonumber 
        \end{eqnarray}
        where $\tilde{x}_{t}$ stacks the $x_{i,t}^{OUT}$, $x_{i,t}^{INF}$ and $\boldsymbol{z}_{t}$. $\beta_t$ stacks the factors $f_{t}^{OUT}$, $f_{t}^{INF}$ and  $\boldsymbol{z}_{t}$ and their lags. $H$ contains the $\lambda_{i,p}^{OUT}$ and $\lambda_{i,p}^{INF}$ and fills the remaining entries with zeros and ones, respectively. \\
        The transition equation reads:
        \vspace{-10pt}
        \begin{eqnarray}
            \beta_{t} = F \beta_{t-1} + \epsilon_{t}, \nonumber
        \end{eqnarray}
        where $F$ stacks the (non-zero) autoregressive VAR parameters $\boldsymbol{A}$ and we assume
        \vspace{-10pt}
        \begin{eqnarray}
            \begin{bmatrix}
                r_{t} \\
                \epsilon_{t}
            \end{bmatrix}
            \stackrel{\text{i.i.d.}}{\sim}  N(\begin{bmatrix}
                0 \\ 0
            \end{bmatrix},
            \begin{bmatrix}
                R_{t} & 0 \\
                0 & Q
            \end{bmatrix}). \nonumber
        \end{eqnarray} \nonumber
        Further, we define
        \vspace{-10pt}
        \begin{eqnarray}
            \beta_{t \mid s} &=& E(\beta_{t} \mid \tilde{X}^{s}, H, R^{s}, Q) \nonumber \\
            V_{t \mid s} &=& Var(\beta_{t} \mid \tilde{X}^{s}, H, R^{s}, Q) \nonumber.
        \end{eqnarray}
        Then, we implement the conventional Kalman (forward) filter:
        \vspace{-10pt}
        \begin{eqnarray}
            \beta_{t \mid t-1} &=& F \beta_{t-1 \mid t-1}, \nonumber \\
            V_{t \mid t-1} &=& F V_{t-1 \mid t-1} F^{'} + Q, \nonumber \\
            K_{t} &=& V_{t \mid t-1} H^{'}(H V_{t \mid t-1}H^{'} + R_{t})', \nonumber \\
            \beta_{t \mid t} &=& \beta_{t \mid t-1} + K_{t}(\tilde{x}_{t}-H \beta_{t \mid t-1}), \nonumber \\
            V_{t \mid t} &=& V_{t \mid t-1} - K_{t}HV_{t \mid t-1}, \nonumber
        \end{eqnarray}
        where $K_{t}$ describes the \emph{Kalman gain}. \\
        Having filtered the factors from $t = 1,\ldots,T$, we can now smooth the factor in backward recursion:
        \vspace{-10pt}
        \begin{eqnarray}
            \beta_{t \mid t+1} &=& \beta_{t \mid t} + V_{t \mid t} F^{'} V_{t+1 \mid t}^{-1}(\beta_{t+1}-F\beta_{t \mid t}), \nonumber \\
            V_{t \mid t+1} &=& V_{t \mid t} - V_{t \mid t} F^{'} V_{t+1 \mid t}^{-1} F V_{t \mid t} \nonumber.
        \end{eqnarray}
        We refer to \citeA{ck94}, \citeA{p05}, \shortciteA{bbe05} for more details on the estimation of common factors with help of the Carter-Kohn algorithm.
        From $\beta_{t \mid t+1}$, we extract $f_{t}^{OUT}$ and $f_{t}^{INF}$ and stack them into $\boldsymbol{y}_{t}$ and go back to \bf{Step 1}.
    \end{itemize}

\emptythanks 

\newcounter{footnotecounter}
    \setcounter{footnotecounter}{\value{footnote}}
    \setcounter{footnote}{0}

\clearpage
    \title{Online Appendix - The Transmission of Monetary Policy via Common Cycles in the Euro Area\thanks{We thank seminar and conference participants at the Ruhr Graduate 
School in Economics in Essen, the \nth{7} International Conference on Applied Theory, Macro and Empirical Finance at the University of Macedonia in Thessaloniki, the \nth{24} IWH-CIREQ-GW-BOKERI Macroeconometric Workshop at the IWH Halle, the Workshop in Empirical Macroeconomics organized by the University of Innsbruck and the Liechtenstein Institute, the Workshop of the SGH Warsaw School of Economics, the \nth{28} International Conference on Macroeconomic Analysis and International Finance, the \nth{22} Conference of the European Economics and Finance Society, the 2024 European Seminar on Bayesian Econometrics, Philipp Ad\"ammer, Joscha Beckmann, Boris Blagov, Robert Czudaj, Sascha Keweloh, Martin Geiger and Mathias Klein for their valuable feedback. Jan Pr\"user gratefully
acknowledges the support of the German Research Foundation (DFG, 468814087).}}
\begin{titlepage}
\setlength{\droptitle}{-3cm}
\maketitle
\begin{abstract}
	\begin{singlespace}
 \noindent
 This is the online appendix for ``The Transmission of Monetary Policy via Common
Cycles in the Euro Area''. The reader is referred to the paper for more detailed information on the model specifications, their estimation and implementation and the corresponding notation. The online appendix is divided into two main parts. The first part is the empirical appendix which
presents additional empirical results on alternative specifications of the unified FAVAR model. The second part entails a data appendix in which we provide a description of the applied data and a section on the data transformation. The second parts also contains a section on description on the construction of the instrument for the monetary policy shock.
	\end{singlespace}
\end{abstract}
\bigskip

\noindent \textbf{Keywords:} Monetary Policy, International Macroeconomics, FAVAR, Proxy, Sign Restrictions

\noindent \textbf{JEL classification:} C32, E31, E32, E52, E58 \bigskip
\end{titlepage}

\pagenumbering{arabic}

\appendix

\renewcommand{\thesection}{Online Appendix \Alph{section}}
\setcounter{table}{0}
\renewcommand{\thetable}{\Alph{section}.\arabic{table}}
\setcounter{figure}{0}
\renewcommand{\thefigure}{O\Alph{section}.\arabic{figure}}

\newpage

\section{Empirical Appendix}

\subsection*{Credible Bands for Country Impulse Responses}

\begin{center}
	\begin{figure}[H]
    \includegraphics[width=0.99\textwidth]{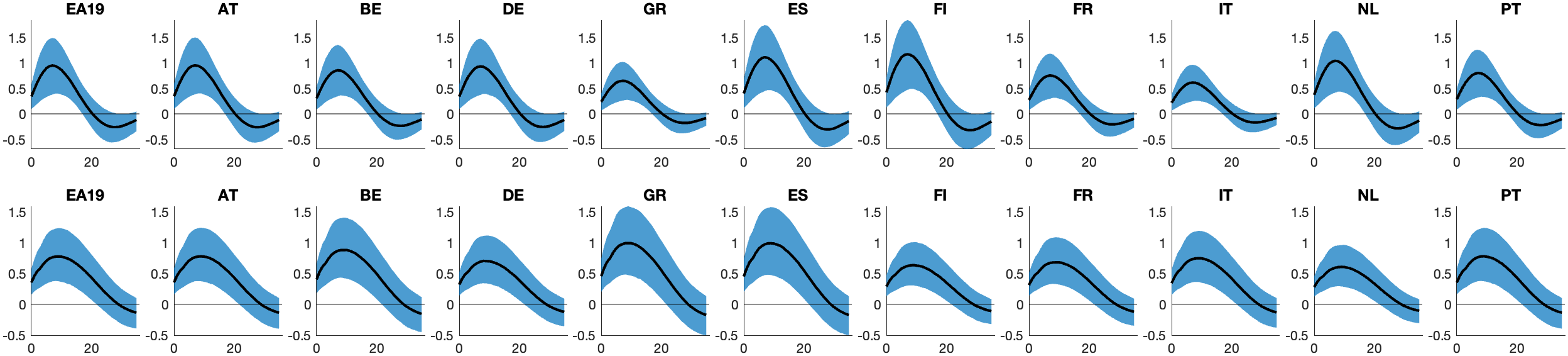}
      \caption{Median impulse responses to an expansionary monetary policy shock. The solid black line depicts the median and the shaded blue area the  68\% credible bands.} \label{fig_OA1}
	\end{figure}
\end{center}

\begin{center}
	\begin{figure}[H]
    \includegraphics[width=0.99\textwidth]{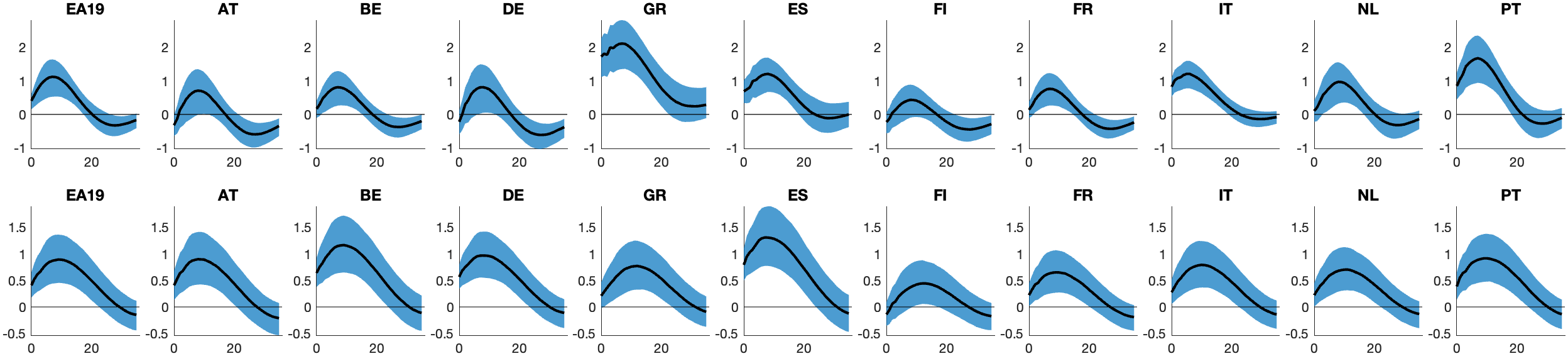}
      \caption{Median impulse responses to an expansionary monetary policy shock. The solid black line depicts the median and the shaded blue area the  68\% credible bands.} \label{fig_OA2}
	\end{figure}
\end{center}

In this section, we present the 68\% credible bands for the country impulse responses for our baseline model specification with muted country-specific financial channel (Figure \ref{fig_OA1}) and non-muted financial channel (Figure \ref{fig_OA2}) as we only showed the median country responses in Figure 4 and Figure 5.

\vspace{-50pt}

\vspace{-50pt}

\clearpage
\subsection*{Agnostic Identification}

In this section, we show the results of the model that employs a more agnostic identification approach as it drops the negative sign restriction on the common cycle for inflation in case of a monetary contraction.

\begin{center}
	\begin{figure}[H]
    \includegraphics[height=8\baselineskip,width=0.99\textwidth]{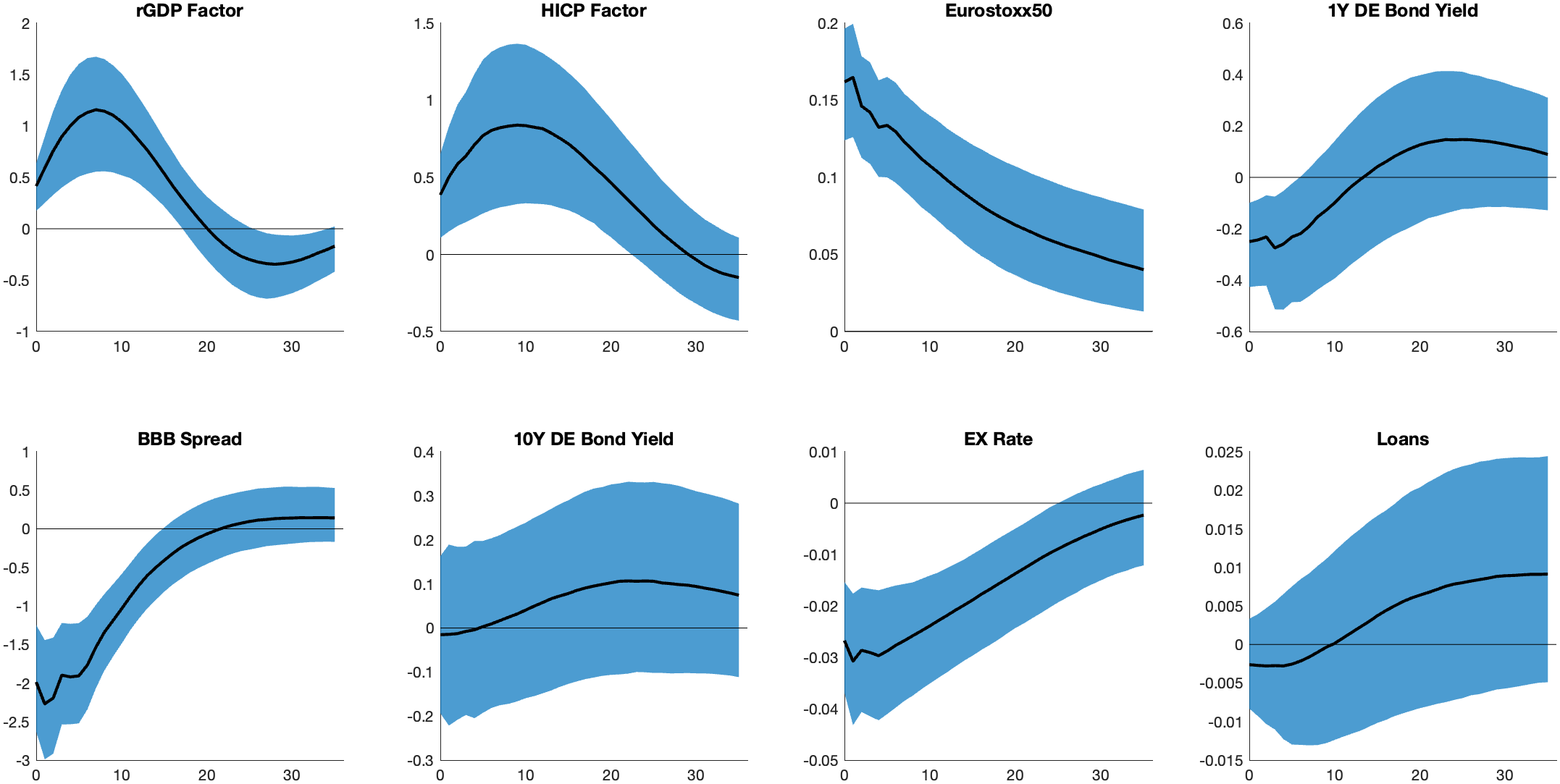}
      \caption{Median impulse responses to an expansionary monetary policy shock (\emph{HICP Factor unrestricted}). The solid black line depicts the median and the shaded blue area the  68\% credible bands.} \label{fig_OA3}
	\end{figure}
\end{center}
\vspace{-50pt}

\begin{center}
    \begin{figure}[H]
    \begin{minipage}{.5\textwidth}
      	\includegraphics[width=0.99\textwidth]{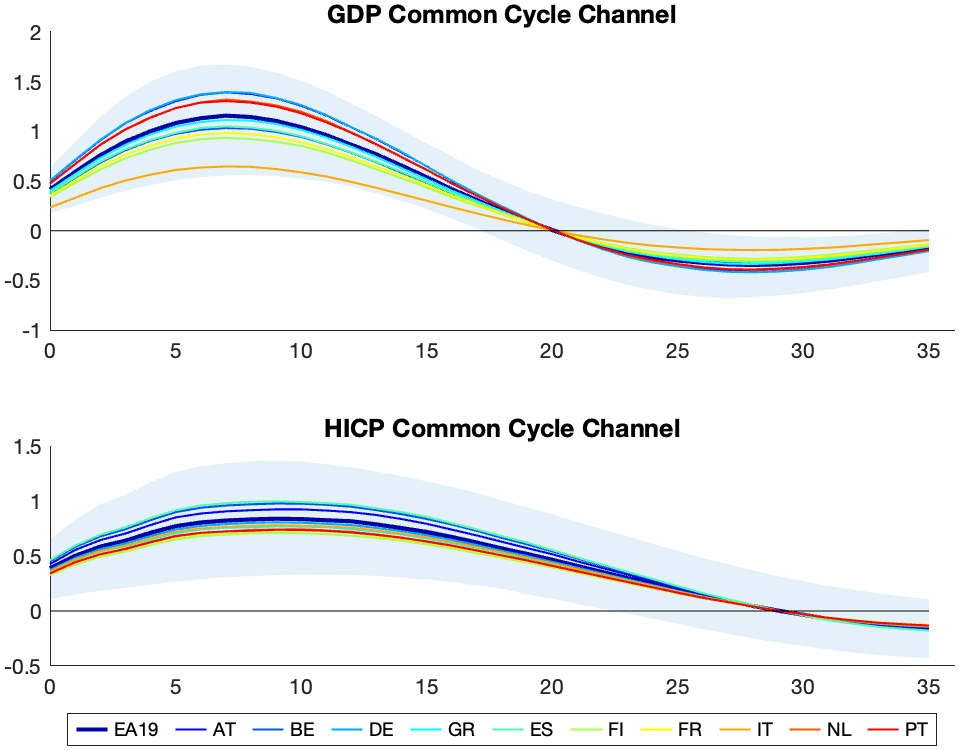}
       \end{minipage}
        \begin{minipage}{.5\textwidth}
      	\includegraphics[width=0.99\textwidth]{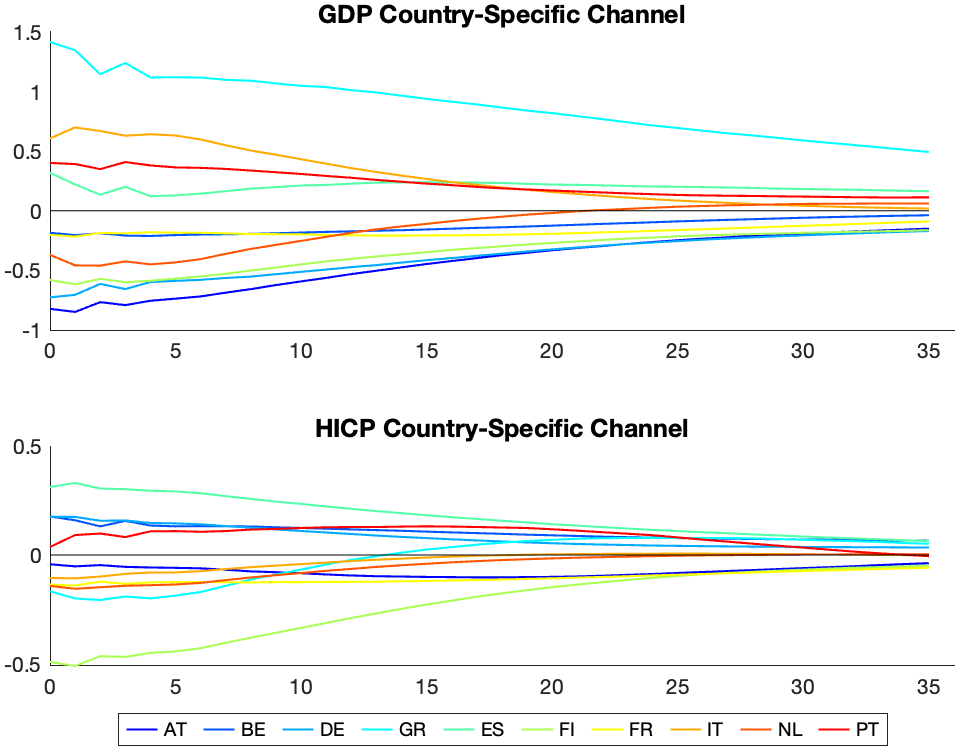}
       \end{minipage}
		  \caption{Country-level median impulse responses of country-specific output growth and inflation
            to an expansionary monetary policy shock (\emph{HICP Factor unrestricted}). The left-hand side plots show the propagation via
            the common cycles and the right-hand side plots show the direct impact via the country-specific
            channels. The solid blue lines depicts the median responses of the euro area aggregate (EA19)
            and the shaded light blue areas the 68\% credible bands.} \label{fig_OA4}
    \end{figure}
\end{center}
\vspace{-50pt}

\clearpage
\subsection*{Poor man's proxy \cite{jk20}}

In this section, we show the results of the model that employs the poor man's proxy proposed by \citeA{jk20}. The proxy is constructed such that changes on announcement days are set to $0$ if the signs of changes in the three-month OIS swap rate and the Eurostoxx50 are identical. Hence, the time series only contains \emph{pure} monetary policy surprises and excludes information shocks.

\begin{center}
	\begin{figure}[H]
    \includegraphics[height=8\baselineskip,width=0.99\textwidth]{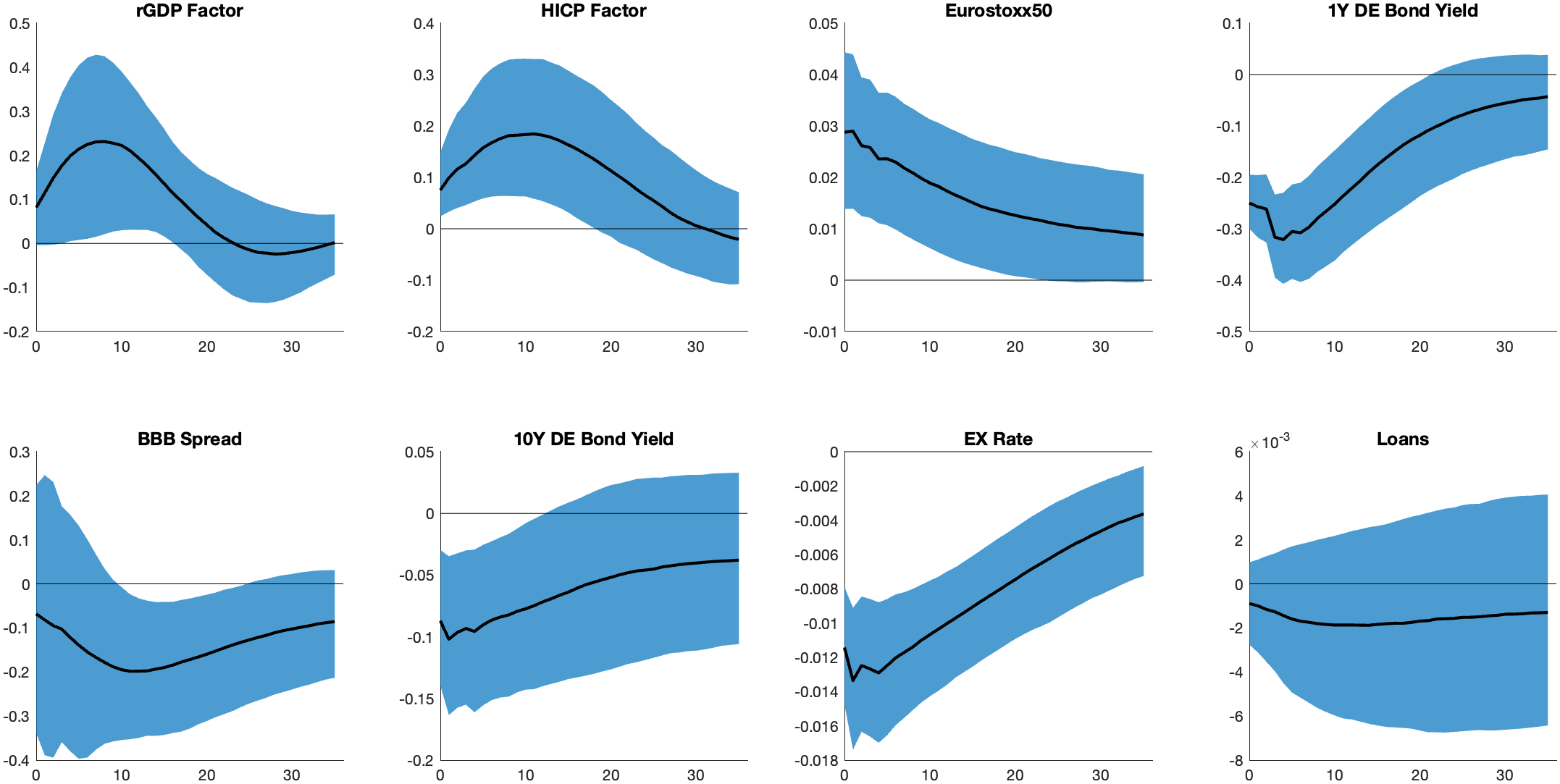}
      \caption{Median impulse responses to an expansionary monetary policy shock (\emph{poor man's proxy}). The solid black line depicts the median and the shaded blue area the  68\% credible bands.} \label{fig_OA5}
	\end{figure}
\end{center}
\vspace{-50pt}

\begin{center}
    \begin{figure}[H]
    \begin{minipage}{.5\textwidth}
      	\includegraphics[width=0.99\textwidth]{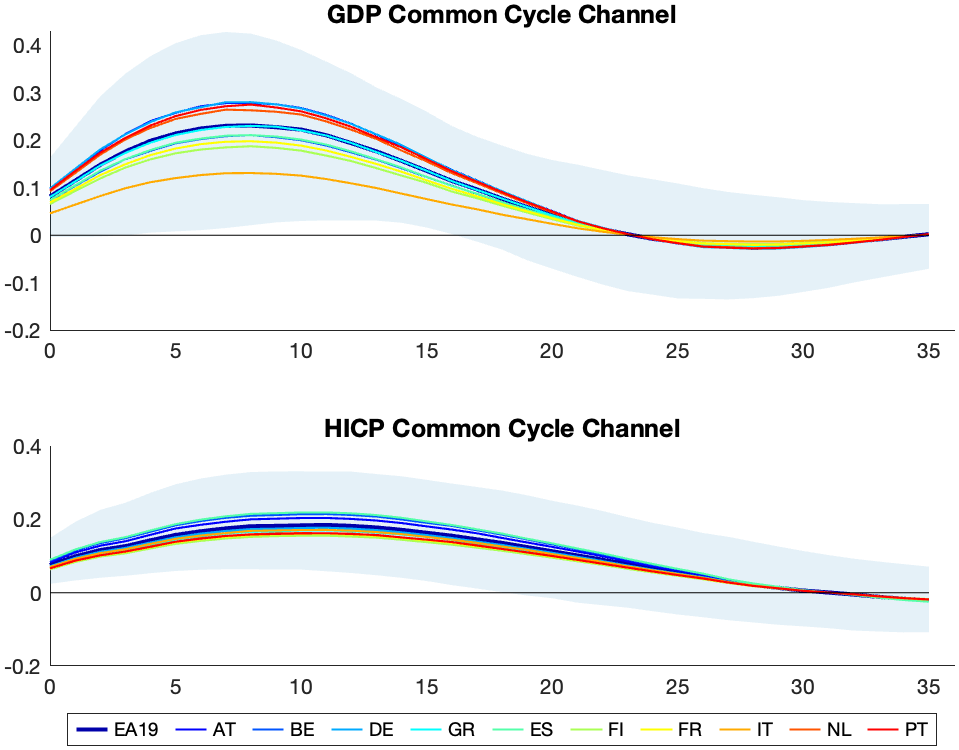}
       \end{minipage}
        \begin{minipage}{.5\textwidth}
      	\includegraphics[width=0.99\textwidth]{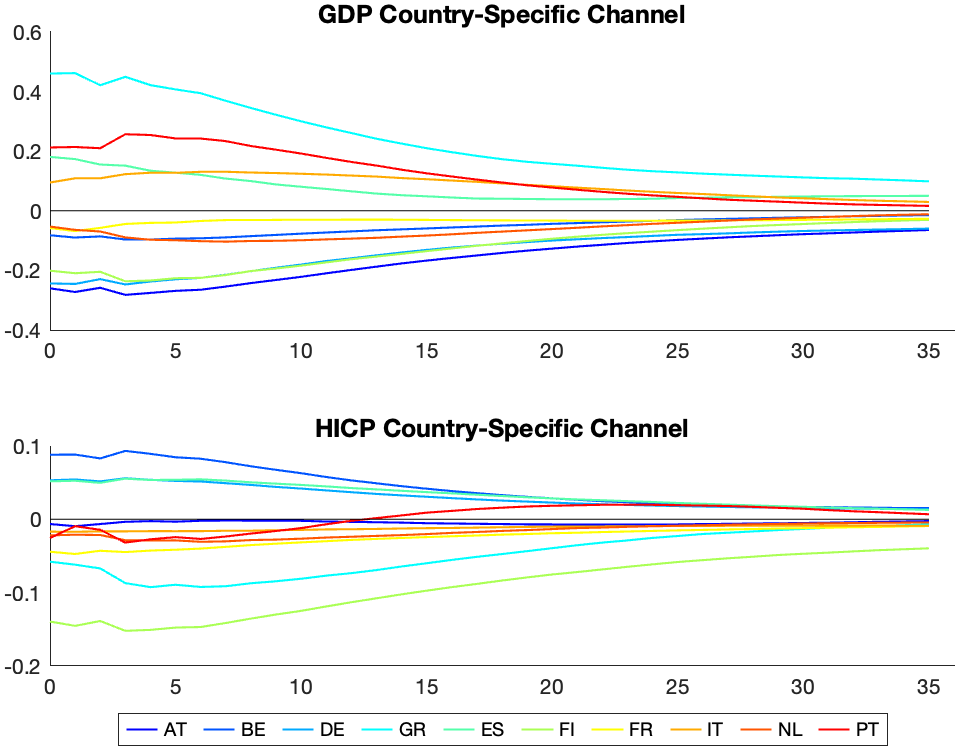}
       \end{minipage}
		  \caption{Country-level median impulse responses of country-specific output growth and inflation
            to an expansionary monetary policy shock (\emph{poor man's proxy} ). The left-hand side plots show the propagation via the common cycles and the right-hand side plots show the direct impact via the country-specific channels. The solid blue lines depicts the median responses of the euro area aggregate (EA19) and the shaded light blue areas the 68\% credible bands.} \label{figOA6}
    \end{figure}
\end{center}
\vspace{-50pt}

\clearpage
\subsection*{Three-Months OIS Rate}

In this section, we show the results of the model that employs the 3-months OIS rate \citeA{jk20}.

\begin{center}
	\begin{figure}[H]
    \includegraphics[height=8\baselineskip,width=0.99\textwidth]{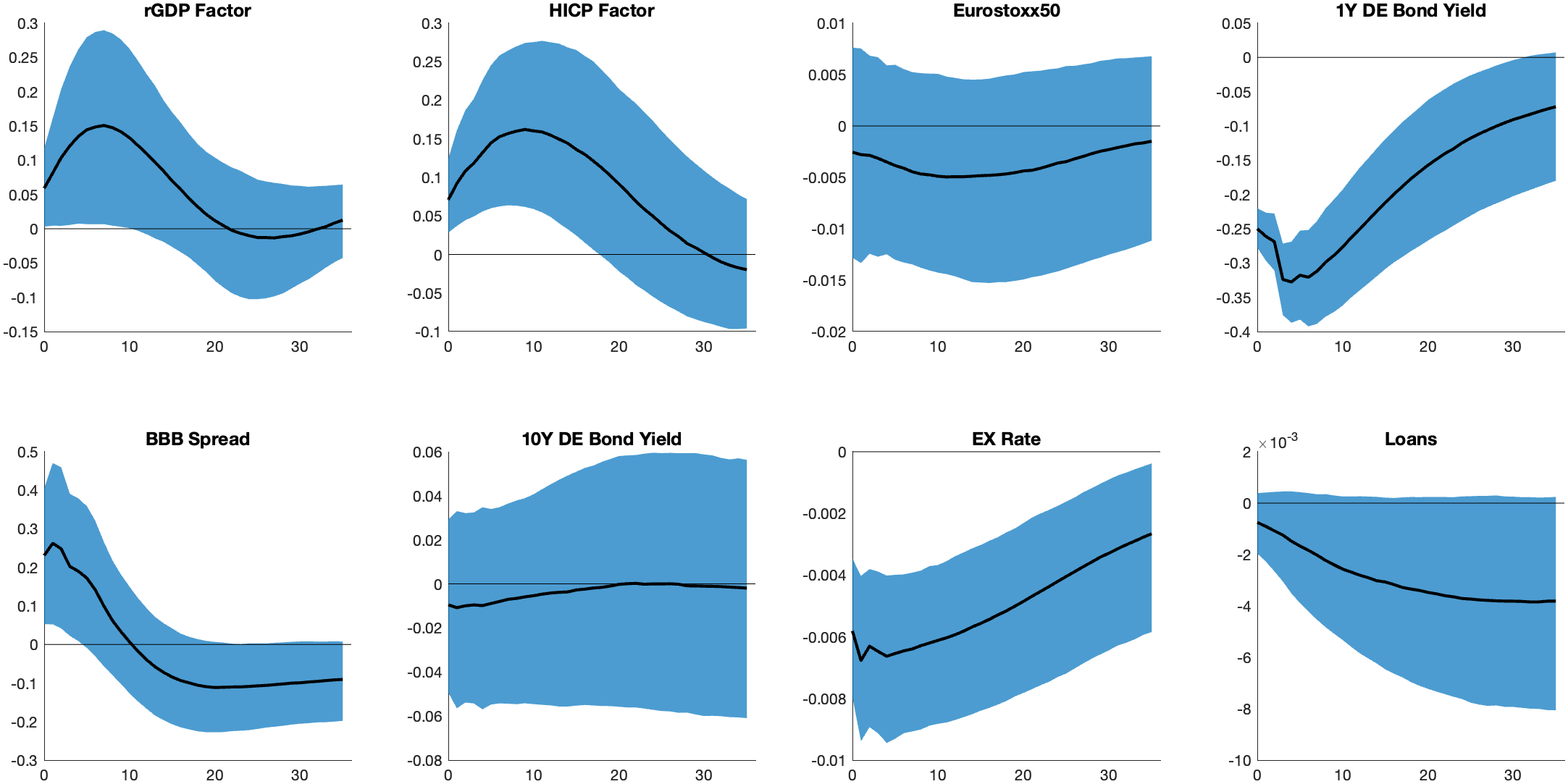}
      \caption{Country-level median impulse responses of country-specific output growth and inflation
            to an expansionary monetary policy shock. The left-hand side plots show the propagation via the common cycles and the right-hand side plots show the direct impact via the country-specific channels. The solid blue lines depicts the median responses of the euro area aggregate (EA19) and the shaded light blue areas the 68\% credible bands.} \label{fig_OA7}
	\end{figure}
\end{center}
\vspace{-50pt}

\begin{center}
    \begin{figure}[H]
    \begin{minipage}{.5\textwidth}
      	\includegraphics[width=0.99\textwidth]{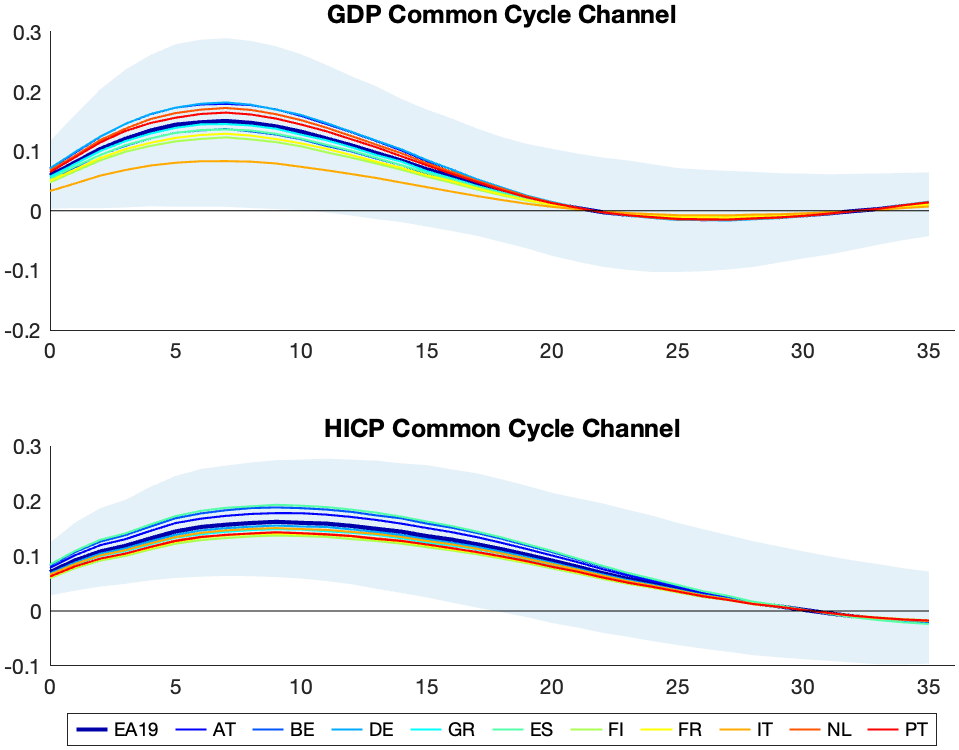}
       \end{minipage}
        \begin{minipage}{.5\textwidth}
      	\includegraphics[width=0.99\textwidth]{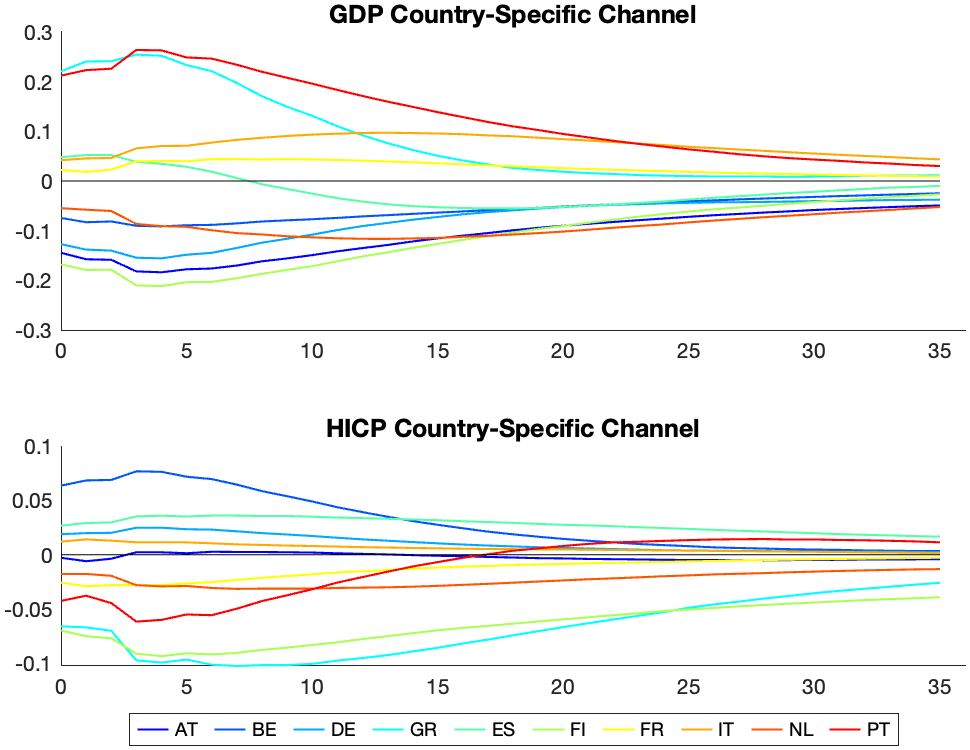}
       \end{minipage}
		  \caption{Country-level median impulse responses of country-specific GDP growth and inflation to an expansionary monetary policy shock. The left-hand side plots show the propagation via the common cycles and the right-hand side plots show the direct impact via financial channels. The solid blue lines depicts the median responses of the euro area aggregate (EA19) and the shaded light blue areas the 68\% credible bands.} \label{fig_OA8}
    \end{figure}
\end{center}
\vspace{-50pt}

\clearpage
\subsection*{Principal Component of 1-Month to 1-Year OIS Rates}

In this section, we show the results of the model that employs principal component of the 1-, 3- and 6-month and the 1-year OIS rates as instruments \citeA{j22}.

\begin{center}
	\begin{figure}[H]
    \includegraphics[height=8\baselineskip,width=0.99\textwidth]{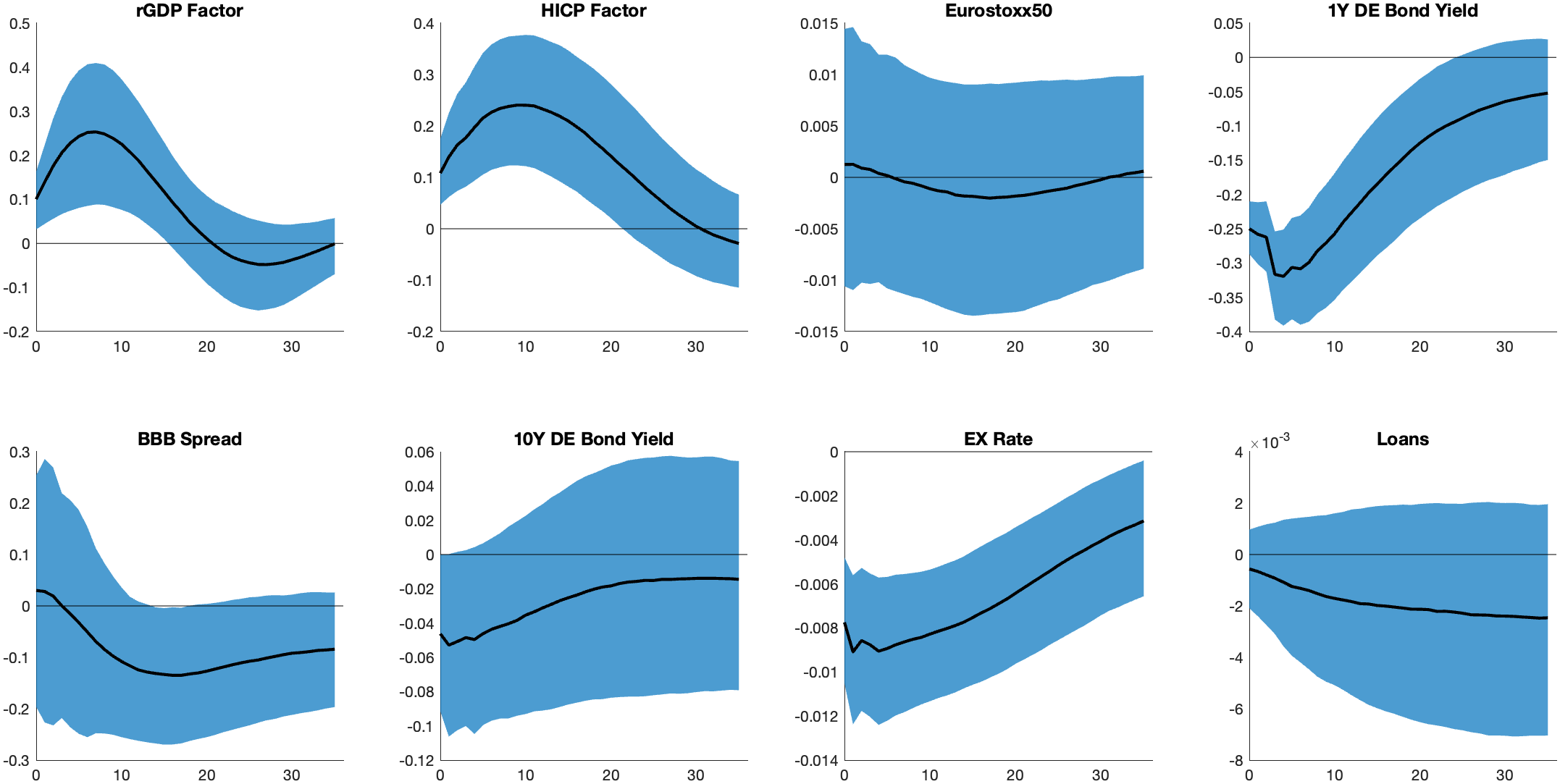}
      \caption{Median impulse responses to an expansionary monetary policy shock. The solid black line depicts the median and the shaded blue area the  68\% credible bands.} \label{fig_OA9}
	\end{figure}
\end{center}
\vspace{-50pt}

\begin{center}
    \begin{figure}[H]
    \begin{minipage}{.5\textwidth}
      	\includegraphics[width=0.99\textwidth]{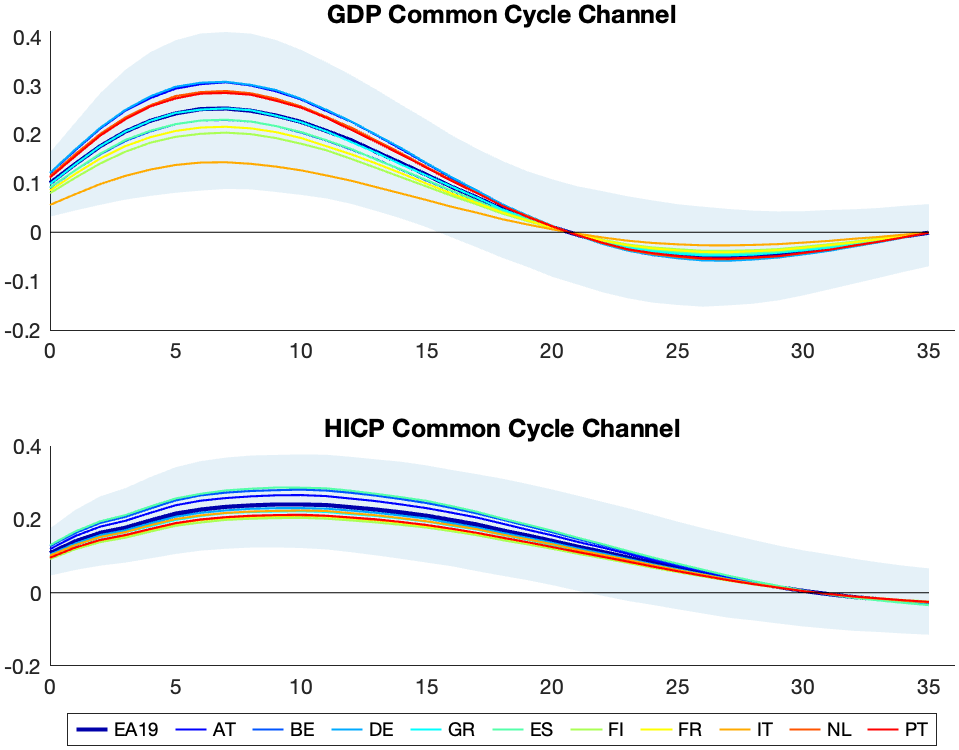}
       \end{minipage}
        \begin{minipage}{.5\textwidth}
      	\includegraphics[width=0.99\textwidth]{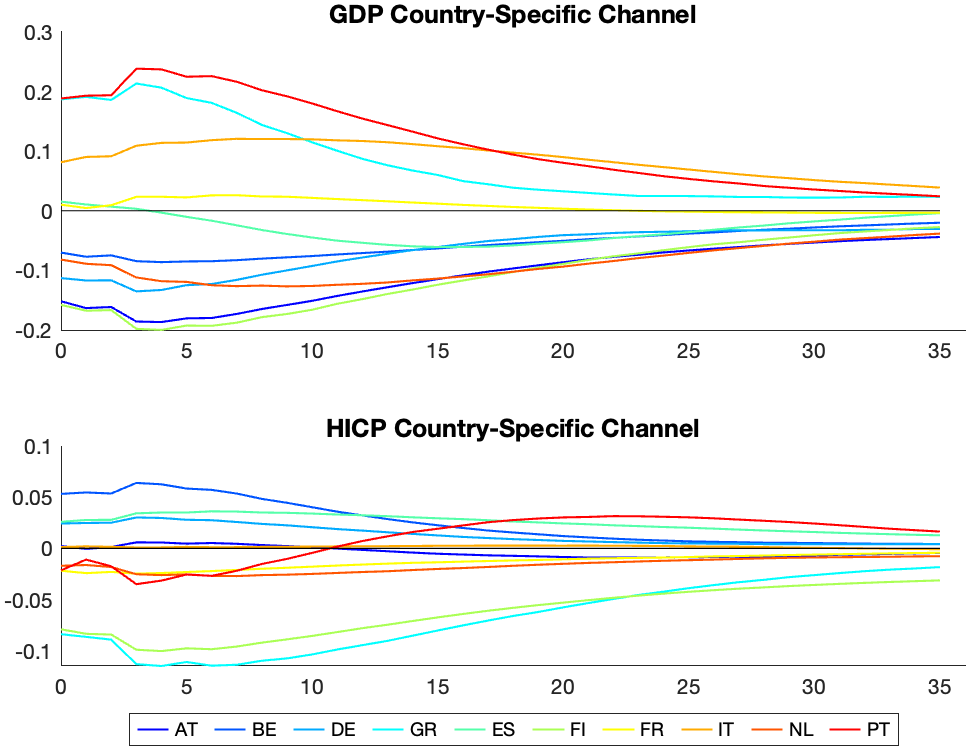}
       \end{minipage}
		  \caption{Country-level median impulse responses of country-specific output growth and inflation
            to an expansionary monetary policy shock. The left-hand side plots show the propagation via the common cycles and the right-hand side plots show the direct impact via the country-specific channels. The solid blue lines depicts the median responses of the euro area aggregate (EA19) and the shaded light blue areas the 68\% credible bands.} \label{fig_OA10}
    \end{figure}
\end{center}
\vspace{-50pt}

\clearpage
\subsection*{Euro Area Average One-Year Government Bond}

In this section, we show the results of the model that employs the euro area average one-year government bond yield as policy rate in the vector $\boldsymbol{z}_{t}$.

\begin{center}
	\begin{figure}[H]
    \includegraphics[height=8\baselineskip,width=0.99\textwidth]{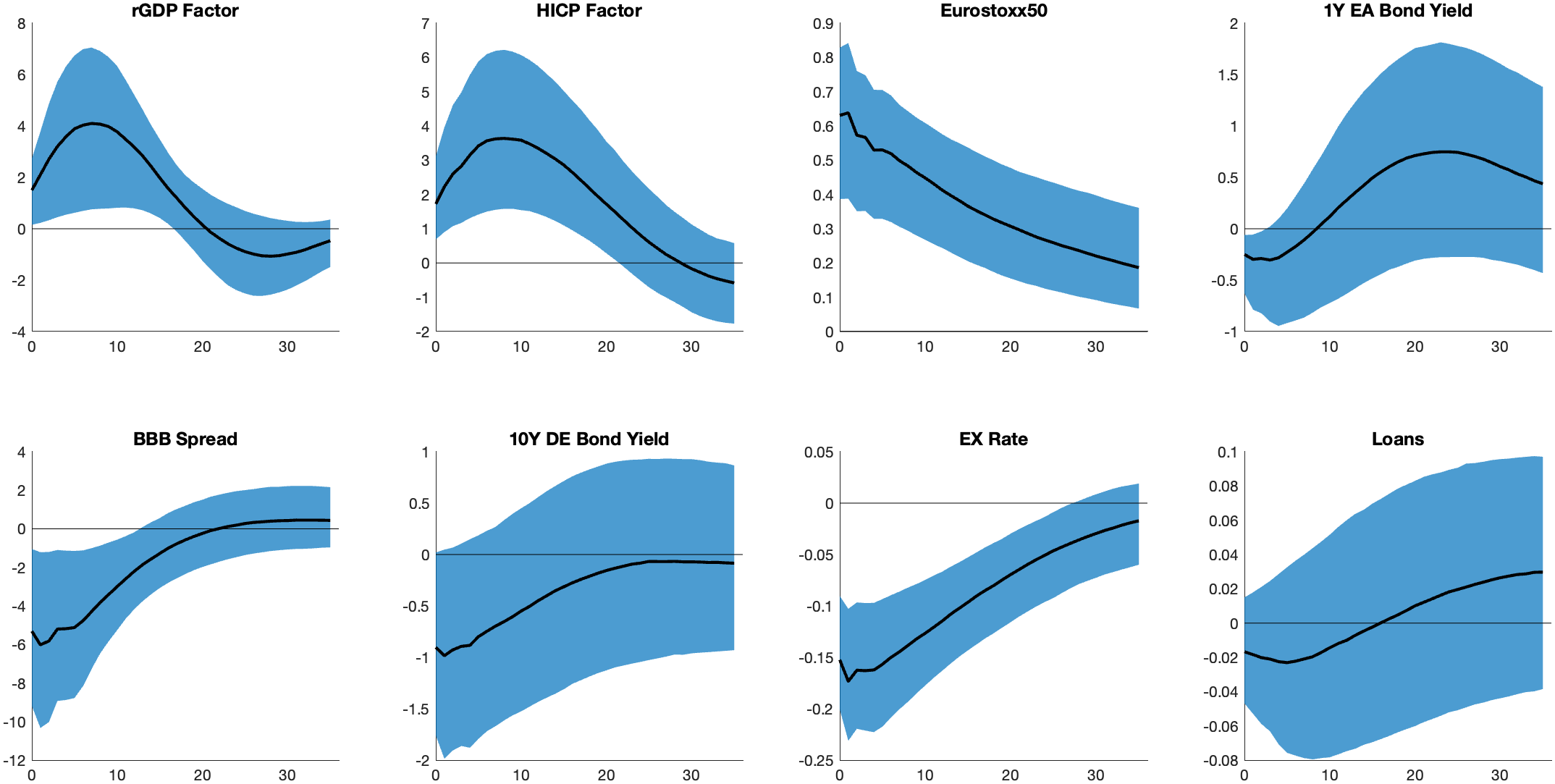}
      \caption{Median impulse responses to an expansionary monetary policy shock (\emph{one-year euro area benchmark government bond yield}). The solid black line depicts the median and the shaded blue area the  68\% credible bands.} \label{fig_OA11}
	\end{figure}
\end{center}
\vspace{-50pt}

\begin{center}
    \begin{figure}[H]
    \begin{minipage}{.5\textwidth}
      	\includegraphics[width=0.99\textwidth]{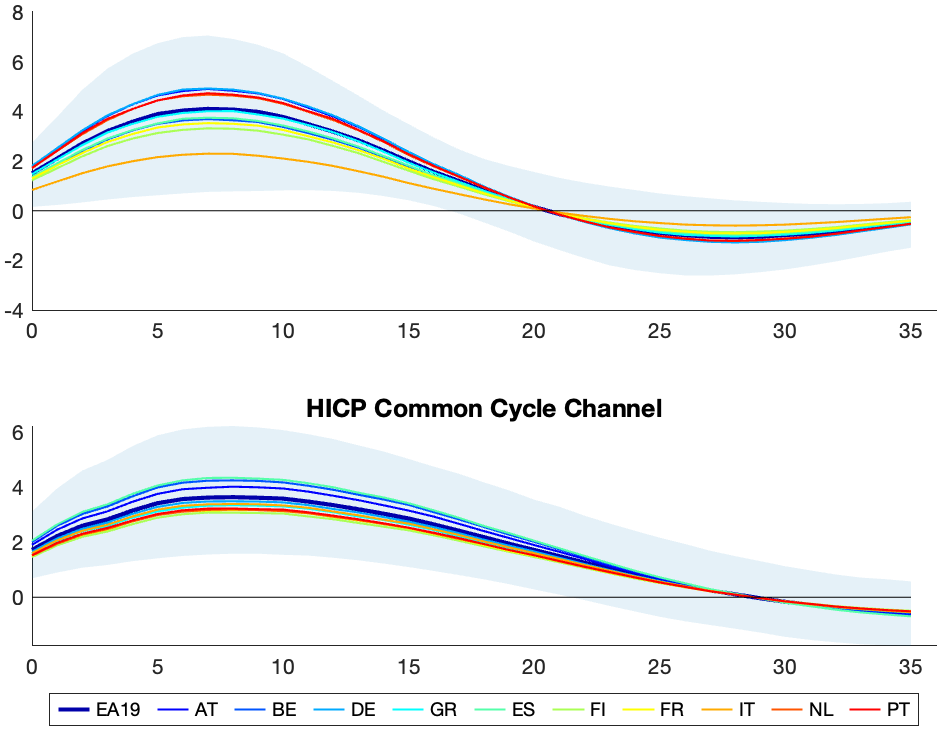}
       \end{minipage}
        \begin{minipage}{.5\textwidth}
      	\includegraphics[width=0.99\textwidth]{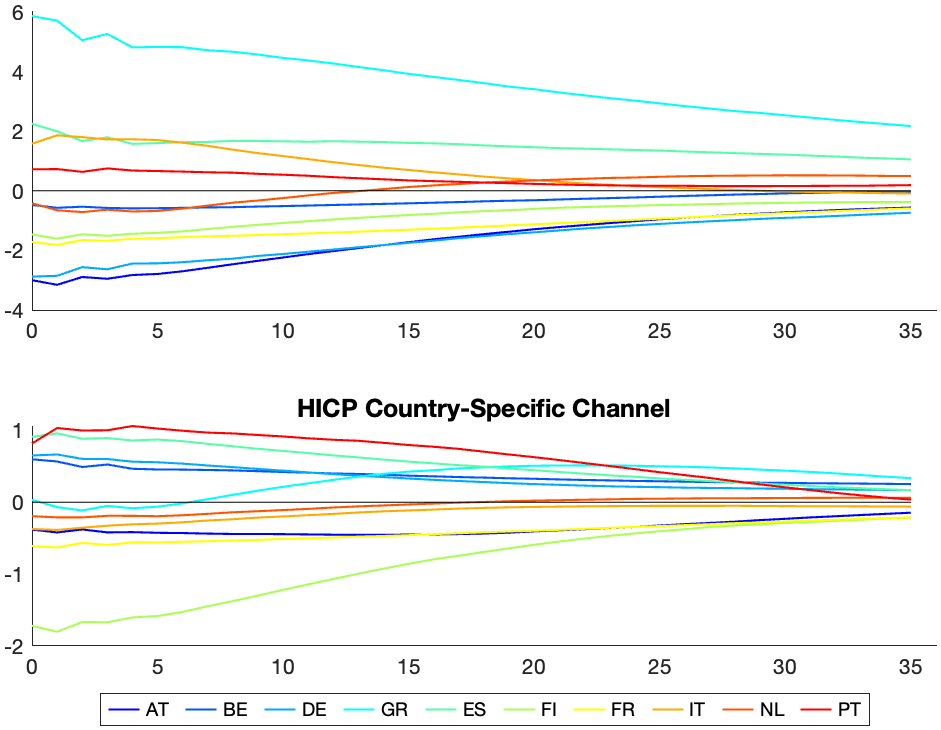}
       \end{minipage}
		  \caption{Country-level median impulse responses of country-specific output growth and inflation
            to an expansionary monetary policy shock (\emph{one-year euro area benchmark government bond yield}). The left-hand side plots show the propagation via the common cycles and the right-hand side plots show the direct impact via the country-specific channels. The solid blue lines depicts the median responses of the euro area aggregate (EA19) and the shaded light blue areas the 68\% credible bands.} \label{fig_OA12}
    \end{figure}
\end{center}
\vspace{-50pt}

\clearpage
\subsection*{Less Informative Proxy}

In this alternative model specification, we set the priors for (non-zero) parameters in the proxy variable equation ($\Phi_{0,1}$, $\Phi_{0,2}$) as follows:
\vspace{-10pt}
\begin{eqnarray}
    \Phi_{0,1} \sim N(\phi_{0,1},V_{0,1}), \quad \Phi_{0,2} \sim N(\phi_{0,2},V_{0,2}), \nonumber
\end{eqnarray}
with $\phi_{0,1} = 0, \phi_{0,2} = 0, V_{0,1} = 1$ and $V_{0,2} = 1$. 

\begin{center}
	\begin{figure}[H]
    \includegraphics[height=8\baselineskip,width=0.99\textwidth]{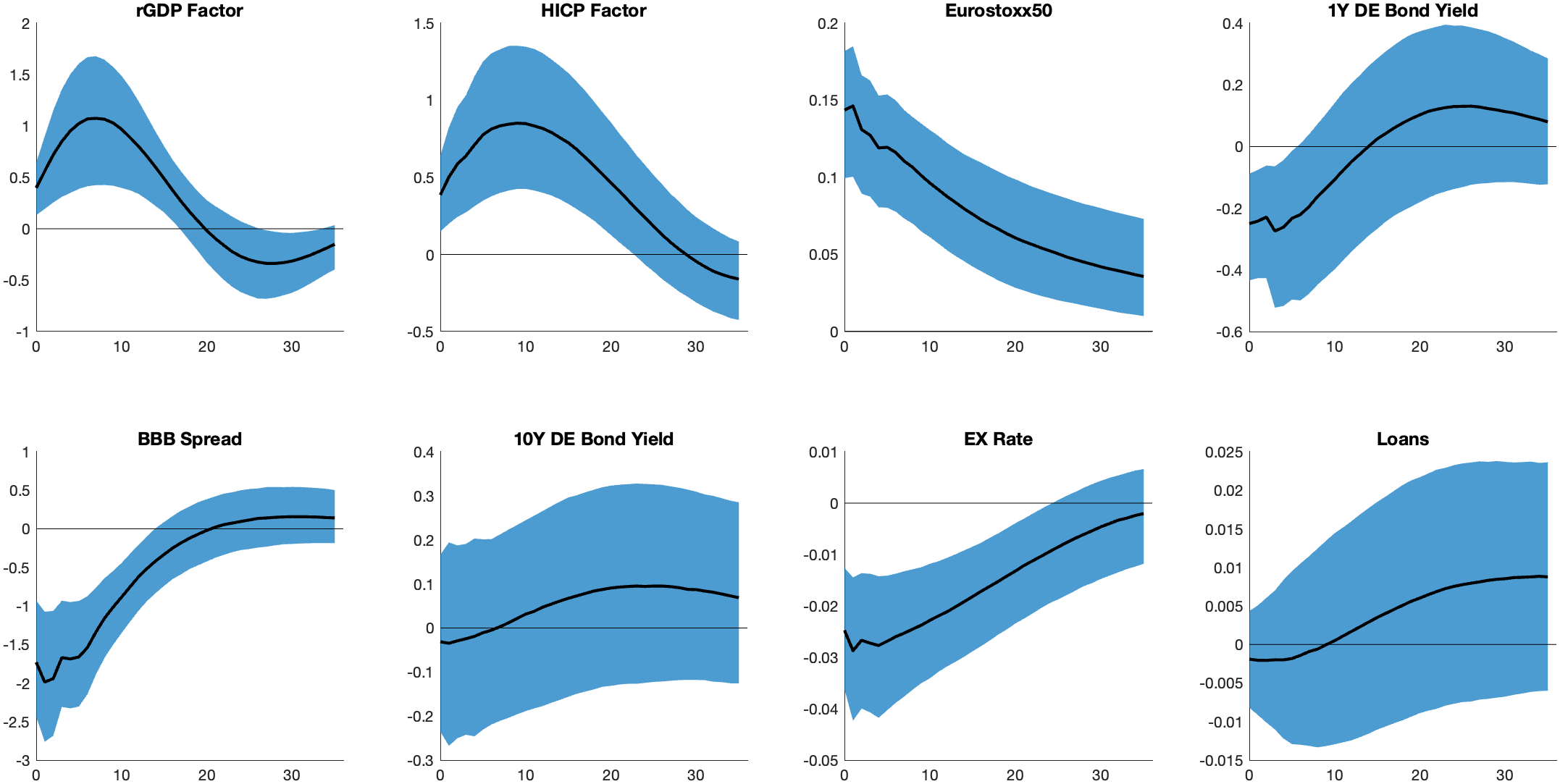}
      \caption{Median impulse responses to an expansionary monetary policy shock. The solid black line depicts the median and the shaded blue area the  68\% credible bands.} \label{fig_OA13}
	\end{figure}
\end{center}
\vspace{-50pt}

\begin{center}
    \begin{figure}[H]
    \begin{minipage}{.5\textwidth}
      	\includegraphics[width=0.99\textwidth]{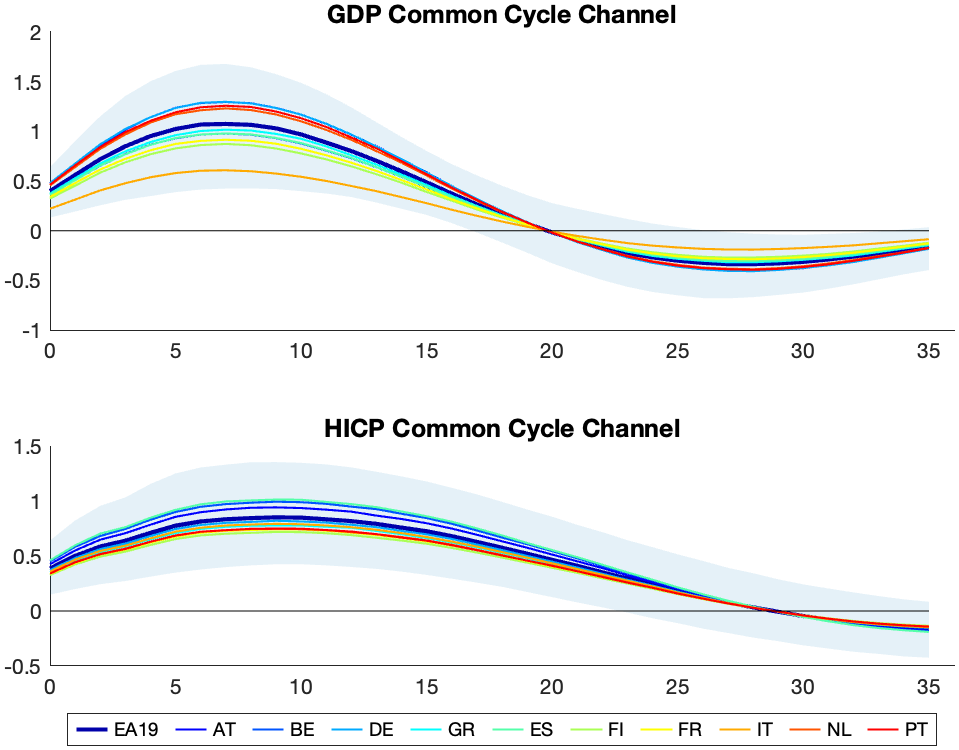}
       \end{minipage}
        \begin{minipage}{.5\textwidth}
      	\includegraphics[width=0.99\textwidth]{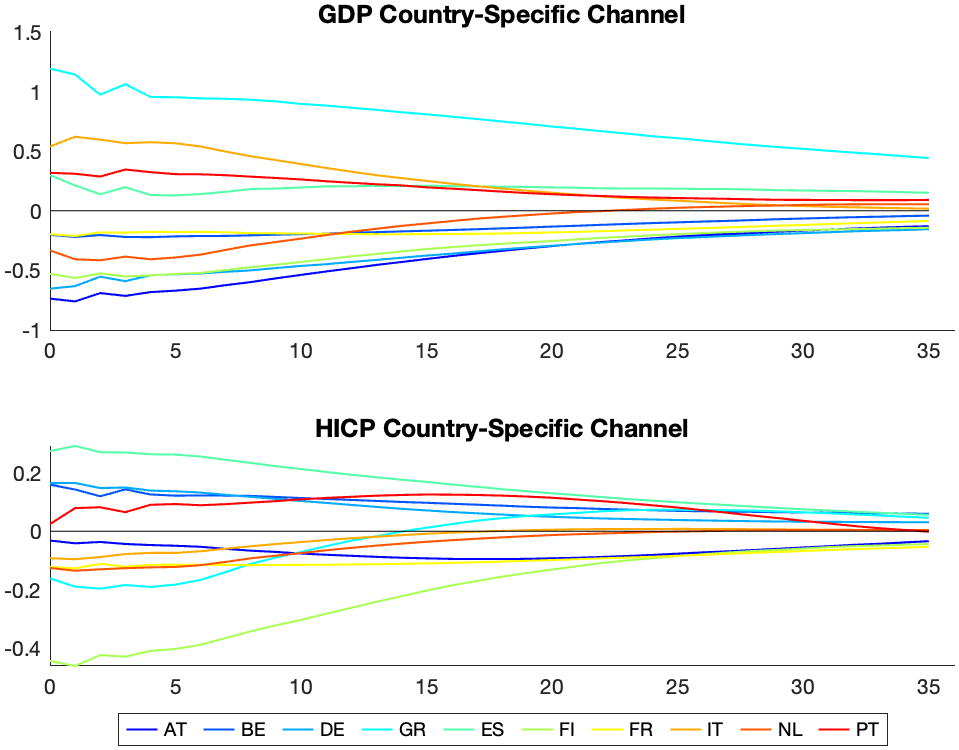}
       \end{minipage}
		  \caption{Country-level median impulse responses of country-specific output growth and inflation
            to an expansionary monetary policy shock. The left-hand side plots show the propagation via the common cycles and the right-hand side plots show the direct impact via the country-specific channels. The solid blue lines depicts the median responses of the euro area aggregate (EA19) and the shaded light blue areas the 68\% credible bands.} \label{fig_OA14}
    \end{figure}
\end{center}
\vspace{-50pt}

\clearpage
\subsection*{Industrial Production}

In this section, we show the results of the model that replaces interpolated GDP growth with growth in industrial production.

\begin{center}
	\begin{figure}[H]
    \includegraphics[width=0.99\textwidth]{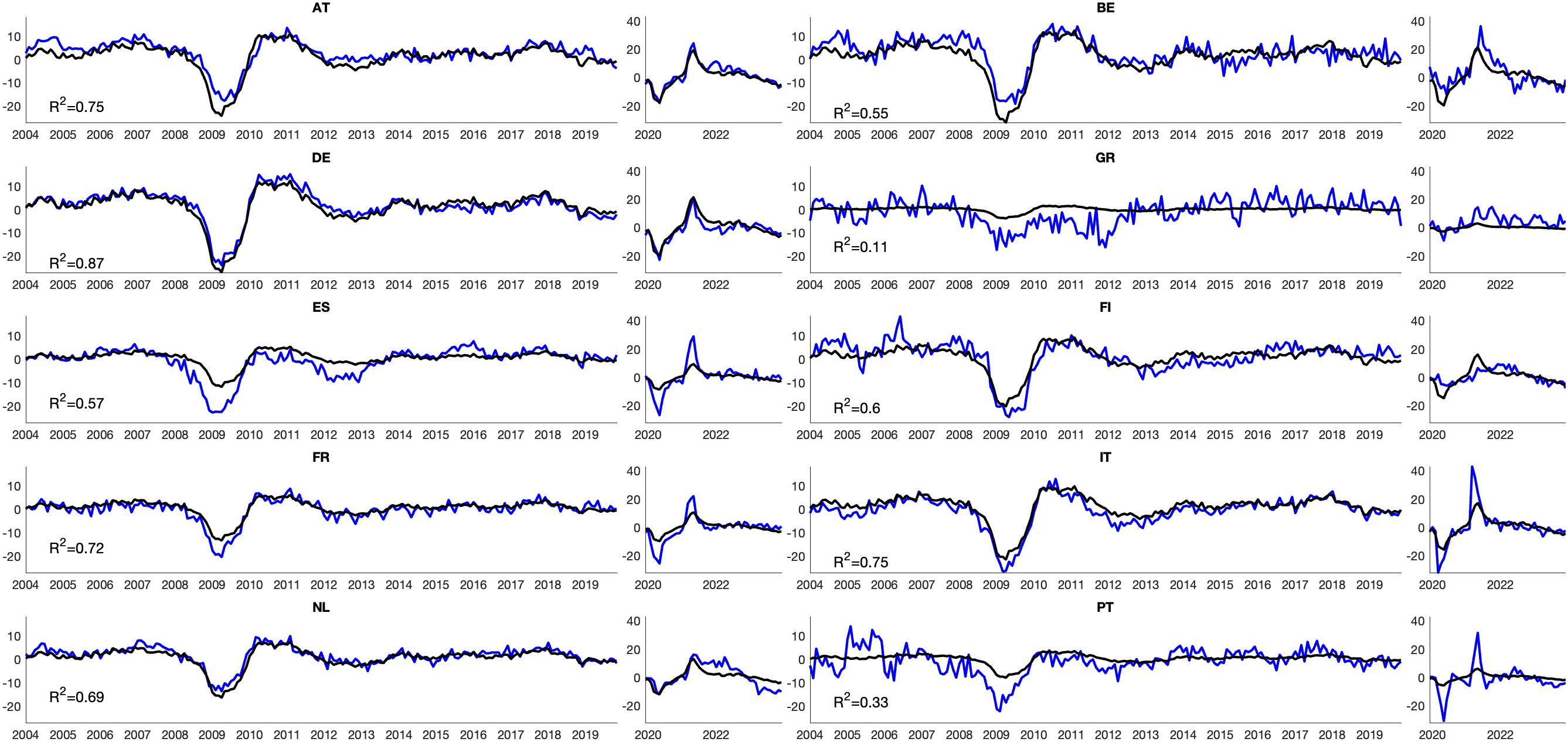}
      \caption{Countries' exposure to common cycle for industrial production growth. The solid blue line depicts industrial production growth and the solid black line portrays the common euro cycle for industrial production growth multiplied by the corresponding country-specific factor loadings. $R^{2}$ denotes the R-squared.} \label{fig_OA15}
	\end{figure}
\end{center}
\vspace{-50pt}

\begin{center}
	\begin{figure}[H]
    \includegraphics[height=8\baselineskip,width=0.99\textwidth]{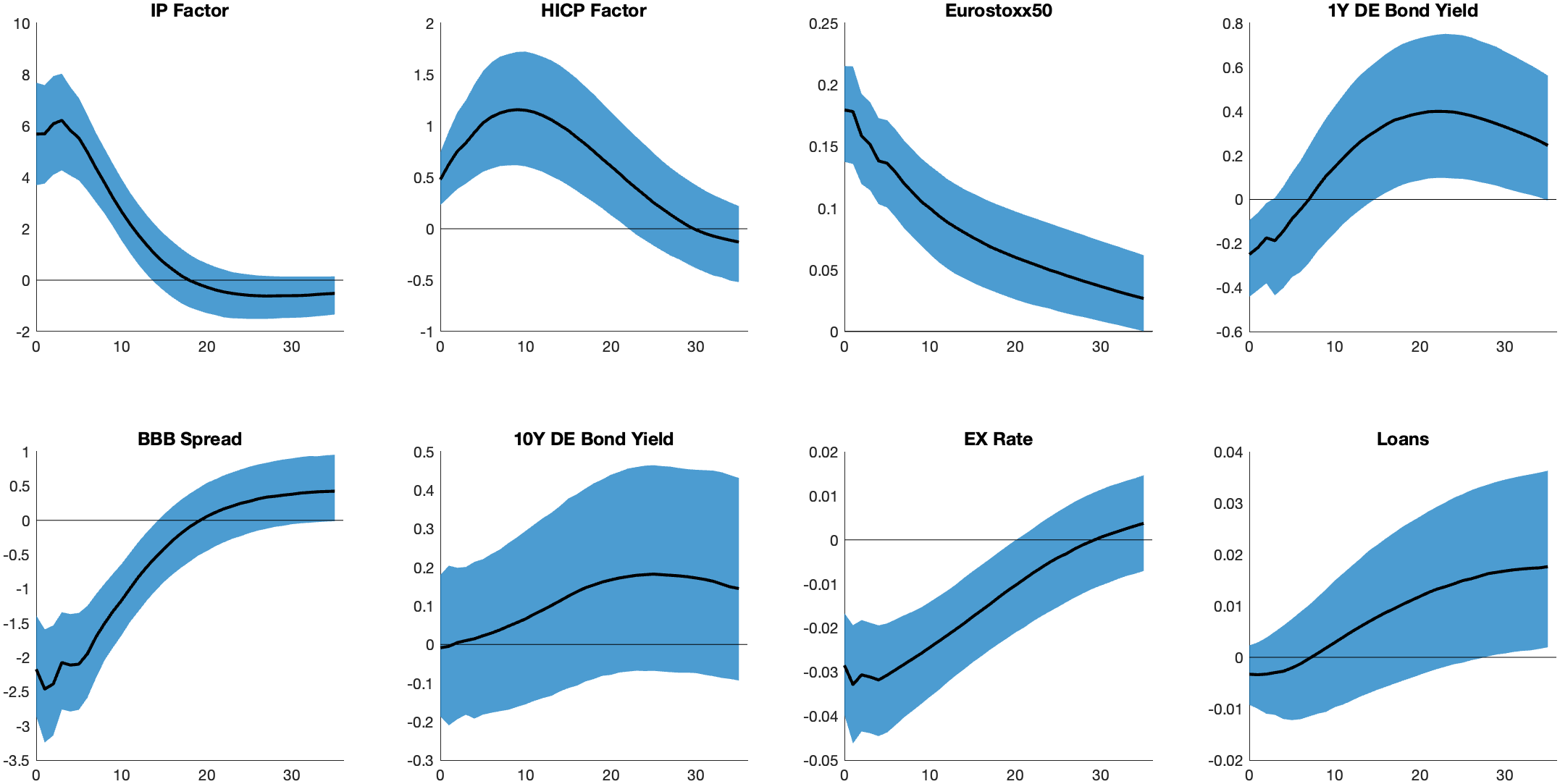}
      \caption{Median impulse responses to an expansionary monetary policy shock. The solid black line depicts the median and the shaded blue area the  68\% credible bands.} \label{fig_OA16}
	\end{figure}
\end{center}
\vspace{-50pt}

\begin{center}
    \begin{figure}[H]
    \begin{minipage}{.5\textwidth}
      	\includegraphics[width=0.99\textwidth]{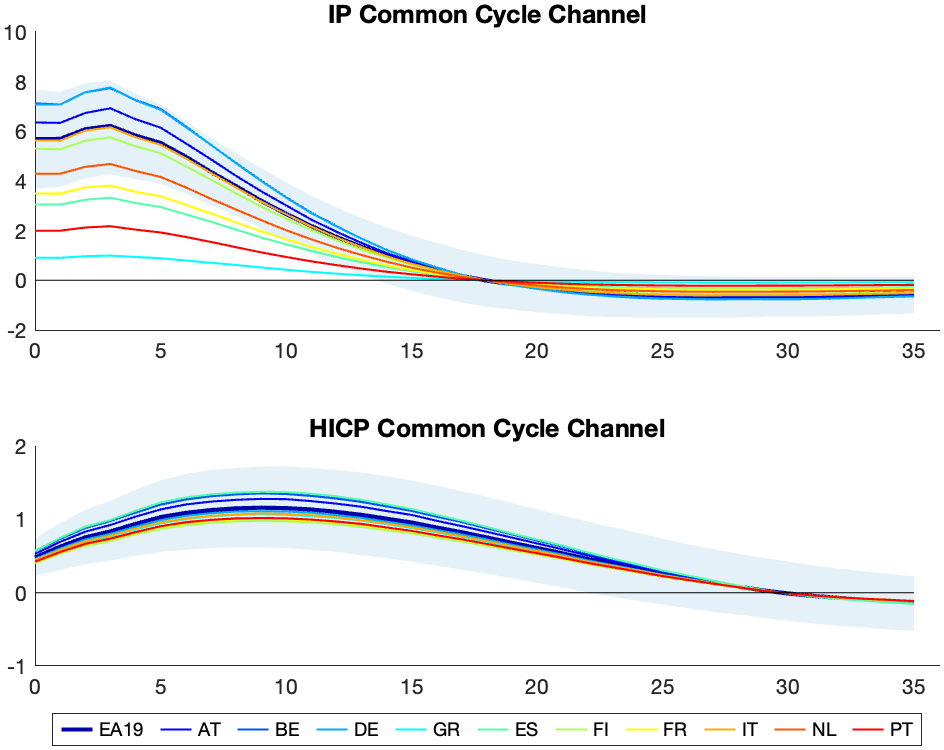}
       \end{minipage}
        \begin{minipage}{.5\textwidth}
      	\includegraphics[width=0.99\textwidth]{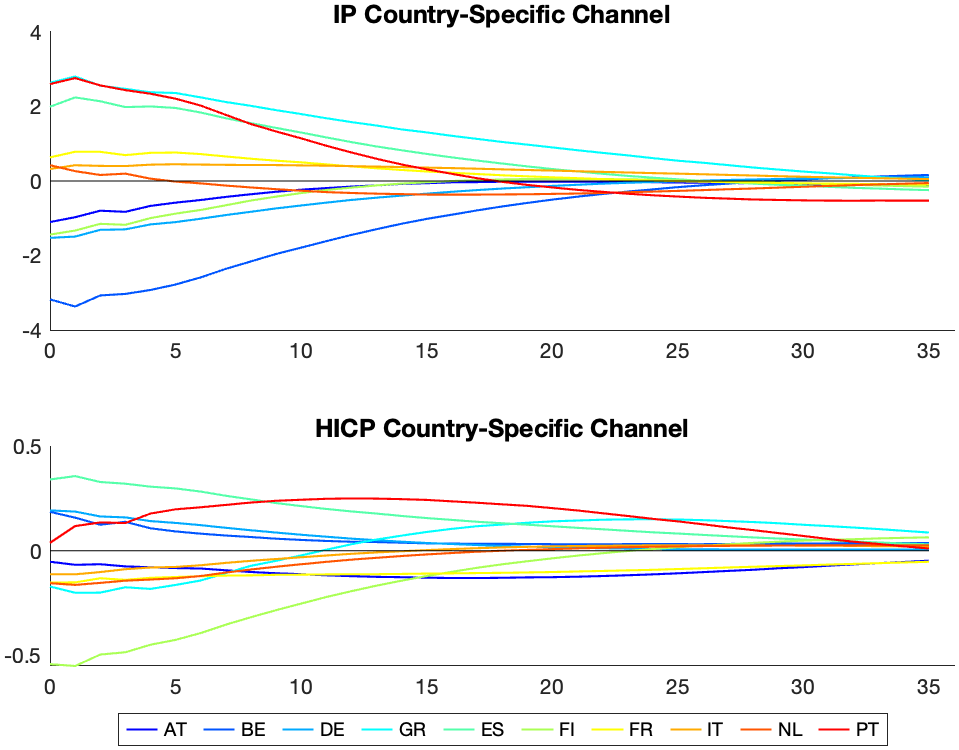}
       \end{minipage}
		  \caption{Country-level median impulse responses of country-specific industrial production growth and inflation
            to an expansionary monetary policy shock. The left-hand side plots show the propagation via the common cycles and the right-hand side plots show the direct impact via the country-specific channels. The solid blue lines depicts the median responses of the euro area aggregate (EA19) and the shaded light blue areas the 68\% credible bands.} \label{fig_OA17}
    \end{figure}
\end{center}
\vspace{-50pt}

\clearpage
\subsection*{Sample until 2019}

In this section, we restrict the sample period to January 2003 to December 2019 and leave out the pandemic and the recent surge in inflation.

\begin{center}
	\begin{figure}[H]
    \includegraphics[height=8\baselineskip,width=0.99\textwidth]{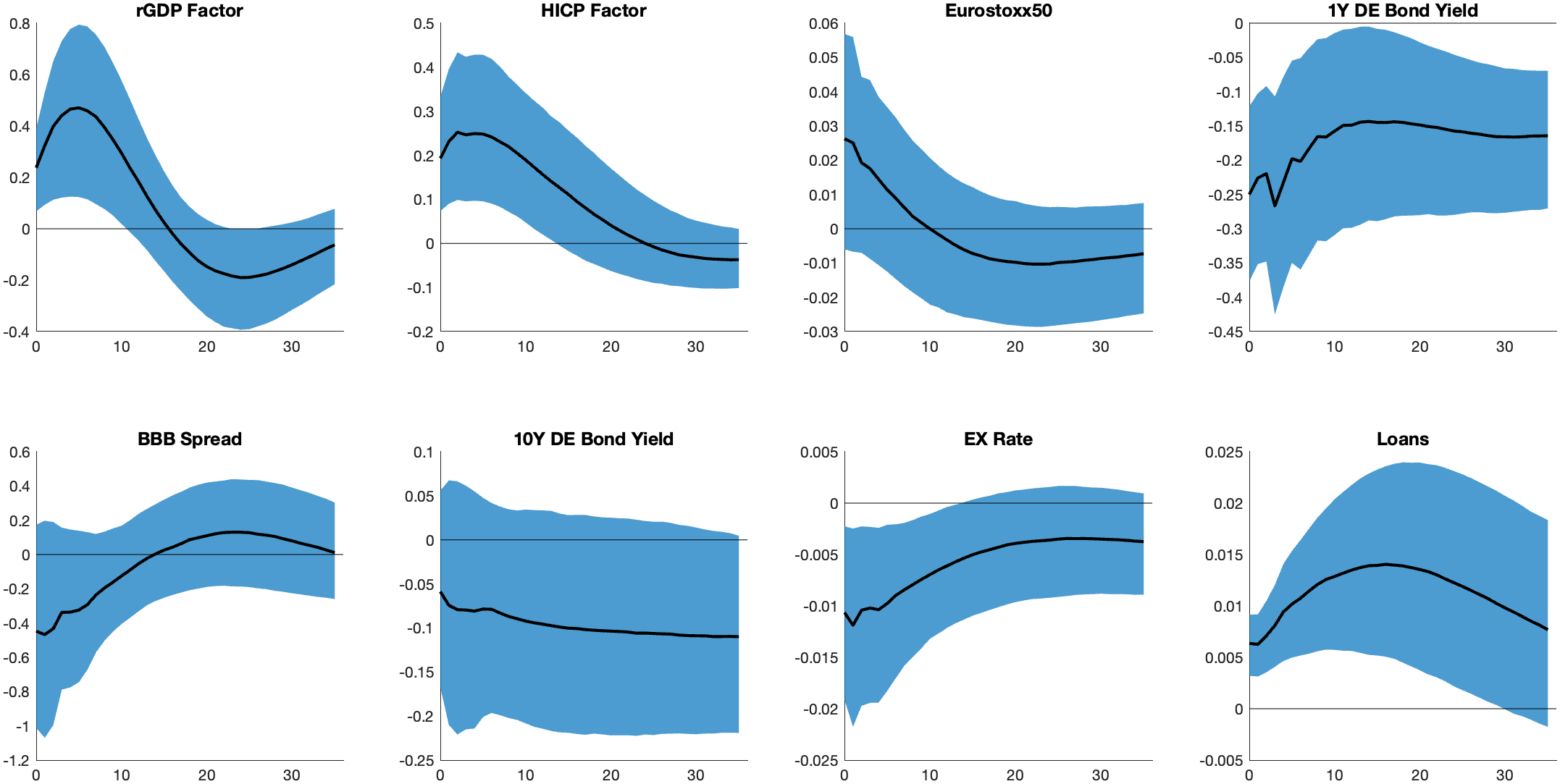}
      \caption{Median impulse responses to an expansionary monetary policy shock. The solid black line depicts the median and the shaded blue area the  68\% credible bands.} \label{fig_OA18}
	\end{figure}
\end{center}
\vspace{-50pt}

\begin{center}
    \begin{figure}[H]
    \begin{minipage}{.5\textwidth}
      	\includegraphics[width=0.99\textwidth]{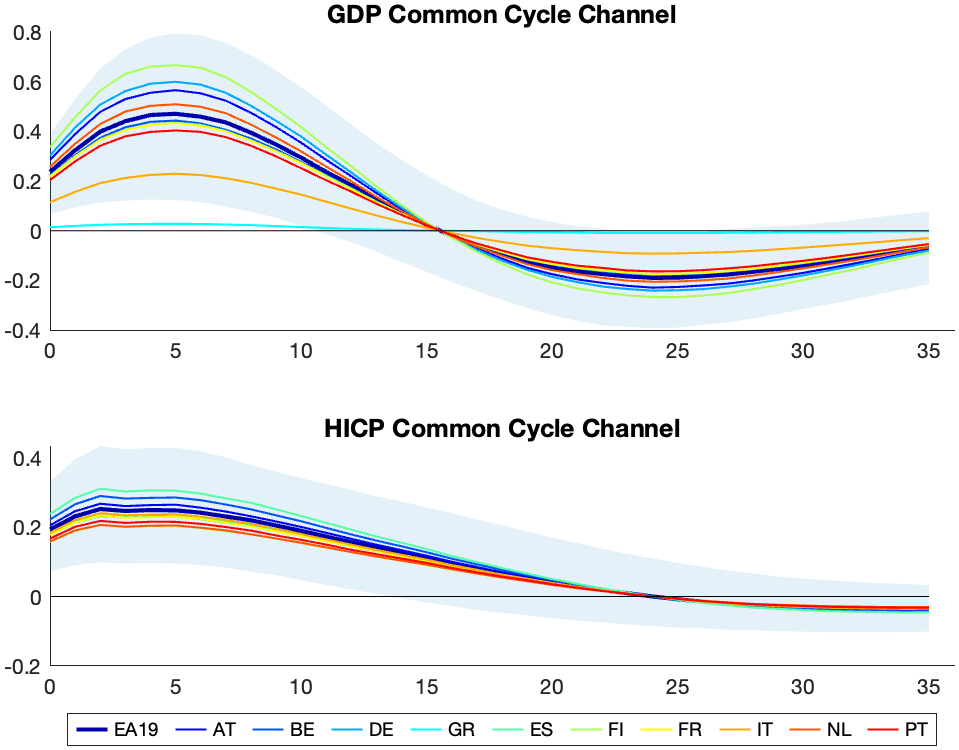}
       \end{minipage}
        \begin{minipage}{.5\textwidth}
      	\includegraphics[width=0.99\textwidth]{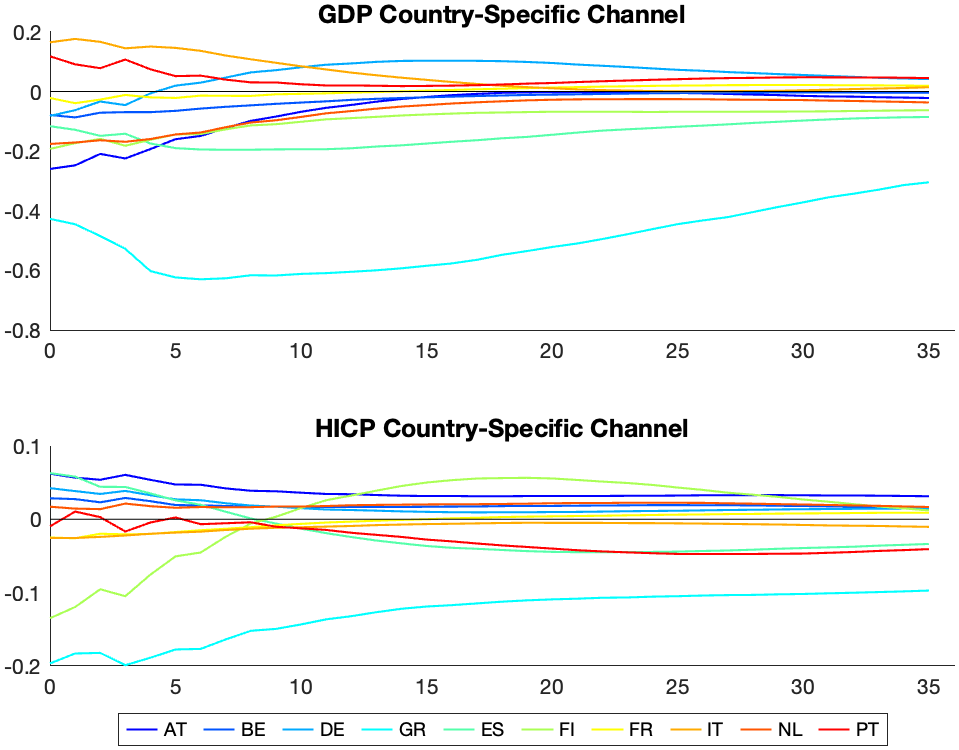}
       \end{minipage}
		  \caption{Country-level median impulse responses of country-specific output growth and inflation
            to an expansionary monetary policy shock. The left-hand side plots show the propagation via the common cycles and the right-hand side plots show the direct impact via the country-specific channels. The solid blue lines depicts the median responses of the euro area aggregate (EA19) and the shaded light blue areas the 68\% credible bands.} \label{fig_OA19}
    \end{figure}
\end{center}
\vspace{-50pt}

\clearpage
\subsection*{Alternative Lag Orders in the DFM}

\begin{center}
	\begin{figure}[H]
    \includegraphics[height=8\baselineskip,width=0.99\textwidth]{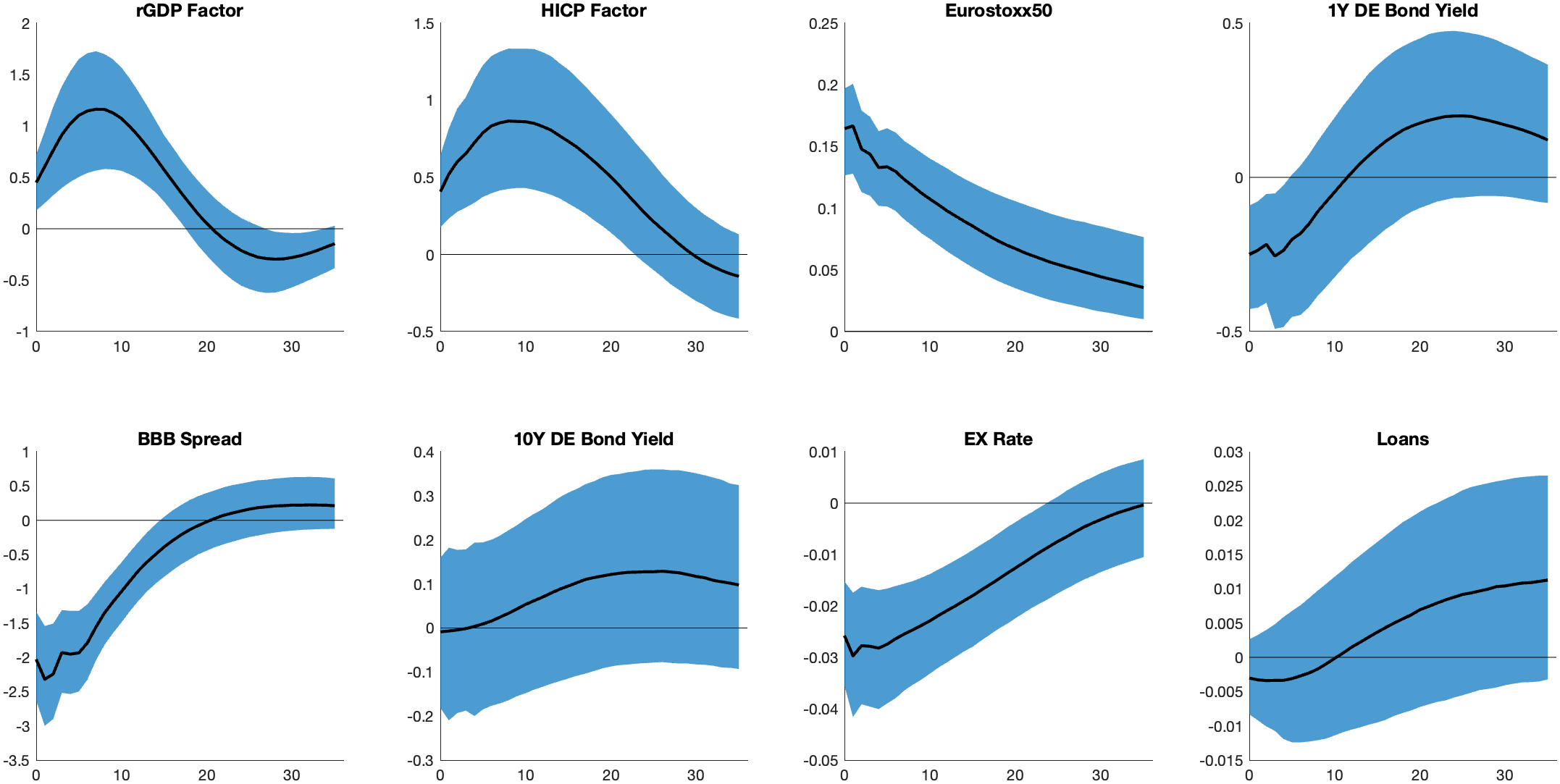}
      \caption{Median impulse responses to an expansionary monetary policy shock. The solid black line depicts the median and the shaded blue area the  68\% credible bands.} \label{fig_OA20}
	\end{figure}
\end{center}

\begin{center}
    \begin{figure}[H]
    \begin{minipage}{.5\textwidth}
      	\includegraphics[width=0.99\textwidth]{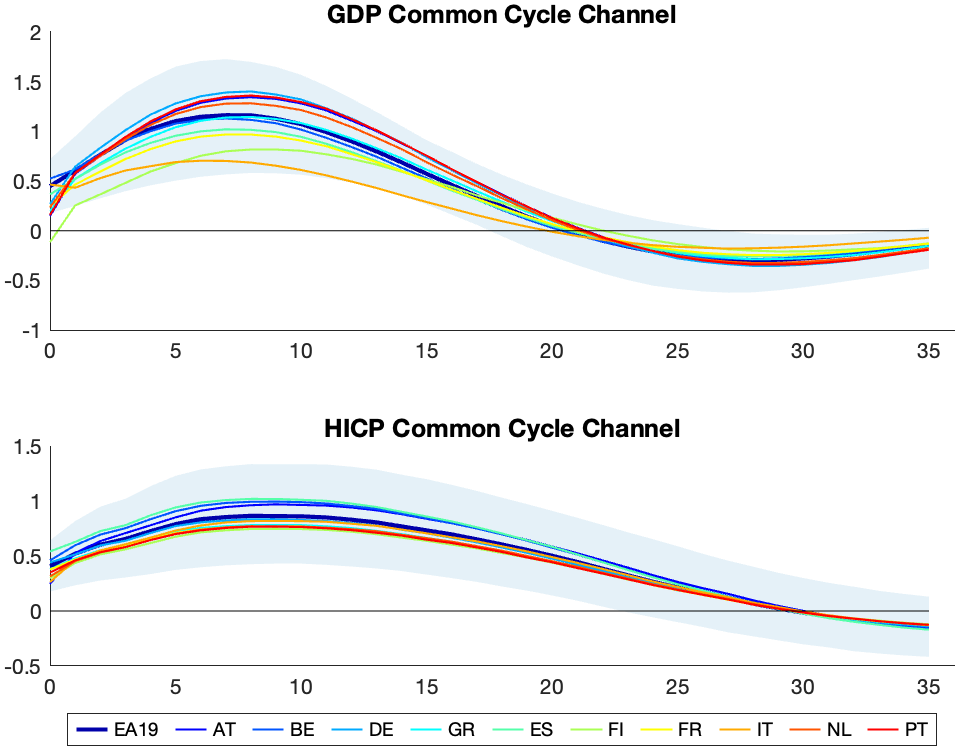}
       \end{minipage}
        \begin{minipage}{.5\textwidth}
      	\includegraphics[width=0.99\textwidth]{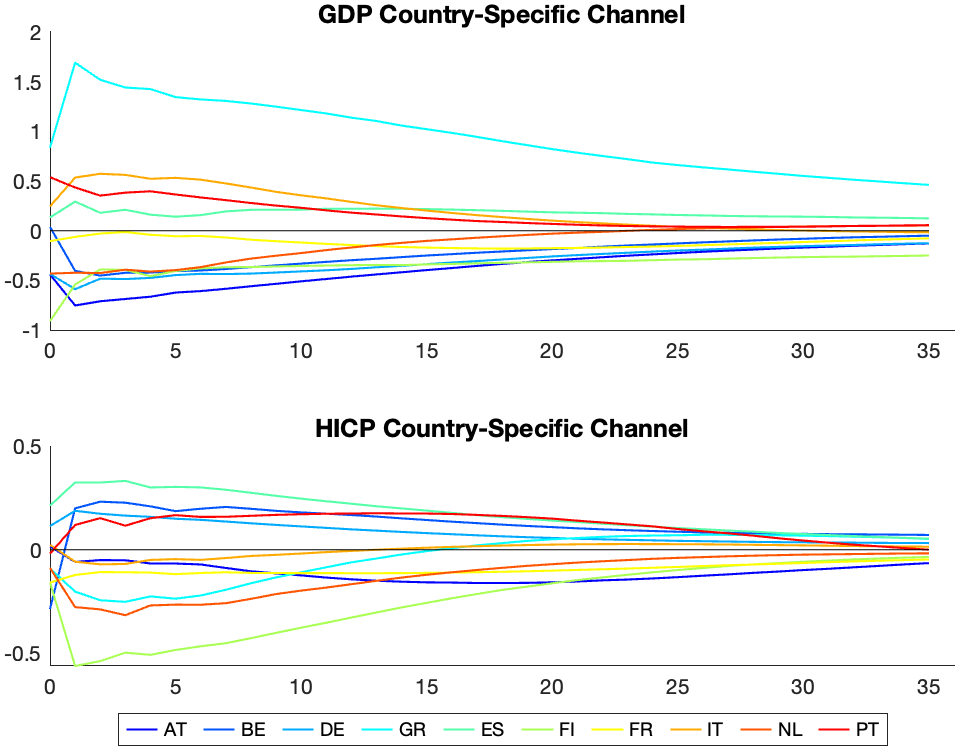}
       \end{minipage}
		  \caption{Country-level median impulse responses of country-specific output growth and inflation
            to an expansionary monetary policy shock (\emph{$P=1$}). The left-hand side plots show the propagation via the common cycles and the right-hand side plots show the direct impact via the country-specific channels. The solid blue lines depicts the median responses of the euro area aggregate (EA19) and the shaded light blue areas the 68\% credible bands.} \label{fig_OA21}
    \end{figure}
\end{center}

\begin{center}
	\begin{figure}[H]
    \includegraphics[height=8\baselineskip,width=0.99\textwidth]{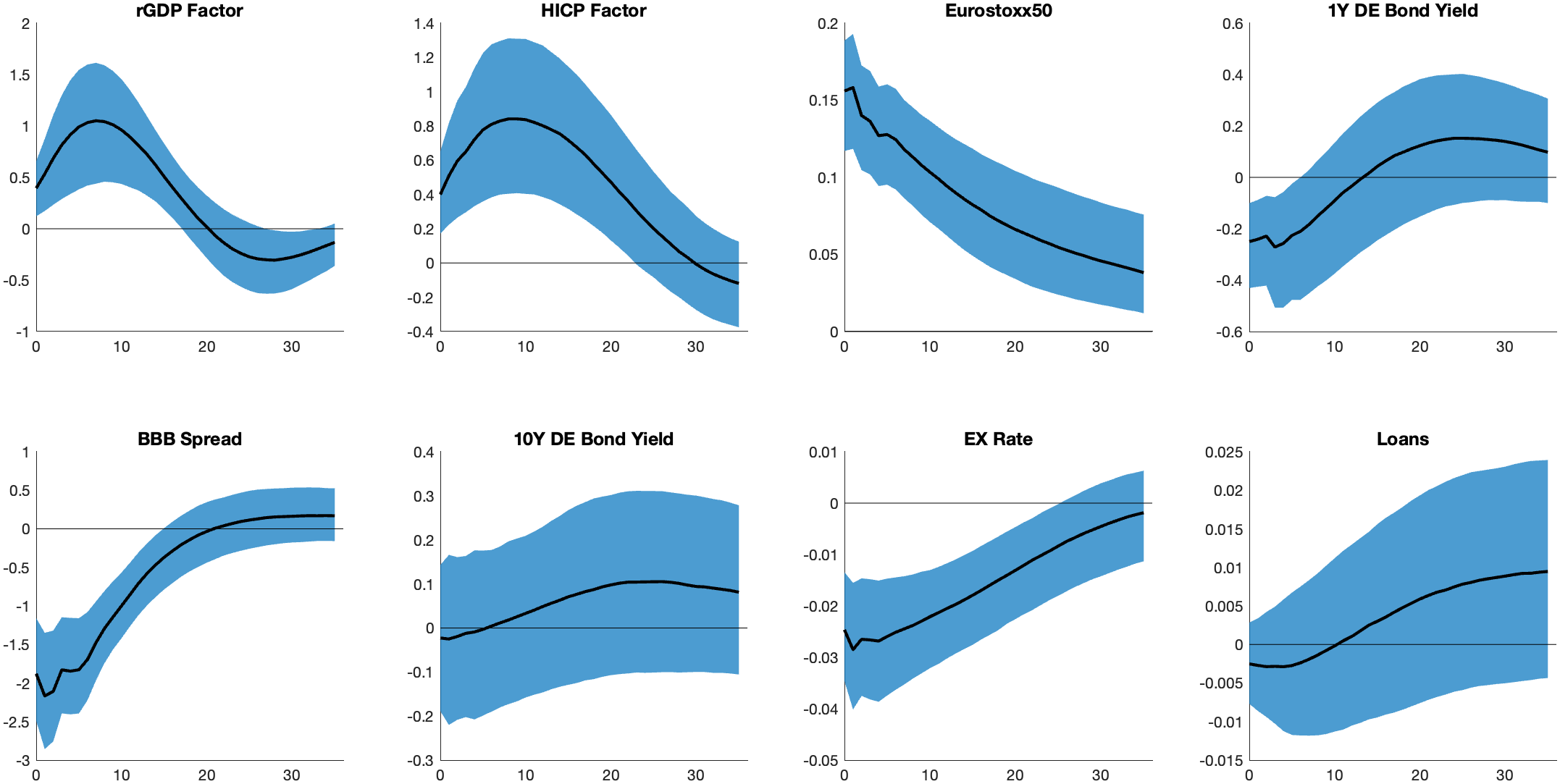}
      \caption{Median impulse responses to an expansionary monetary policy shock. The solid black line depicts the median and the shaded blue area the  68\% credible bands.} \label{fig_OA22}
	\end{figure}
\end{center}

\begin{center}
    \begin{figure}[H]
    \begin{minipage}{.5\textwidth}
      	\includegraphics[width=0.99\textwidth]{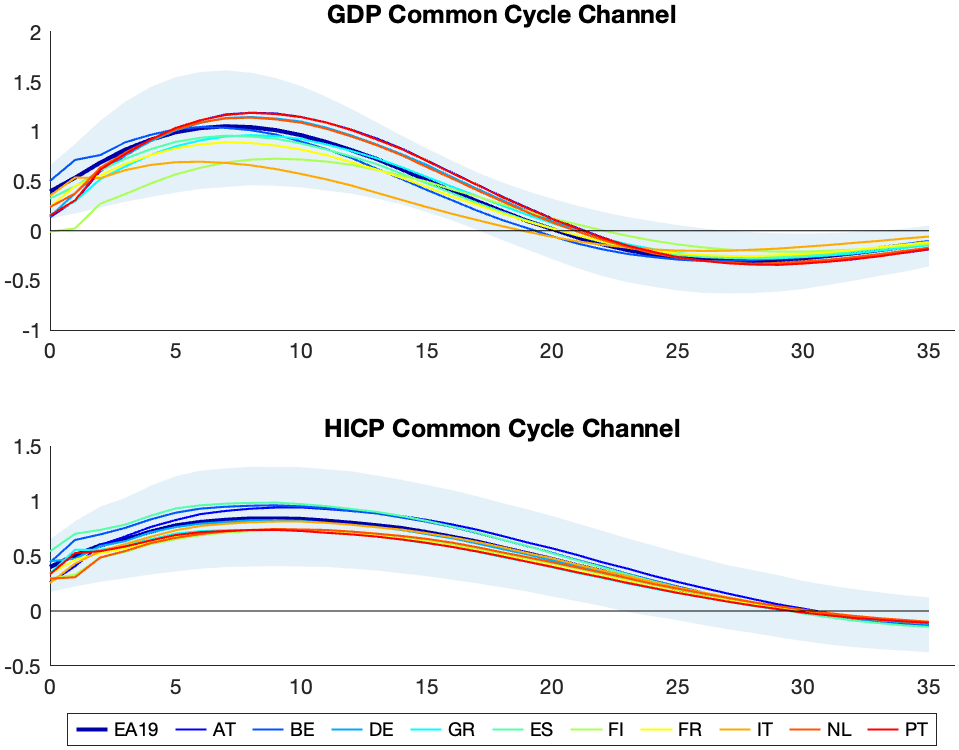}
       \end{minipage}
        \begin{minipage}{.5\textwidth}
      	\includegraphics[width=0.99\textwidth]{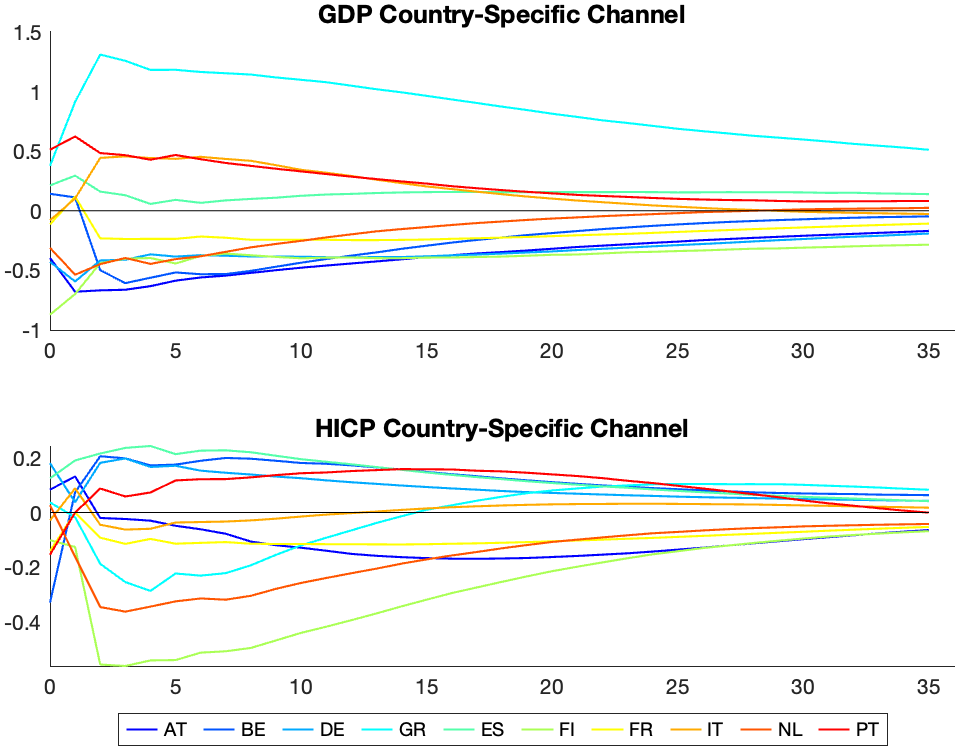}
       \end{minipage}
		  \caption{Country-level median impulse responses of country-specific output growth and inflation
            to an expansionary monetary policy shock (\emph{$P=2$}). The left-hand side plots show the propagation via the common cycles and the right-hand side plots show the direct impact via the country-specific channels. The solid blue lines depicts the median responses of the euro area aggregate (EA19) and the shaded light blue areas the 68\% credible bands.} \label{fig_OA23}
    \end{figure}
\end{center}

\begin{center}
	\begin{figure}[H]
    \includegraphics[height=8\baselineskip,width=0.99\textwidth]{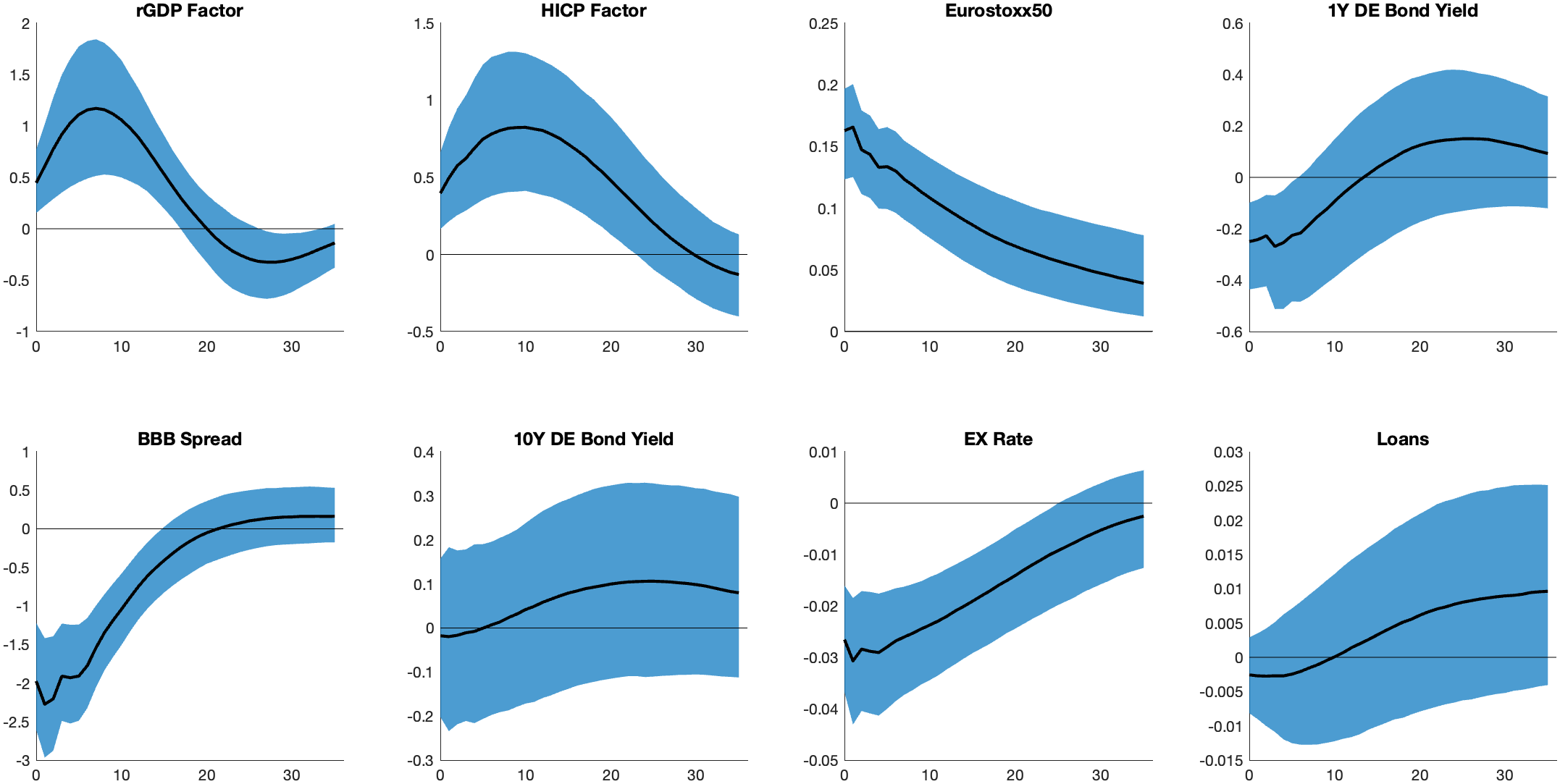}
      \caption{Median impulse responses to an expansionary monetary policy shock (\emph{$P=3$}). The solid black line depicts the median and the shaded blue area the  68\% credible bands.} \label{fig_OA24}
	\end{figure}
\end{center}

\begin{center}
    \begin{figure}[H]
    \begin{minipage}{.5\textwidth}
      	\includegraphics[width=0.99\textwidth]{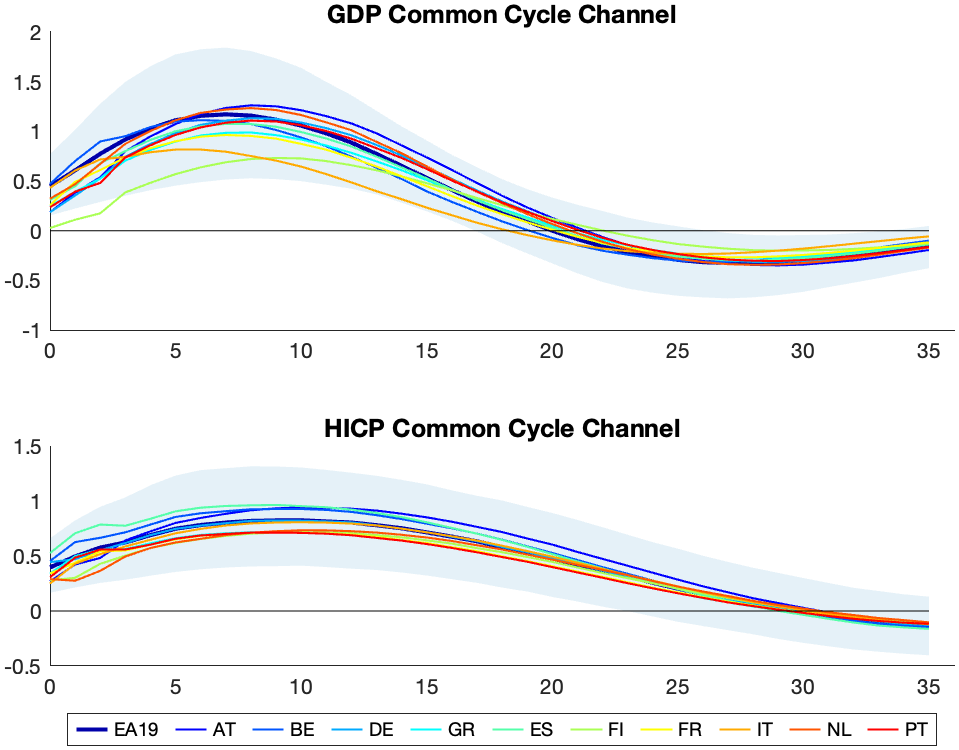}
       \end{minipage}
        \begin{minipage}{.5\textwidth}
      	\includegraphics[width=0.99\textwidth]{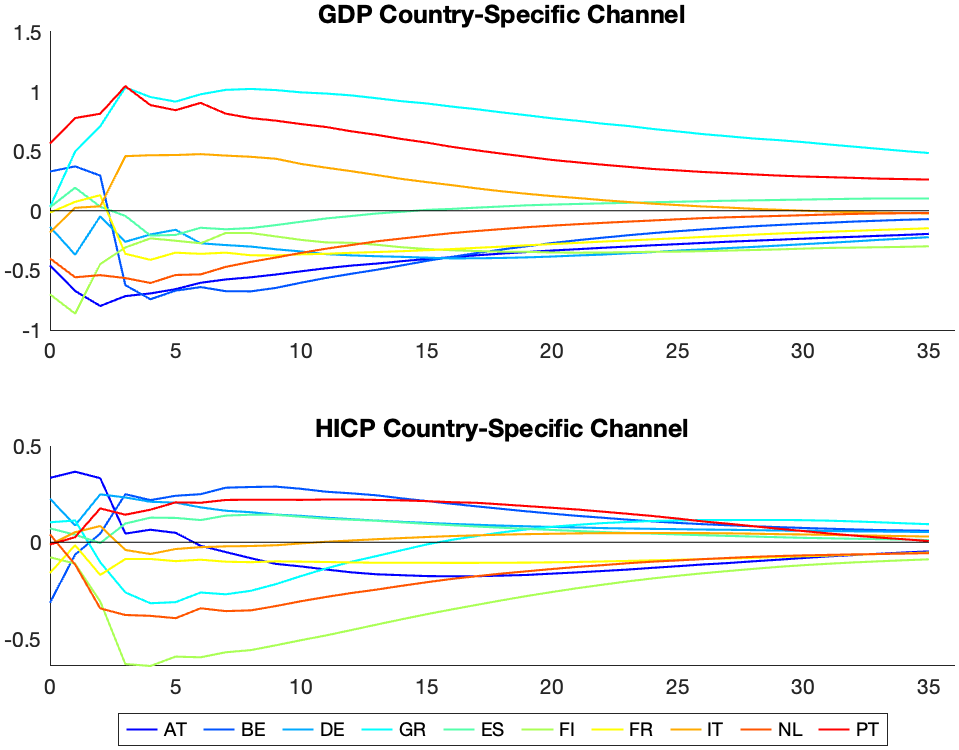}
       \end{minipage}
		  \caption{Country-level median impulse responses of country-specific output growth and inflation
            to an expansionary monetary policy shock (\emph{$P=3$}). The left-hand side plots show the propagation via the common cycles and the right-hand side plots show the direct impact via the country-specific channels. The solid blue lines depicts the median responses of the euro area aggregate (EA19) and the shaded light blue areas the 68\% credible bands.} \label{fig_OA25}
    \end{figure}
\end{center}

In this alternative model specification, we extend the factor equation (3) by including lags of the factors and of the variables $\boldsymbol{z}_{t}$:
\begin{eqnarray}
    x_{i,t}^{OUT} &=& \sum_{p=0}^{P} \lambda_{i,p}^{OUT} f_{t-p}^{OUT} + \sum_{p=0}^{P} \boldsymbol{\lambda}_{i,p}^{OUT,\boldsymbol{z}} \boldsymbol{z}_{t} + e_{i,t}^{OUT}, \nonumber \\
    x_{i,t}^{INF} &=& \sum_{p=0}^{P} \lambda_{i,p}^{INF} f_{t-p}^{INF} + \sum_{p=0}^{P} \boldsymbol{\lambda}_{i,p}^{INF,\boldsymbol{z}} \boldsymbol{z}_{t} + e_{i,t}^{INF}. \nonumber
\end{eqnarray}
The inclusion of lags of the cycles is important to allow for disproportionate country-level responses to shocks which hit the common cycles and for potentially richer dynamics via $\lambda_{i,p}^{OUT}$ and $\lambda_{i,p}^{INF}$ \shortcite{adp21}. Despite, euro member countries may lead or follow the common cycles. Analogously, we allow for lagged effects of the financial channels via $\boldsymbol{\lambda}_{i,p}^{OUT,\boldsymbol{z}}$ and $\boldsymbol{\lambda}_{i,p}^{INF,\boldsymbol{z}}$. To continuously satisfy statistical identification of the cycles, we set the lagged factor loadings of the Euro area aggregate to 0. The below results show that the main results of our baseline model with $P=0$ are robust to the lag orders of $P=1$ (Figures \ref{fig_OA20} and \ref{fig_OA21}), $P=2$ (Figures \ref{fig_OA22} and \ref{fig_OA23}) and $P=3$ (Figures \ref{fig_OA24} and \ref{fig_OA25}). Therefore, we set $P=0$ in our baseline specification to ensure a parsimonious model specification.

\vspace{-50pt}

\vspace{-50pt}

\vspace{-50pt}

\vspace{-50pt}

\vspace{-50pt}

\vspace{-50pt}

\clearpage
\subsection*{Alternative Lag Orders in the SVAR}

\begin{center}
	\begin{figure}[H]
    \includegraphics[height=8\baselineskip,width=0.99\textwidth]{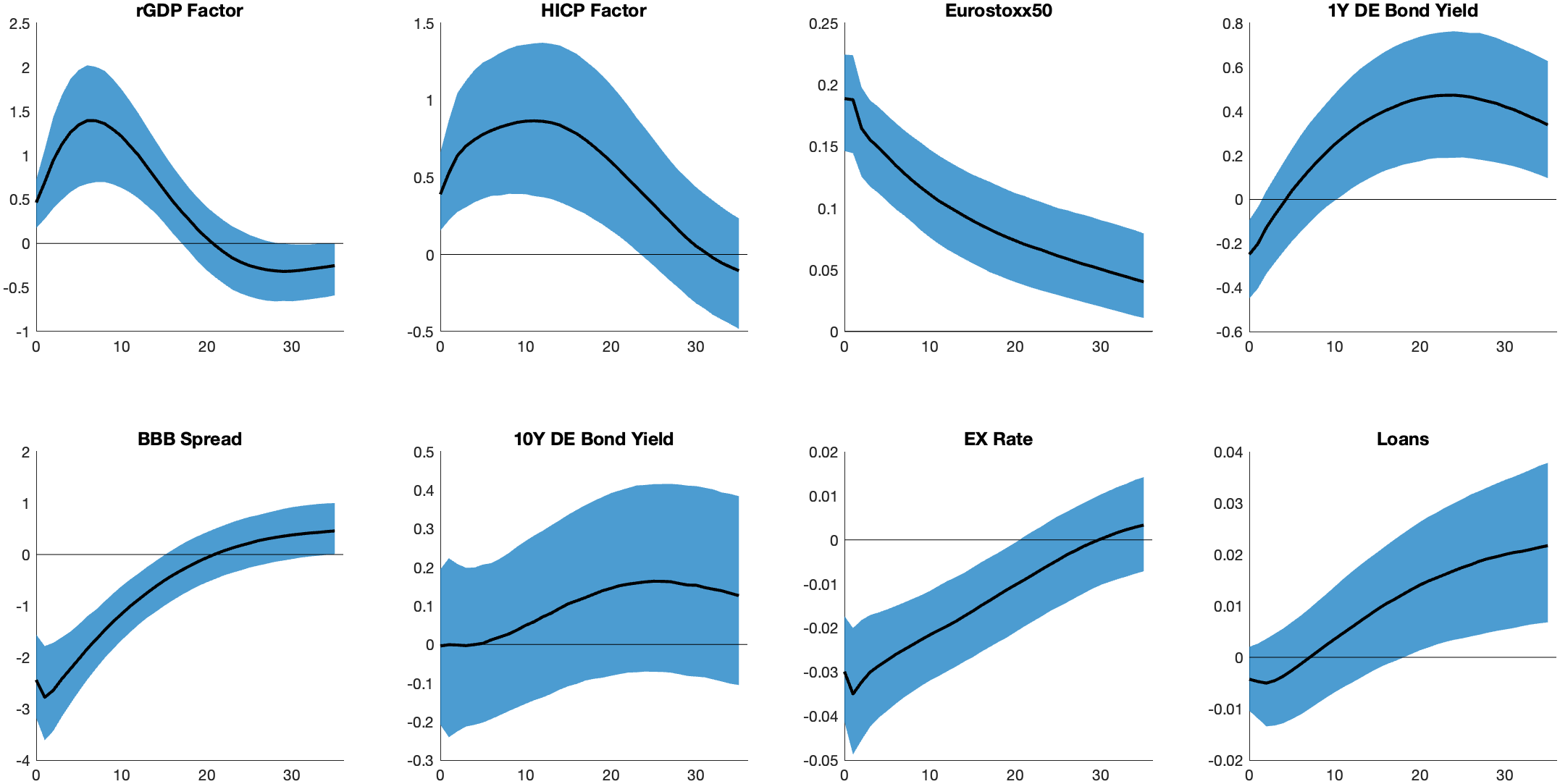}
      \caption{Median impulse responses to an expansionary monetary policy shock. The solid black line depicts the median and the shaded blue area the  68\% credible bands.} \label{fig_OA26}
	\end{figure}
\end{center}

\begin{center}
    \begin{figure}[H]
    \begin{minipage}{.5\textwidth}
      	\includegraphics[width=0.99\textwidth]{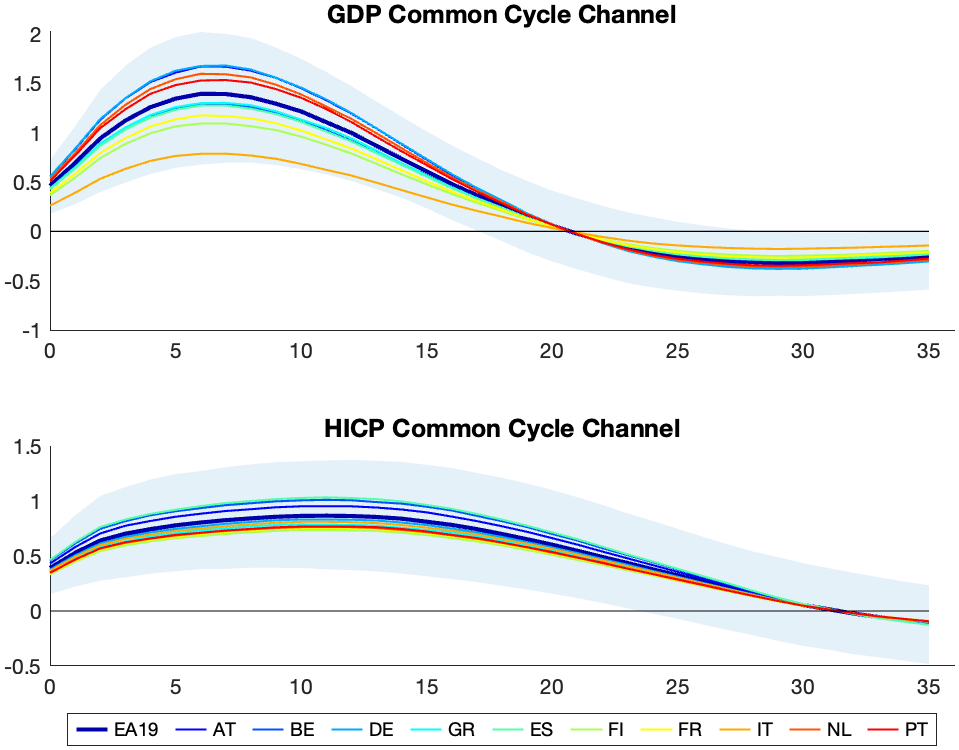}
       \end{minipage}
        \begin{minipage}{.5\textwidth}
      	\includegraphics[width=0.99\textwidth]{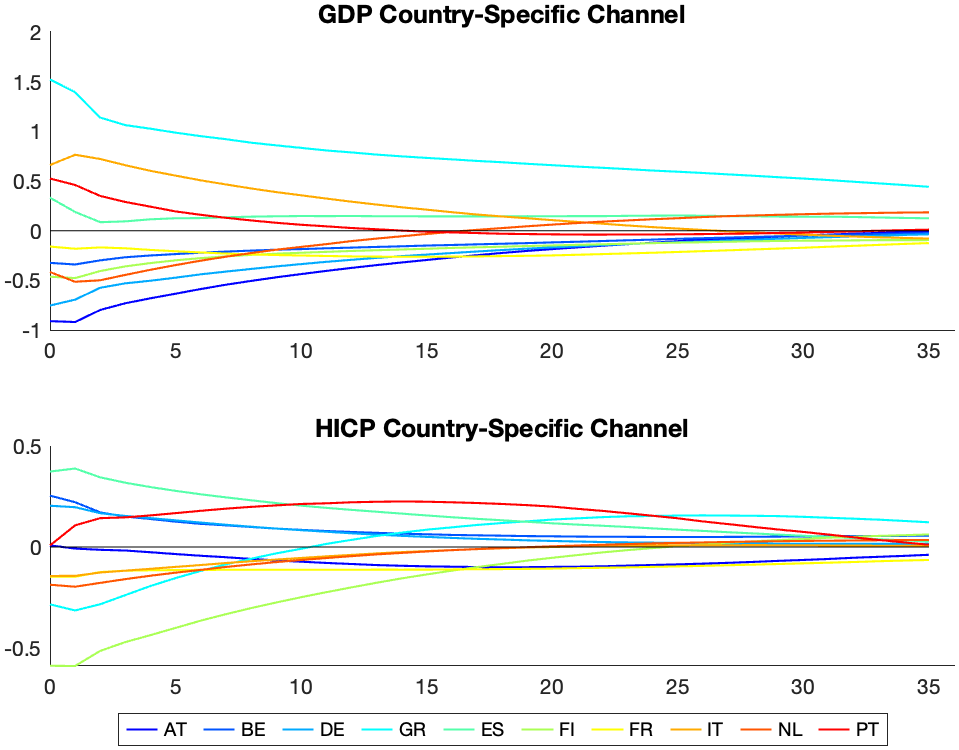}
       \end{minipage}
		  \caption{Country-level median impulse responses of country-specific output growth and inflation
            to an expansionary monetary policy shock (\emph{$L=3$}). The left-hand side plots show the propagation via the common cycles and the right-hand side plots show the direct impact via the country-specific channels. The solid blue lines depicts the median responses of the euro area aggregate (EA19) and the shaded light blue areas the 68\% credible bands.} \label{fig_OA27}
    \end{figure}
\end{center}

\begin{center}
	\begin{figure}[H]
    \includegraphics[height=8\baselineskip,width=0.99\textwidth]{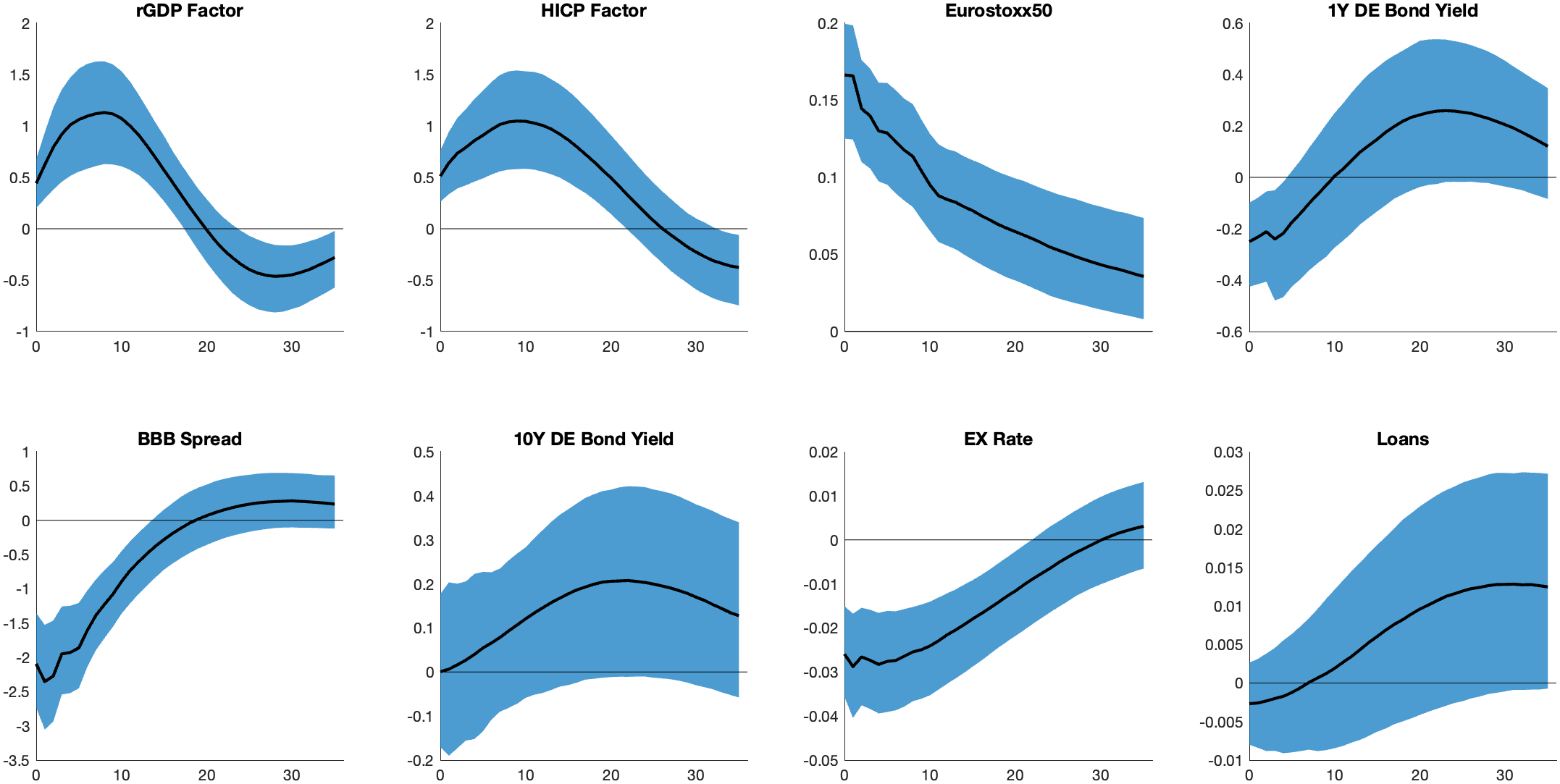}
      \caption{Median impulse responses to an expansionary monetary policy shock. The solid black line depicts the median and the shaded blue area the  68\% credible bands.} \label{fig_OA28}
	\end{figure}
\end{center}

\begin{center}
    \begin{figure}[H]
    \begin{minipage}{.5\textwidth}
      	\includegraphics[width=0.99\textwidth]{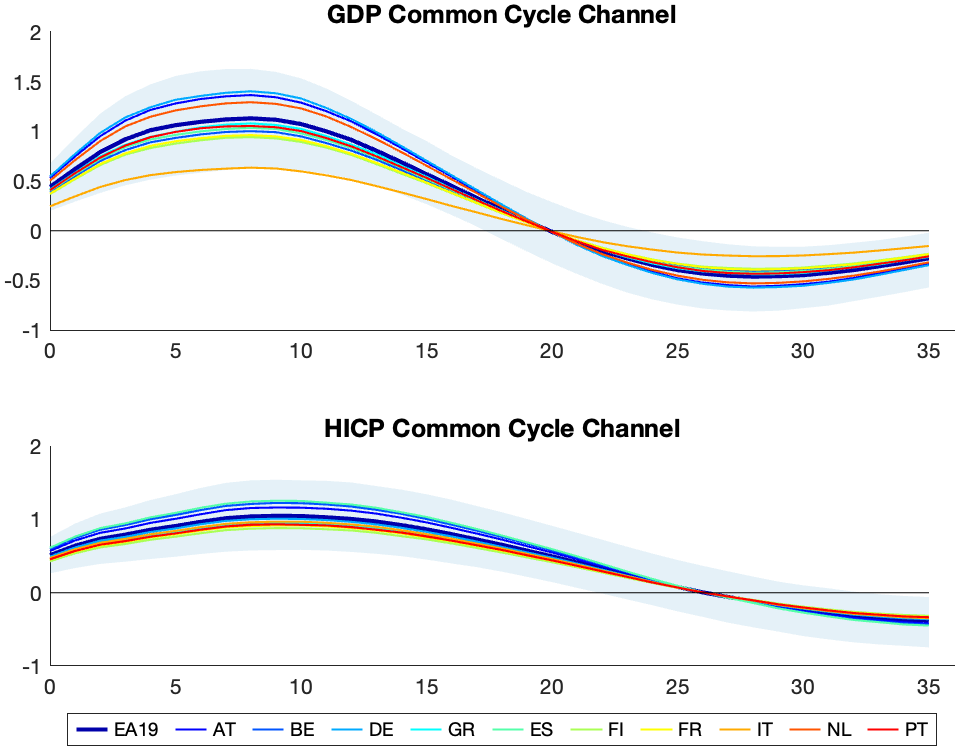}
       \end{minipage}
        \begin{minipage}{.5\textwidth}
      	\includegraphics[width=0.99\textwidth]{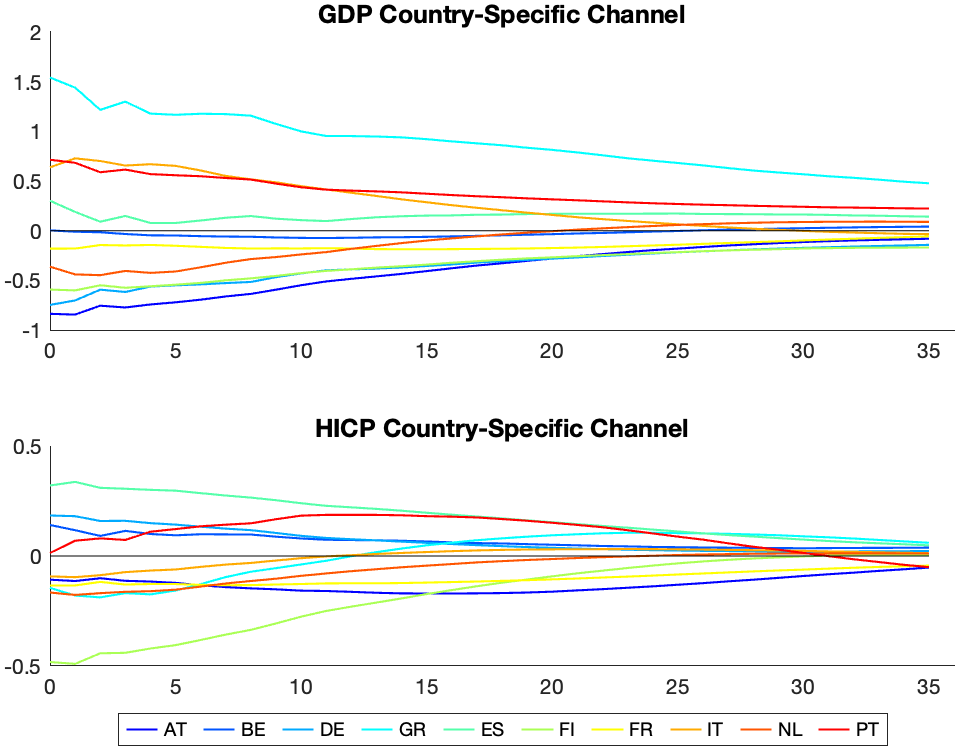}
       \end{minipage}
		  \caption{Country-level median impulse responses of country-specific output growth and inflation
            to an expansionary monetary policy shock (\emph{$L=12$}). The left-hand side plots show the propagation via the common cycles and the right-hand side plots show the direct impact via the country-specific channels. The solid blue lines depicts the median responses of the euro area aggregate (EA19) and the shaded light blue areas the 68\% credible bands.} \label{fig_OA29}
    \end{figure}
\end{center}

In this alternative model specification, we set $L=3$ (Figures \ref{fig_OA26} and \ref{fig_OA27}) and $L=12$ (Figures \ref{fig_OA28} and \ref{fig_OA29}).

\vspace{-50pt}

\vspace{-50pt}

\vspace{-50pt}

\vspace{-50pt}

\clearpage
\subsection*{Alternative Annual Growth Rates}

\begin{center}
	\begin{figure}[H]
    \includegraphics[height=8\baselineskip,width=0.99\textwidth]{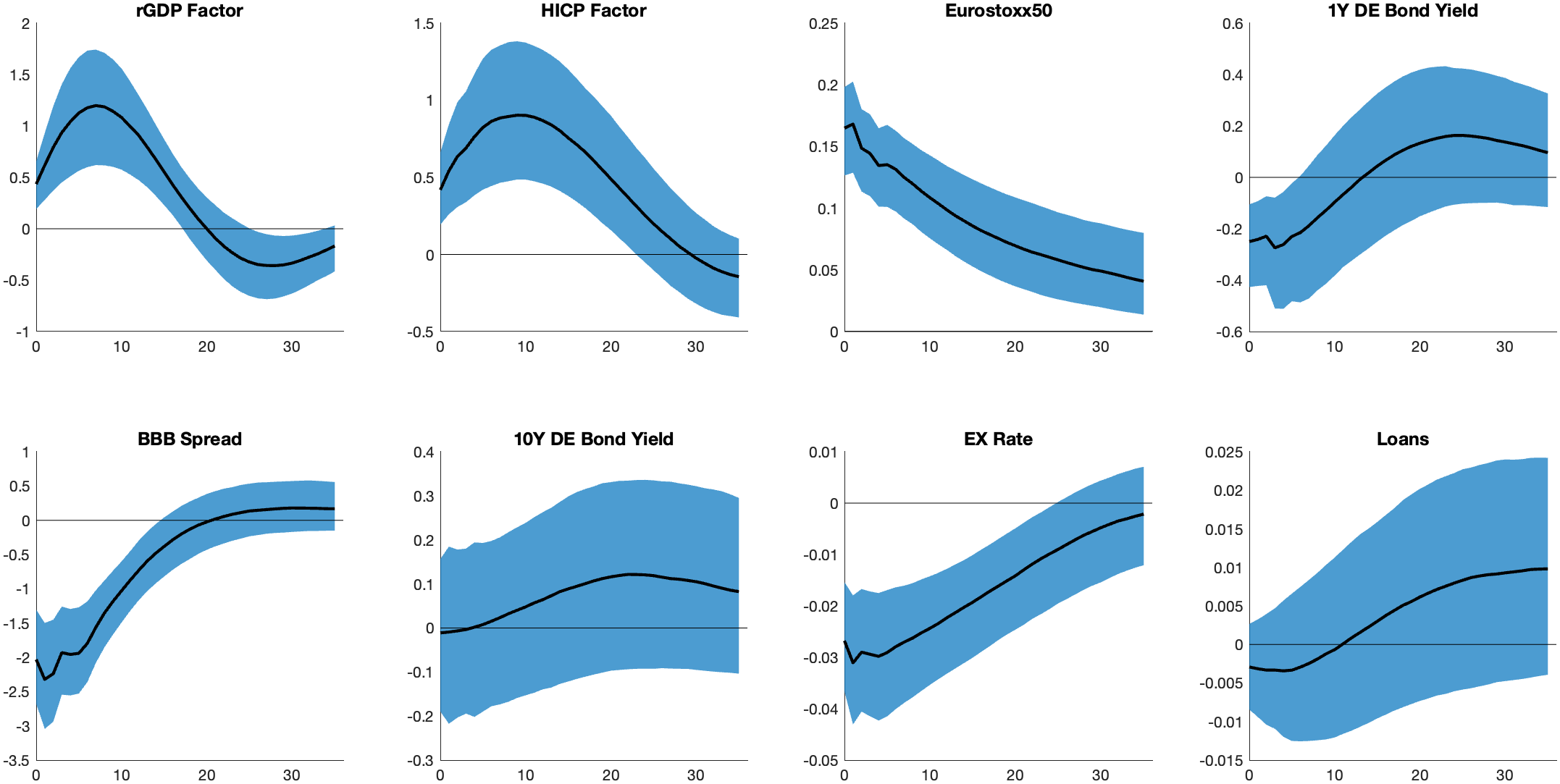}
      \caption{Median impulse responses to an expansionary monetary policy shock. The solid black line depicts the median and the shaded blue area the  68\% credible bands.} \label{fig_OA30}
	\end{figure}
\end{center}

\begin{center}
    \begin{figure}[H]
    \begin{minipage}{.5\textwidth}
      	\includegraphics[width=0.99\textwidth]{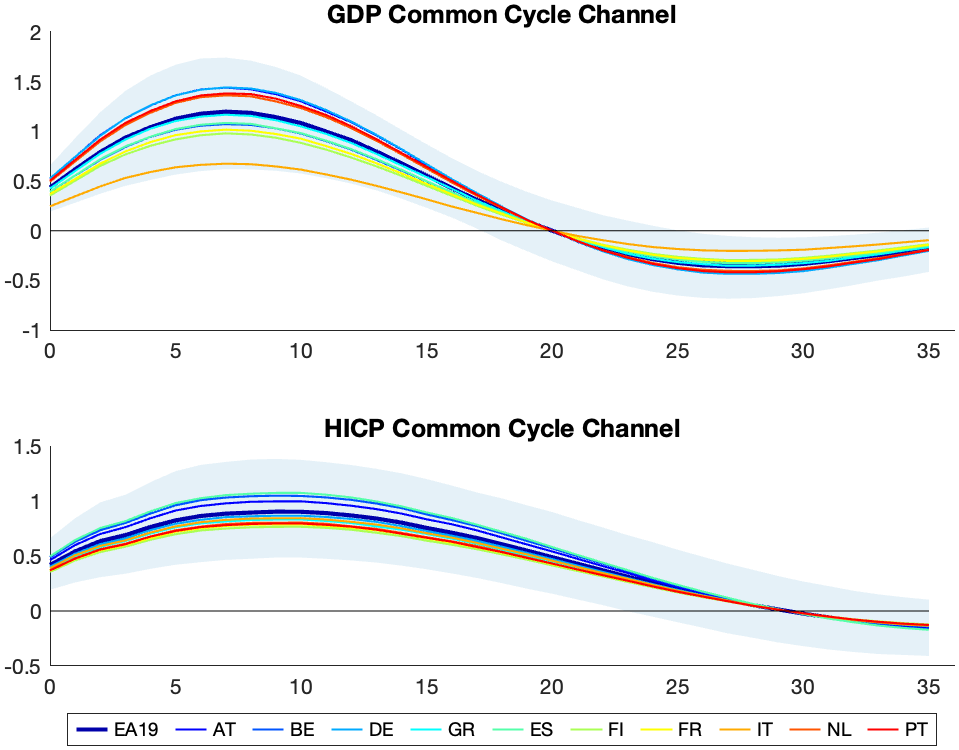}
       \end{minipage}
        \begin{minipage}{.5\textwidth}
      	\includegraphics[width=0.99\textwidth]{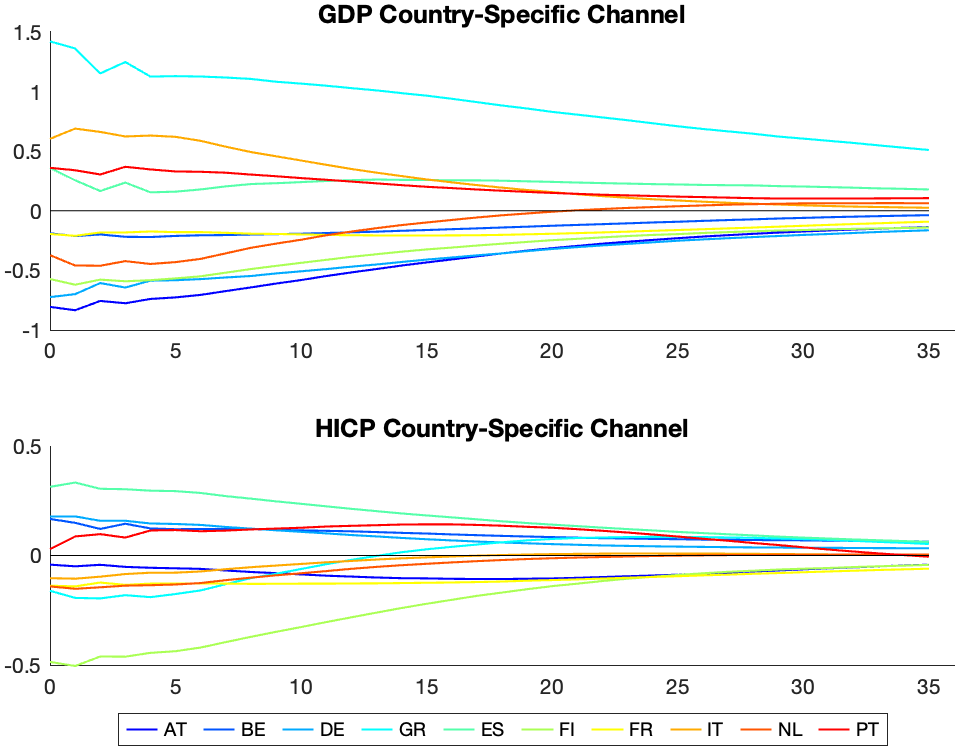}
       \end{minipage}
		  \caption{Country-level median impulse responses of country-specific output growth and inflation
            to an expansionary monetary policy shock. The left-hand side plots show the propagation via the common cycles and the right-hand side plots show the direct impact via the country-specific channels. The solid blue lines depicts the median responses of the euro area aggregate (EA19) and the shaded light blue areas the 68\% credible bands.} \label{fig_OA31}
    \end{figure}
\end{center}

\begin{center}
	\begin{figure}[H]
    \includegraphics[height=8\baselineskip,width=0.99\textwidth]{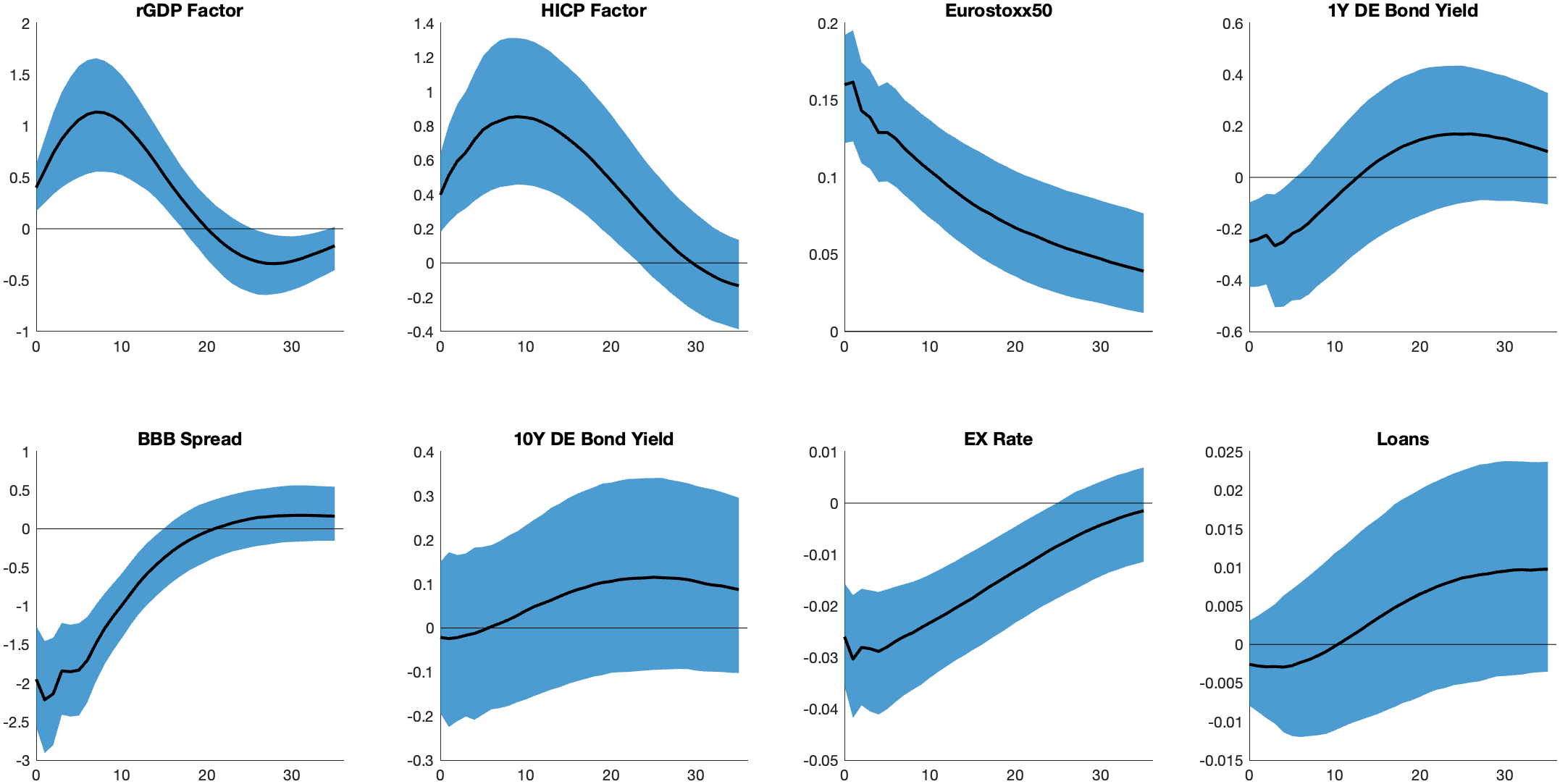}
      \caption{Median impulse responses to an expansionary monetary policy shock. The solid black line depicts the median and the shaded blue area the  68\% credible bands.} \label{fig_OA32}
	\end{figure}
\end{center}

\begin{center}
    \begin{figure}[H]
    \begin{minipage}{.5\textwidth}
      	\includegraphics[width=0.99\textwidth]{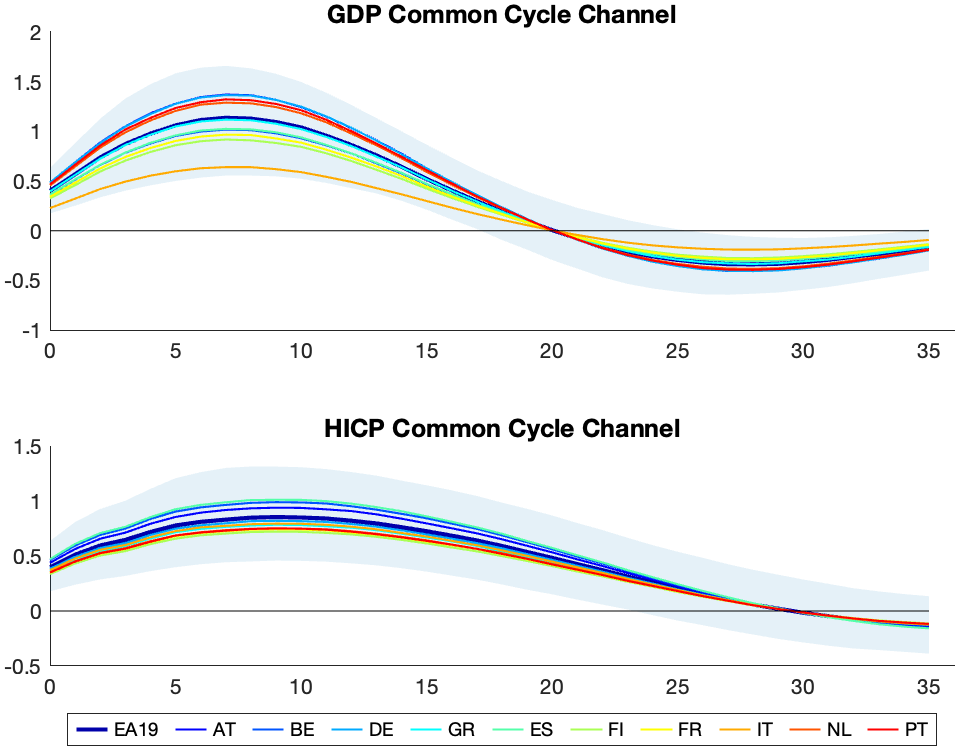}
       \end{minipage}
        \begin{minipage}{.5\textwidth}
      	\includegraphics[width=0.99\textwidth]{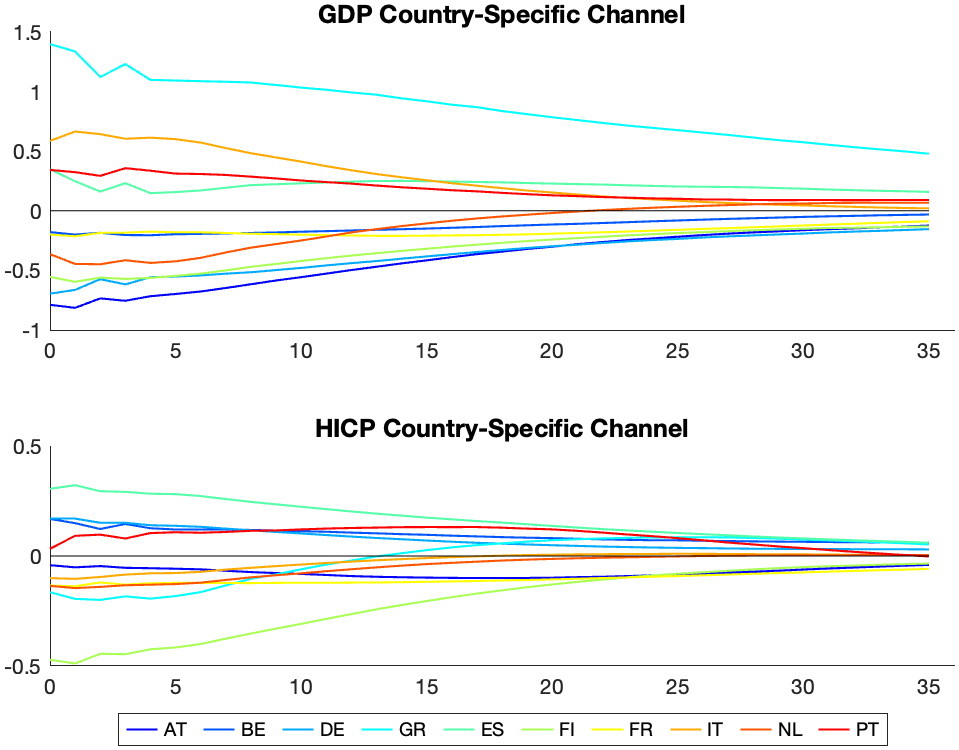}
       \end{minipage}
		  \caption{Country-level median impulse responses of country-specific output growth and inflation
            to an expansionary monetary policy shock. The left-hand side plots show the propagation via the common cycles and the right-hand side plots show the direct impact via the country-specific channels. The solid blue lines depicts the median responses of the euro area aggregate (EA19) and the shaded light blue areas the 68\% credible bands.} \label{fig_OA33}
    \end{figure}
\end{center}

In this section, we present the results of the model that uses the following annual growth rates for GDP and inflation.
\begin{itemize}
    \item[a.] The logarithmic annual growth rate (Figures \ref{fig_OA30} and \ref{fig_OA31}): 
    \begin{eqnarray}
        x_{it}^{j} &=& (log(X_{i13:T}^{j})-log(X_{i1:T-12}^{j})) \times 100 \nonumber
    \end{eqnarray}
    \item[b.] The annual growth rate proposed by \citeA{bh23} (Figures \ref{fig_OA32} and \ref{fig_OA33}): 
    \begin{eqnarray}
        x_{it}^{j} &=& \frac{X_{i13:T}^{j}-X_{i1:T-12}^{j}}{0.5 \times (X_{i13:T}^{j}-X_{i1:T-12}^{j})}  \times 100 \nonumber
    \end{eqnarray}
\end{itemize}

\vspace{-50pt}

\vspace{-50pt}

\vspace{-50pt}

\vspace{-50pt}

\clearpage
\subsection*{Original Time Series without Outlier Adjustment}

\begin{center}
	\begin{figure}[H]
    \includegraphics[height=8\baselineskip,width=0.99\textwidth]{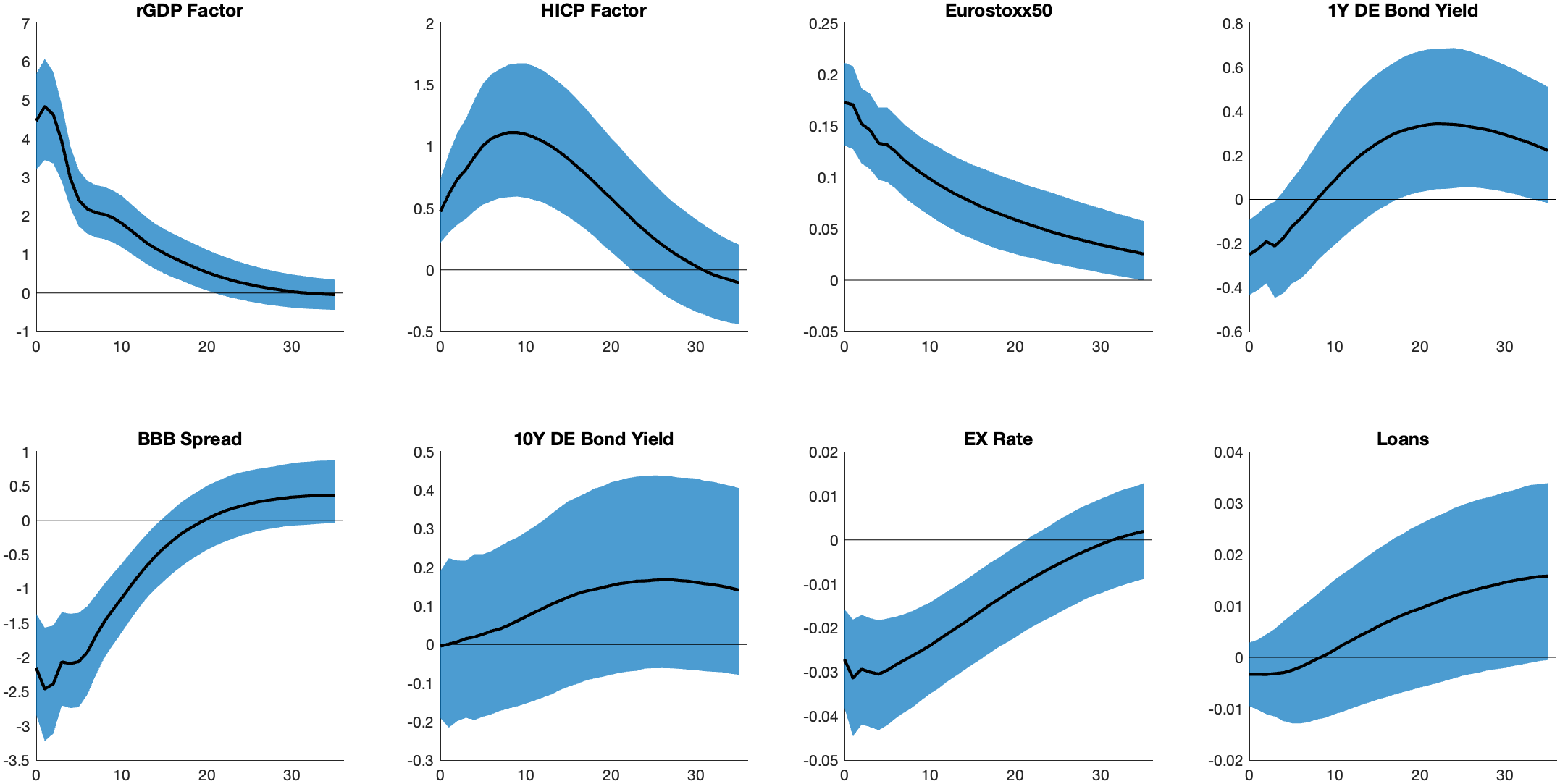}
      \caption{Median impulse responses to an expansionary monetary policy shock. The solid black line depicts the median and the shaded blue area the  68\% credible bands.} \label{fig_OA34}
	\end{figure}
\end{center}

\begin{center}
    \begin{figure}[H]
    \begin{minipage}{.5\textwidth}
      	\includegraphics[width=0.99\textwidth]{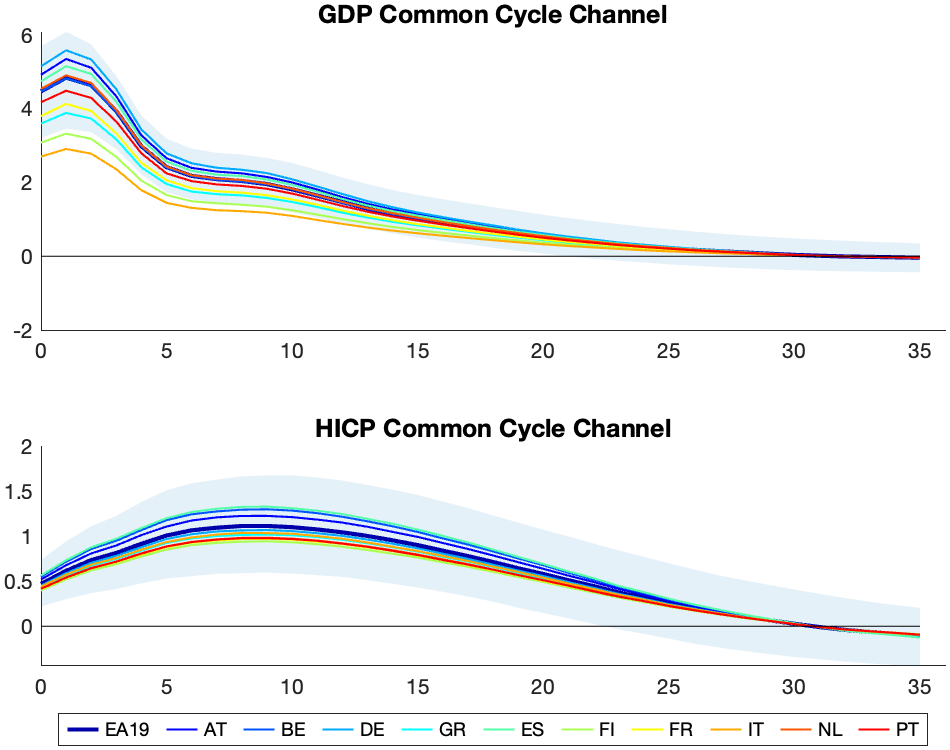}
       \end{minipage}
        \begin{minipage}{.5\textwidth}
      	\includegraphics[width=0.99\textwidth]{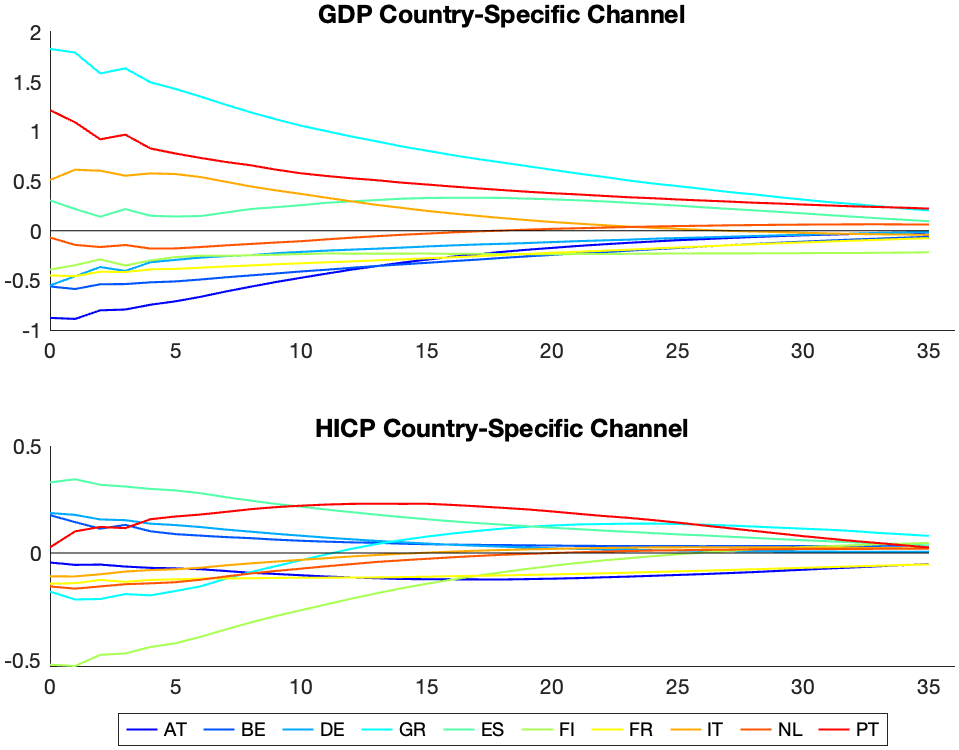}
       \end{minipage}
		  \caption{Country-level median impulse responses of country-specific output growth and inflation
            to an expansionary monetary policy shock. The left-hand side plots show the propagation via the common cycles and the right-hand side plots show the direct impact via the country-specific channels. The solid blue lines depicts the median responses of the euro area aggregate (EA19) and the shaded light blue areas the 68\% credible bands.} \label{fig_OA35}
    \end{figure}
\end{center}

\begin{center}
	\begin{figure}[H]
    \includegraphics[height=8\baselineskip,width=0.99\textwidth]{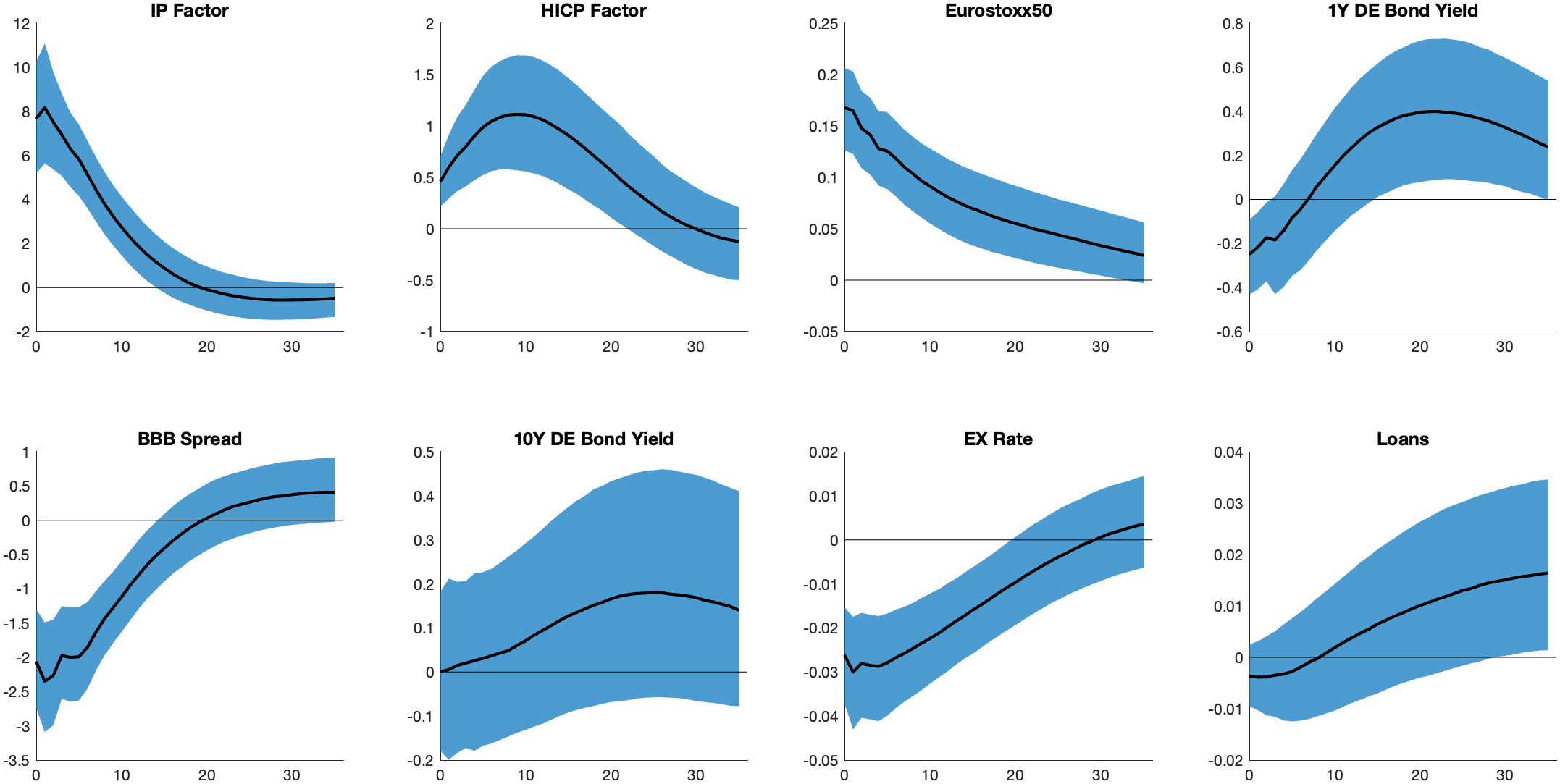}
      \caption{Median impulse responses to an expansionary monetary policy shock. The solid black line depicts the median and the shaded blue area the  68\% credible bands.} \label{fig_OA36}
	\end{figure}
\end{center}

\begin{center}
    \begin{figure}[H]
    \begin{minipage}{.5\textwidth}
      	\includegraphics[width=0.99\textwidth]{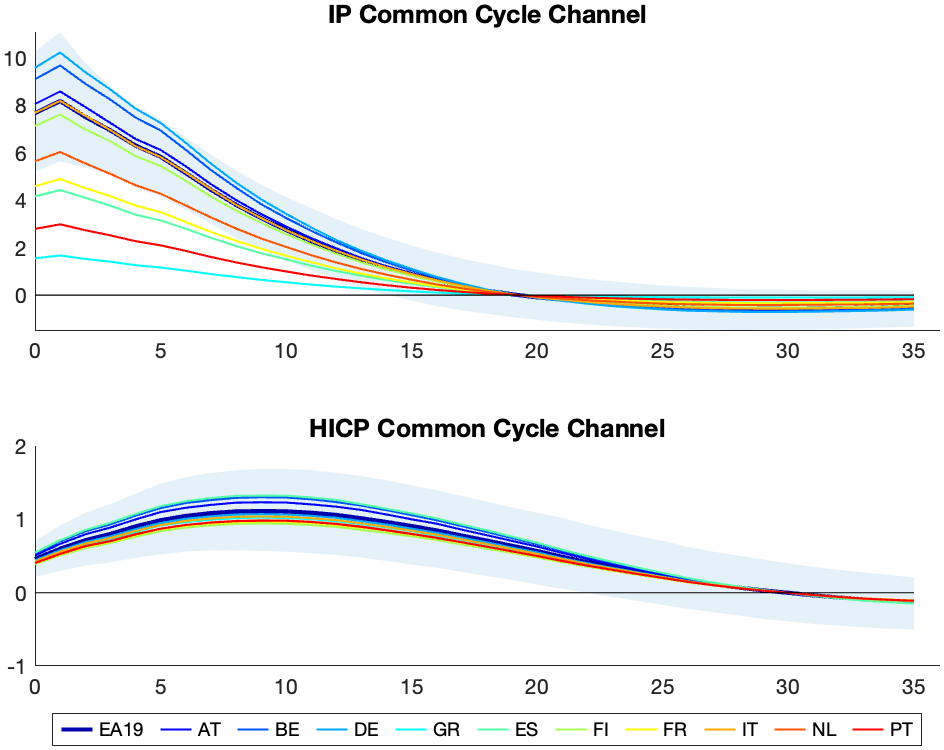}
       \end{minipage}
        \begin{minipage}{.5\textwidth}
      	\includegraphics[width=0.99\textwidth]{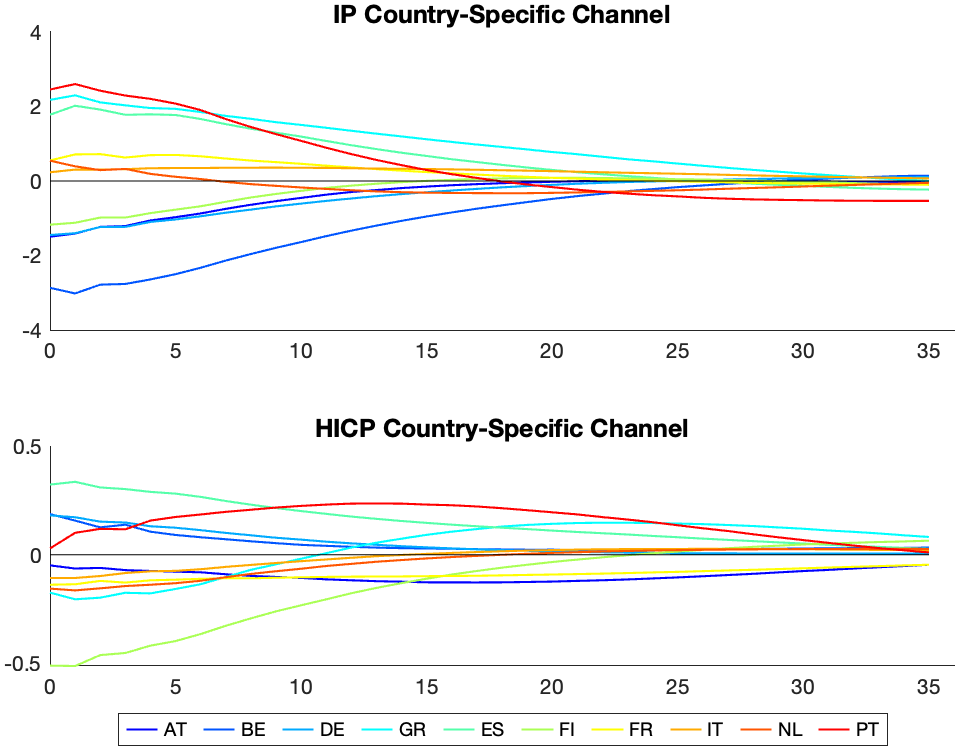}
       \end{minipage}
		  \caption{Country-level median impulse responses of country-specific output growth and inflation
            to an expansionary monetary policy shock. The left-hand side plots show the propagation via the common cycles and the right-hand side plots show the direct impact via the country-specific channels. The solid blue lines depicts the median responses of the euro area aggregate (EA19) and the shaded light blue areas the 68\% credible bands.} \label{fig_OA37}
    \end{figure}
\end{center}

In this section, we present the results for the models that use time series for interpolated GDP (Figures \ref{fig_OA34} and \ref{fig_OA35}) and industrial production (Figures \ref{fig_OA36} and \ref{fig_OA37}) that are not adjusted for outliers.

\vspace{-50pt}

\vspace{-50pt}

\vspace{-50pt}

\vspace{-50pt}

\clearpage
\subsection*{Smoothing and Outlier Adjustment}

\begin{center}
	\begin{figure}[H]
    \includegraphics[height=8\baselineskip,width=0.99\textwidth]{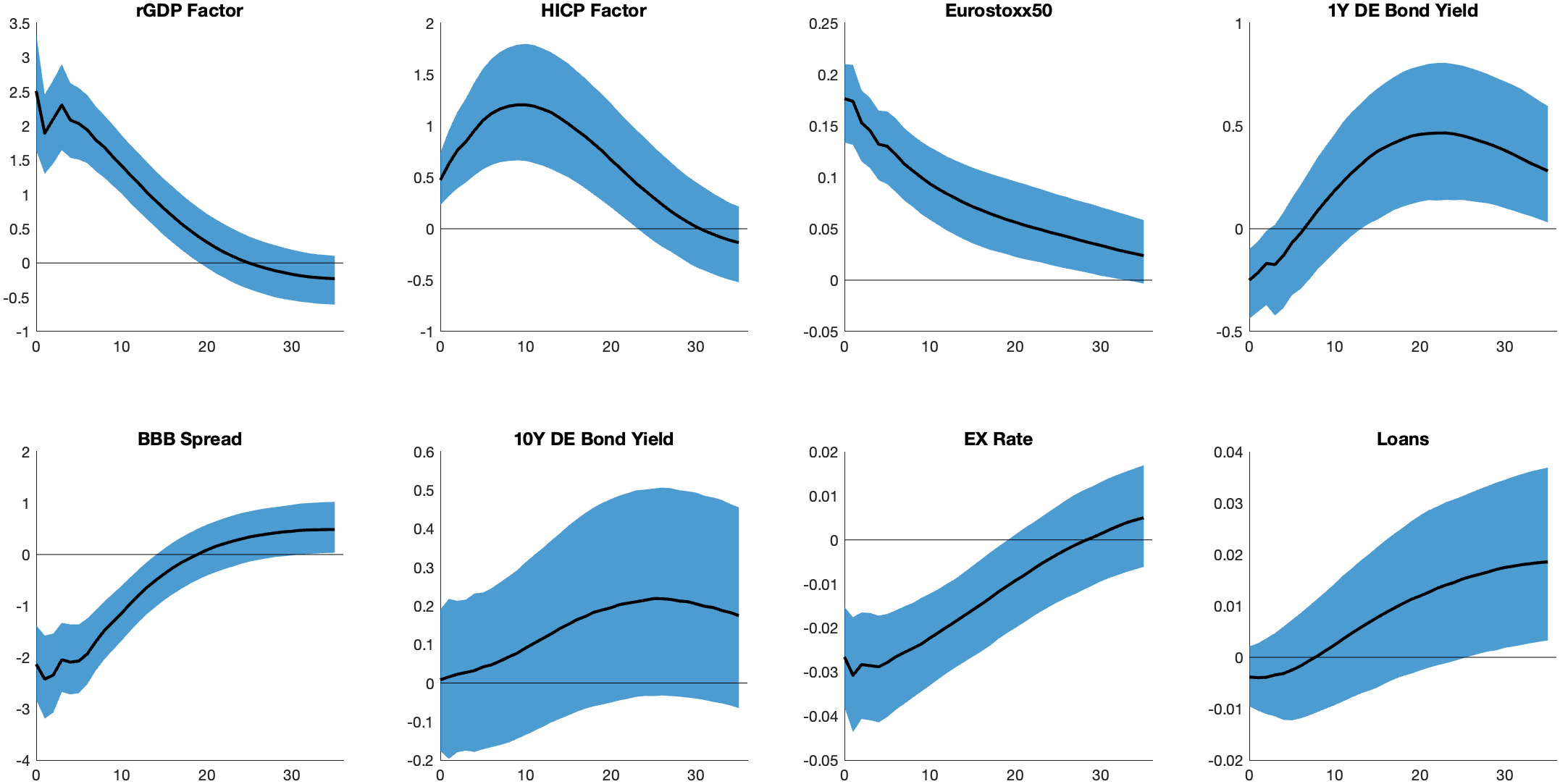}
      \caption{Median impulse responses to an expansionary monetary policy shock. The solid black line depicts the median and the shaded blue area the  68\% credible bands.} \label{fig_OA38}
	\end{figure}
\end{center}

\begin{center}
    \begin{figure}[H]
    \begin{minipage}{.5\textwidth}
      	\includegraphics[width=0.99\textwidth]{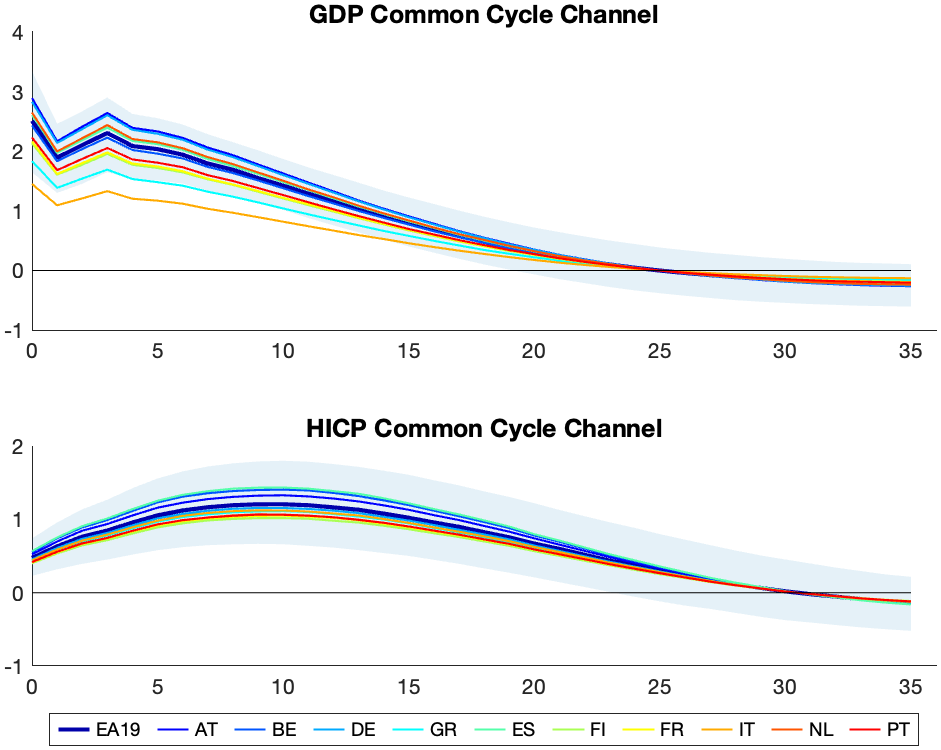}
       \end{minipage}
        \begin{minipage}{.5\textwidth}
      	\includegraphics[width=0.99\textwidth]{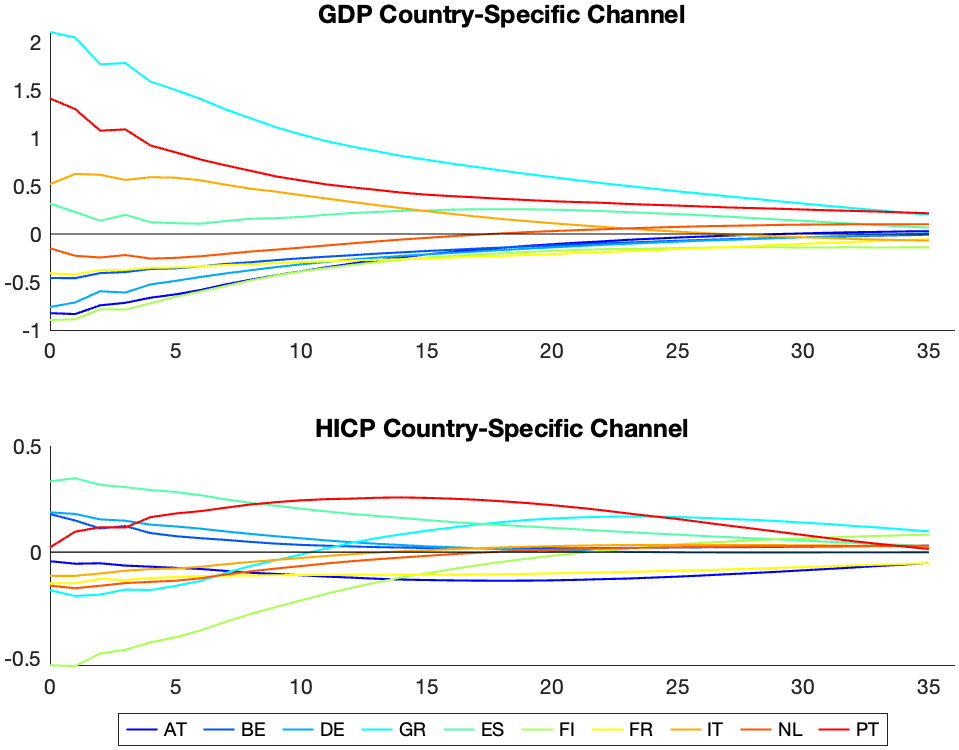}
       \end{minipage}
		  \caption{Country-level median impulse responses of country-specific output growth and inflation
            to an expansionary monetary policy shock. The left-hand side plots show the propagation via the common cycles and the right-hand side plots show the direct impact via the country-specific channels. The solid blue lines depicts the median responses of the euro area aggregate (EA19) and the shaded light blue areas the 68\% credible bands.} \label{fig_OA39}
    \end{figure}
\end{center}

\begin{center}
	\begin{figure}[H]
    \includegraphics[height=8\baselineskip,width=0.99\textwidth]{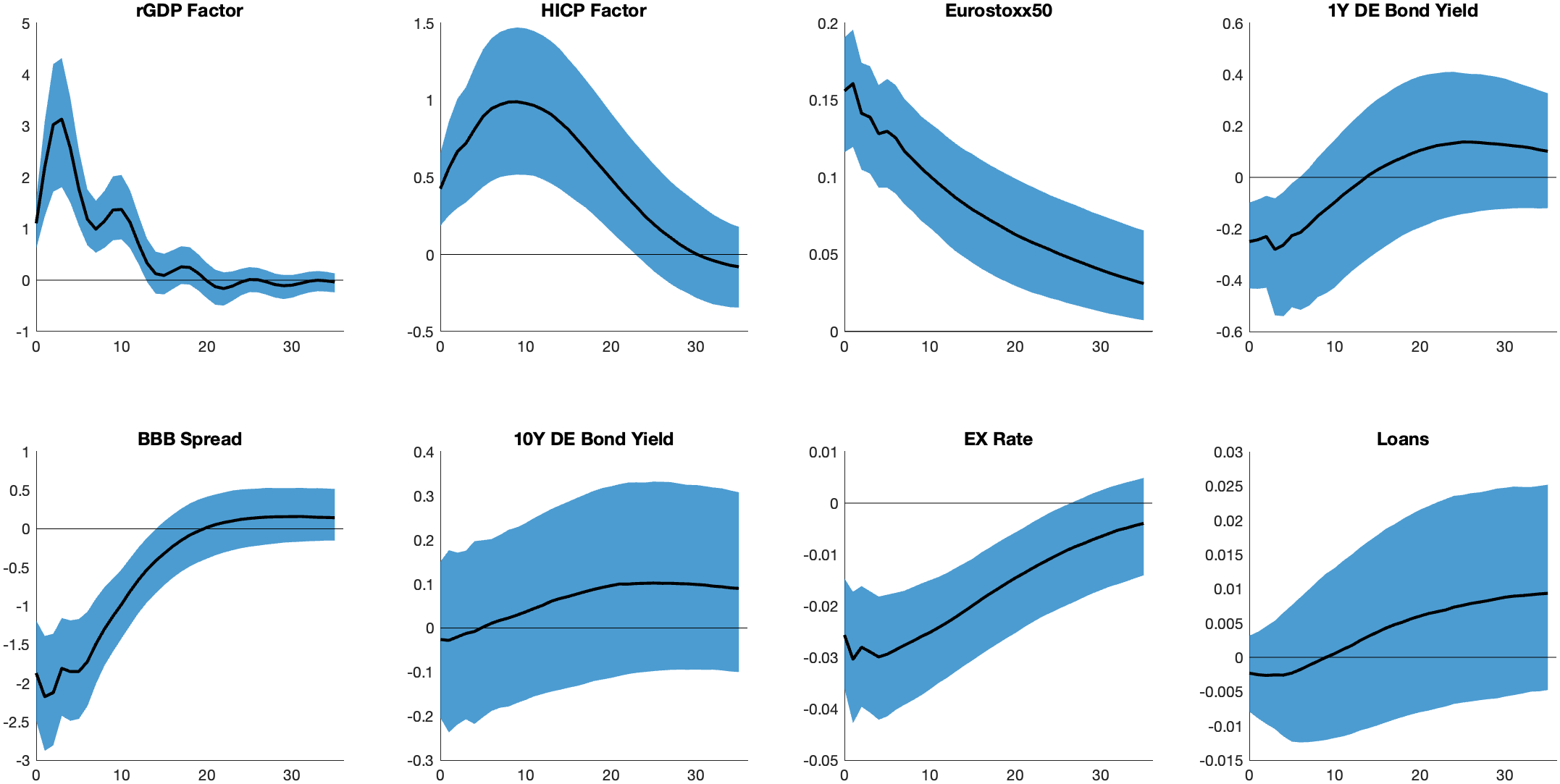}
      \caption{Median impulse responses to an expansionary monetary policy shock. The solid black line depicts the median and the shaded blue area the  68\% credible bands.} \label{fig_OA40}
	\end{figure}
\end{center}

\begin{center}
    \begin{figure}[H]
    \begin{minipage}{.5\textwidth}
      	\includegraphics[width=0.99\textwidth]{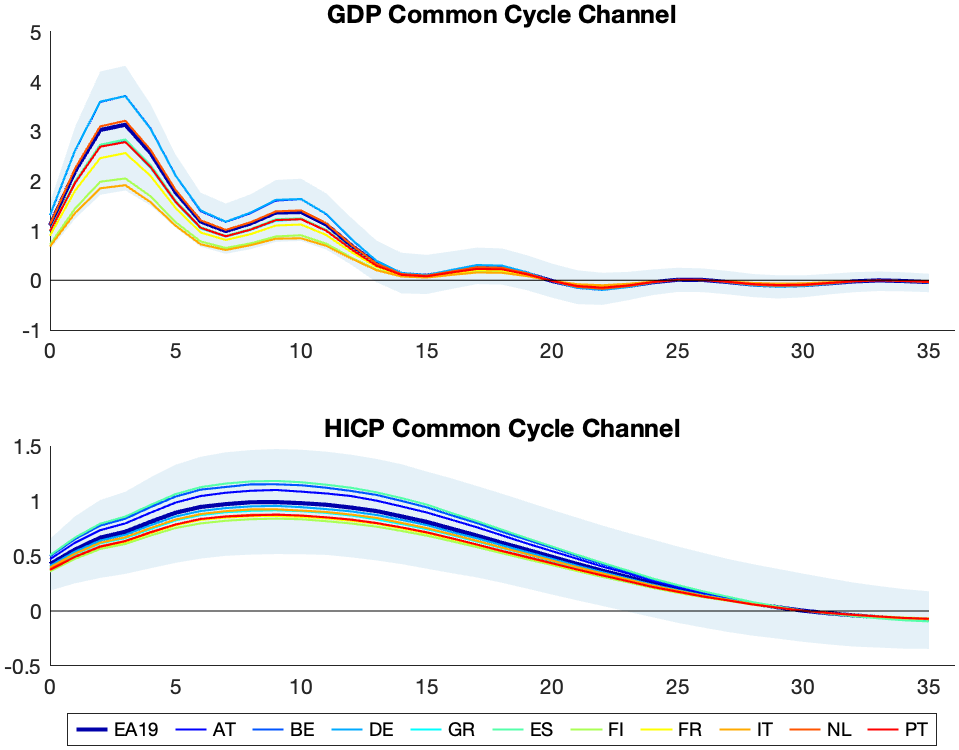}
       \end{minipage}
        \begin{minipage}{.5\textwidth}
      	\includegraphics[width=0.99\textwidth]{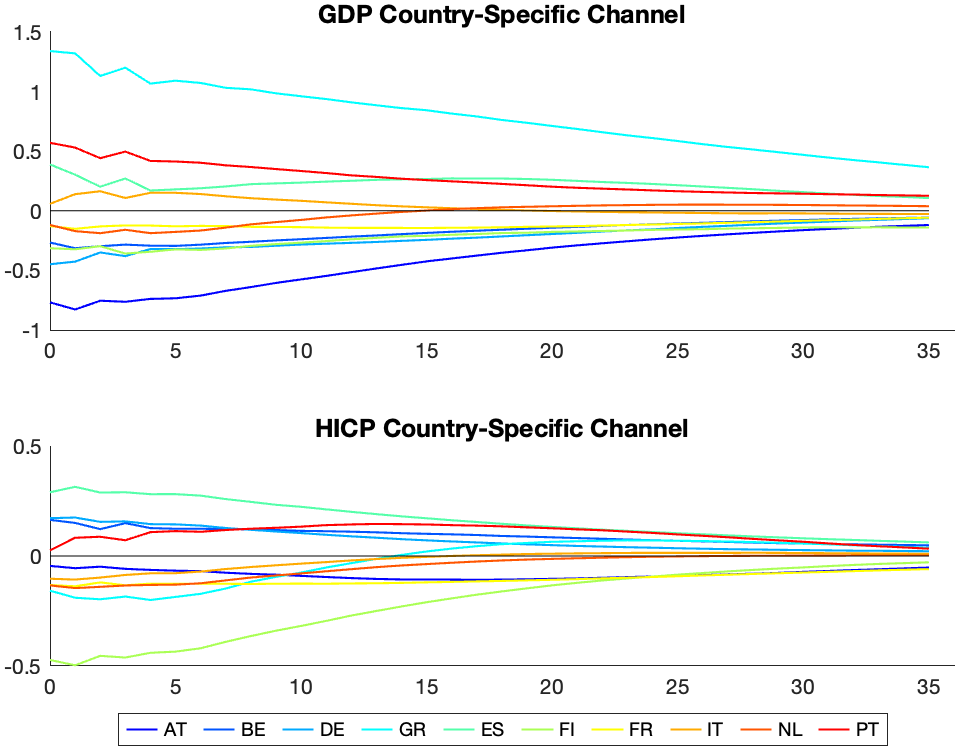}
       \end{minipage}
		  \caption{Country-level median impulse responses of country-specific output growth and inflation
            to an expansionary monetary policy shock. The left-hand side plots show the propagation via the common cycles and the right-hand side plots show the direct impact via the country-specific channels. The solid blue lines depicts the median responses of the euro area aggregate (EA19) and the shaded light blue areas the 68\% credible bands.} \label{fig_OA41}
    \end{figure}
\end{center}

\begin{center}
	\begin{figure}[H]
    \includegraphics[height=8\baselineskip,width=0.99\textwidth]{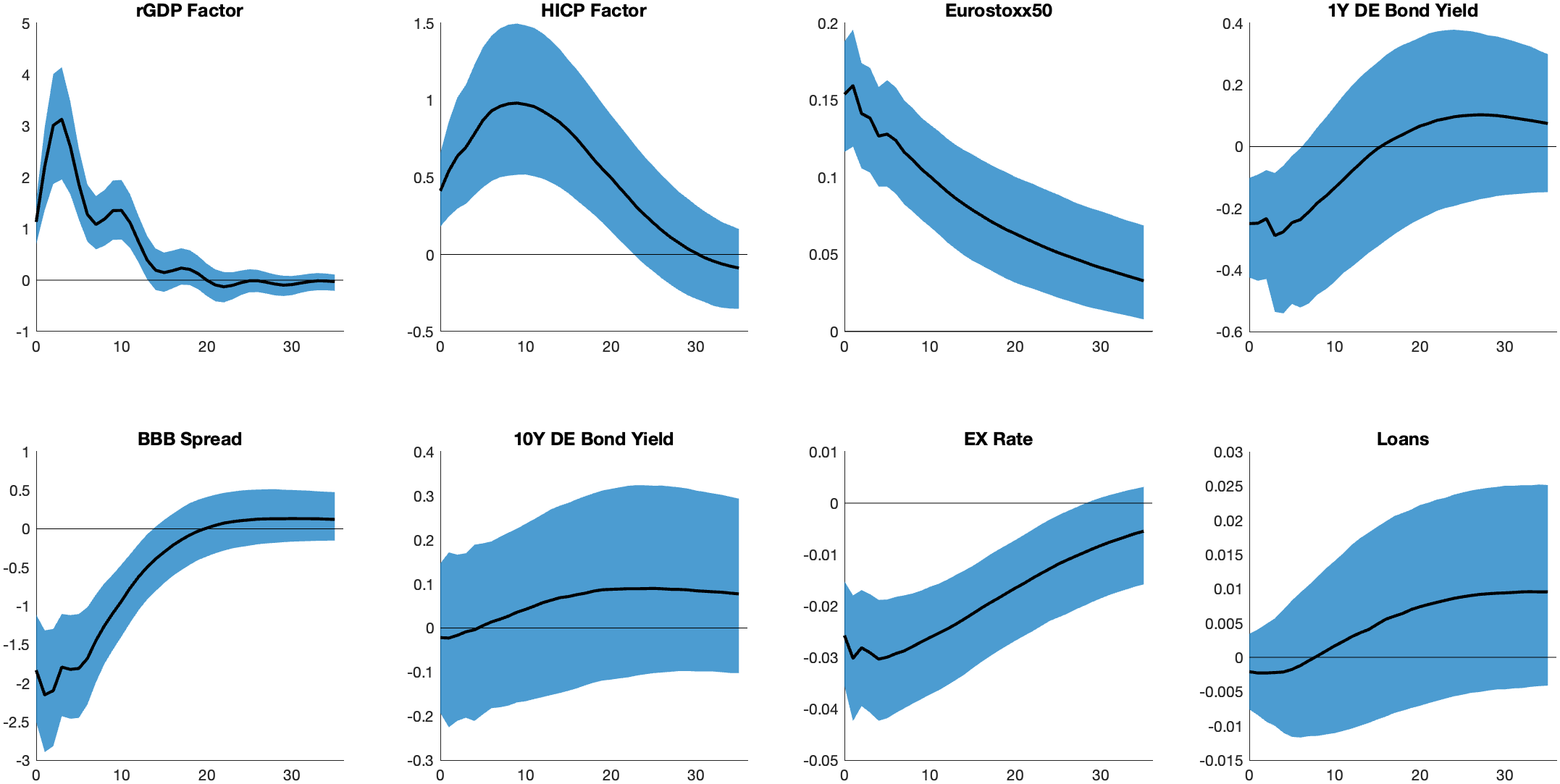}
      \caption{Median impulse responses to an expansionary monetary policy shock. The solid black line depicts the median and the shaded blue area the  68\% credible bands.} \label{fig_OA42}
	\end{figure}
\end{center}

\begin{center}
    \begin{figure}[H]
    \begin{minipage}{.5\textwidth}
      	\includegraphics[width=0.99\textwidth]{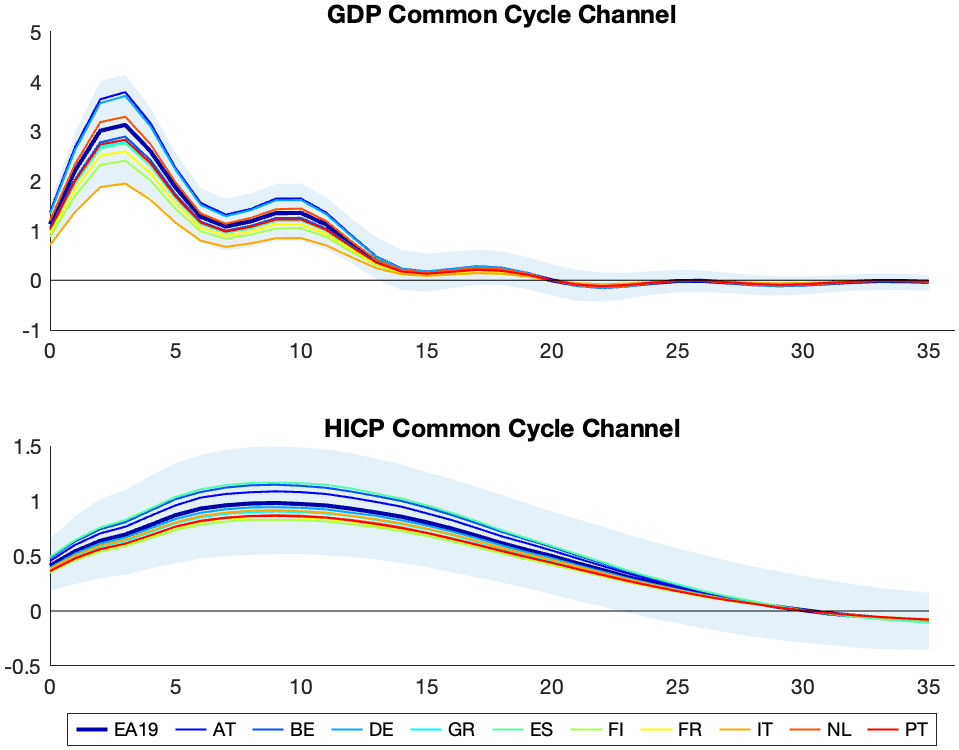}
       \end{minipage}
        \begin{minipage}{.5\textwidth}
      	\includegraphics[width=0.99\textwidth]{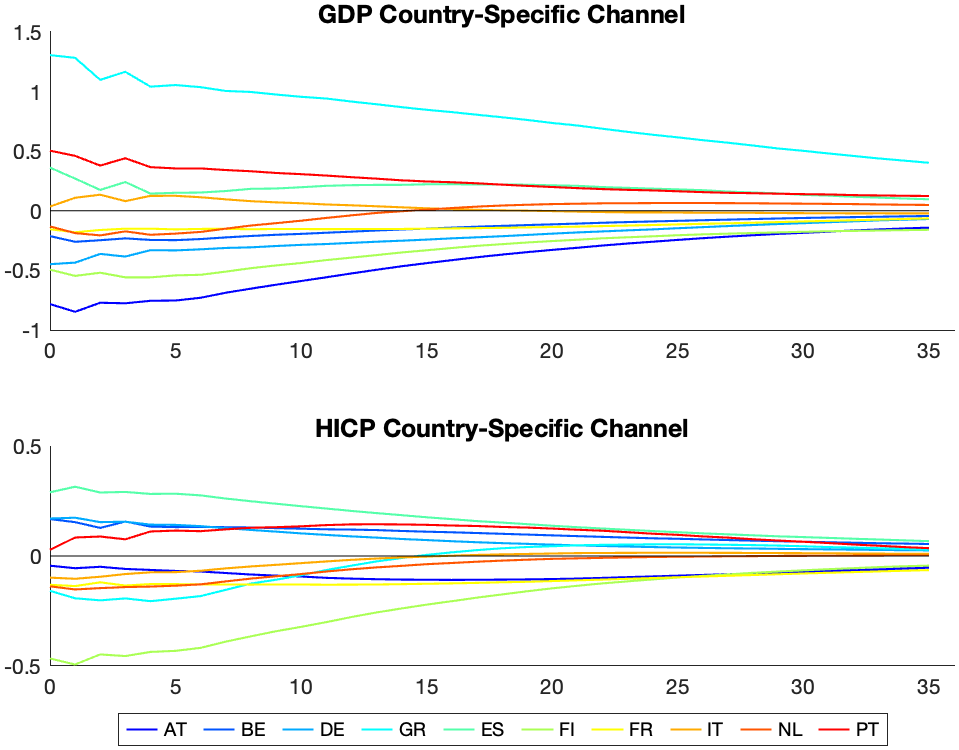}
       \end{minipage}
		  \caption{Country-level median impulse responses of country-specific output growth and inflation
            to an expansionary monetary policy shock. The left-hand side plots show the propagation via the common cycles and the right-hand side plots show the direct impact via the country-specific channels. The solid blue lines depicts the median responses of the euro area aggregate (EA19) and the shaded light blue areas the 68\% credible bands.} \label{fig_OA43}
    \end{figure}
\end{center}

In this section, we present the results for the models that use interpolated GDP adjusted for both additive outliers and temporal changes (Figures \ref{fig_OA38} and \ref{fig_OA39}), smoothed and interpolated GDP adjusted for additive outliers (Figures \ref{fig_OA40} and \ref{fig_OA41}) and smoothed and interpolated GDP adjusted for both additive outliers and temporal changes (Figures \ref{fig_OA42} and \ref{fig_OA43}). Here, we follow \citeA{jk20} and fit a cubic smoothing spline to the time series.

\vspace{-50pt}

\vspace{-50pt}

\vspace{-50pt}

\vspace{-50pt}

\vspace{-50pt}

\vspace{-50pt}
\newpage
\subsection*{Joint Impulse Responses following \citeA{ik22}}

\begin{center}
	\begin{figure}[H]
    \includegraphics[width=0.99\textwidth]{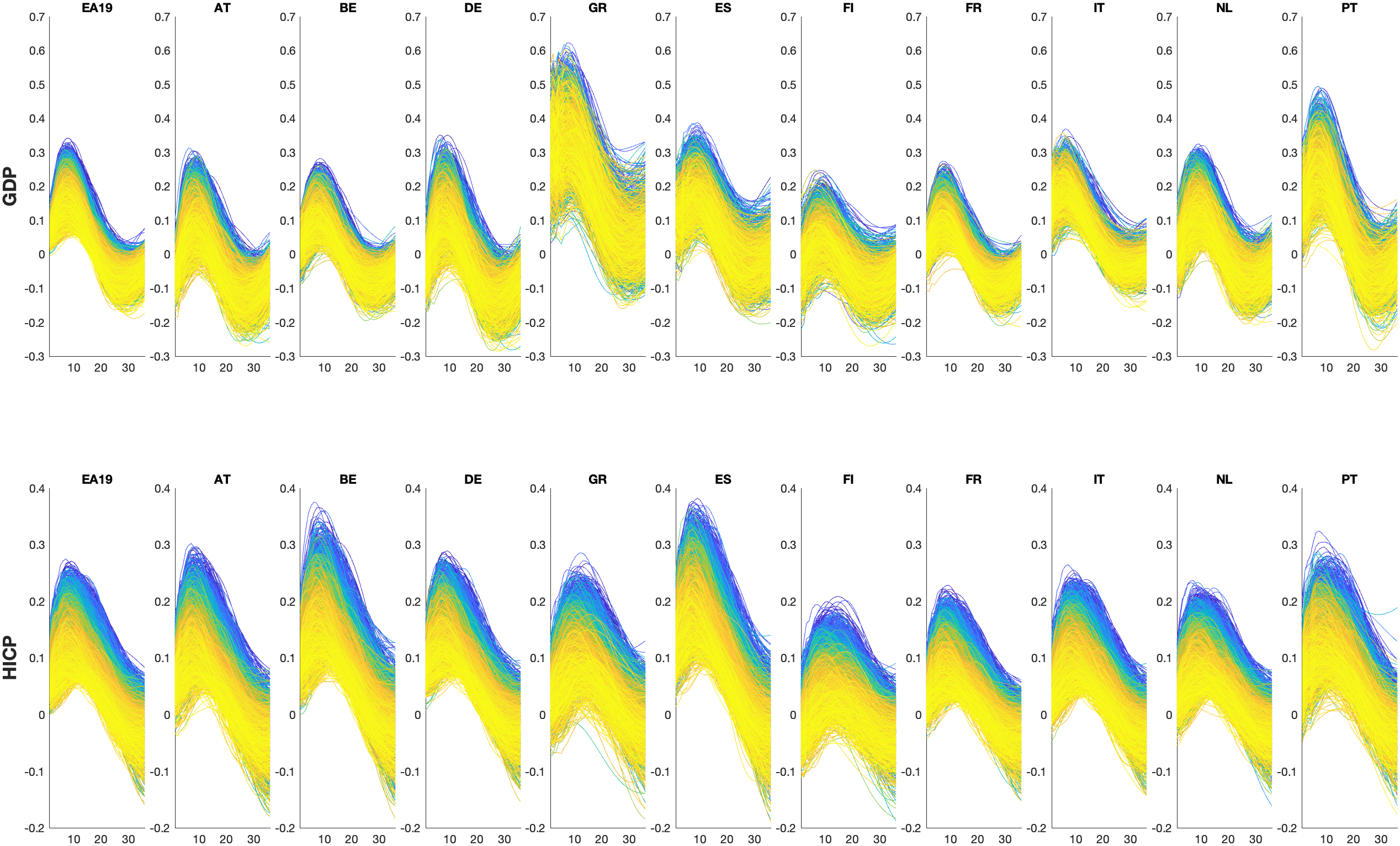}
      \caption{Joint impulse responses to an expansionary monetary policy shock. The depicted impulse responses constitute the 68\% joint credible set under a quadratic loss function. The country impulse responses are ordered after the maximum sum of the responses of GDP and inflation for the euro area aggregate over the horizon of 36 months.} \label{fig_OA44}
	\end{figure}
\end{center}

\begin{center}
	\begin{figure}[H]
    \includegraphics[width=0.99\textwidth]{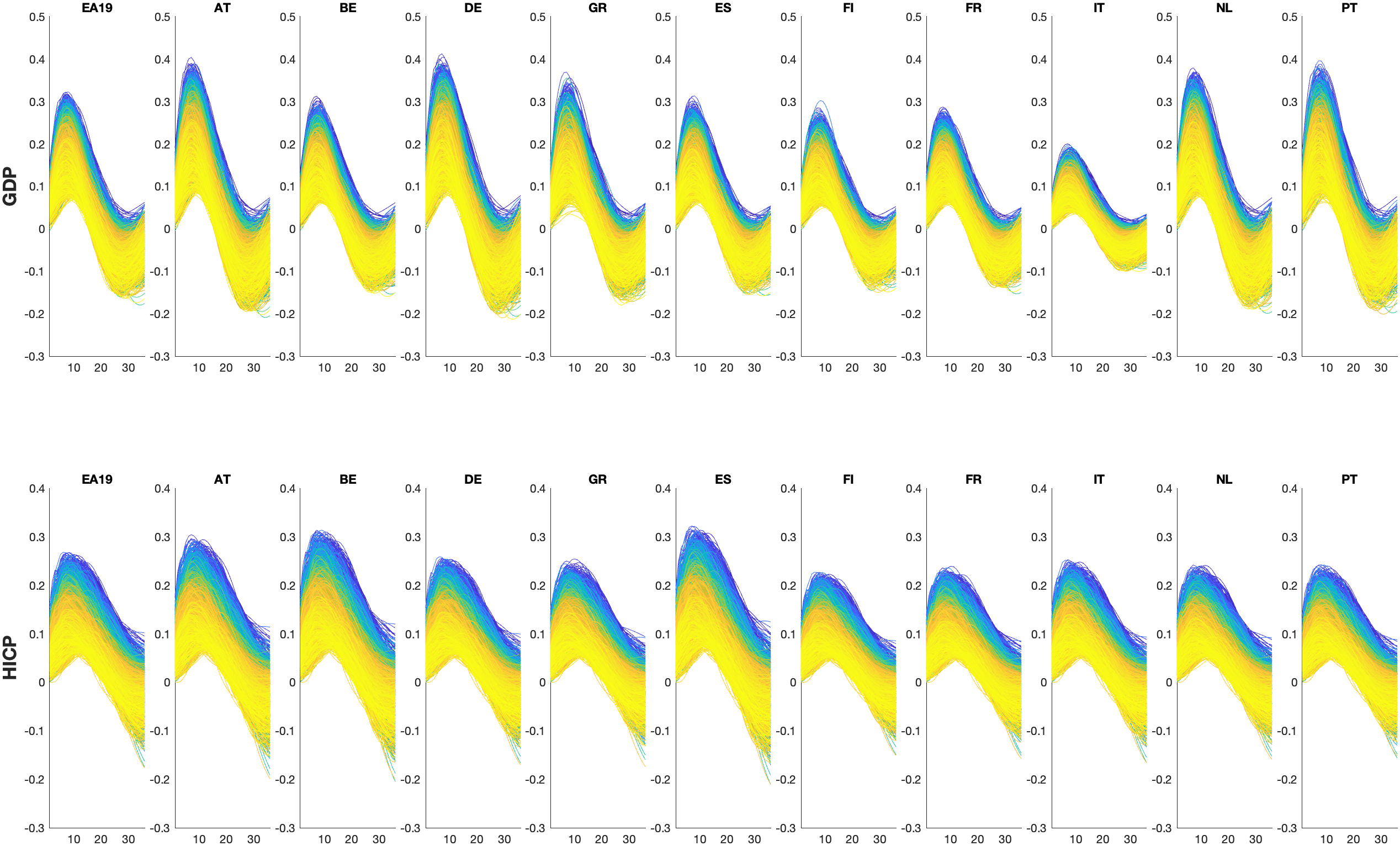}
      \caption{Joint impulse responses to an expansionary monetary policy shock via the common cycles. The depicted impulse responses constitute the 68\% joint credible set under a quadratic loss function. The country impulse responses are ordered after the maximum sum of the responses of GDP and inflation for the euro area aggregate over the horizon of 36 months.} \label{fig_OA45}
	\end{figure}
\end{center}

In this section, we present the joint impulse responses following the approach by \citeA{ik22}. Instead of estimating pointwise posterior medians and other percentiles which possibly are missing in the obtained set of impulse responses, we retrieve a joint credible set of responses that accounts for the joint uncertainty in the interval bands of the country responses. Here, we show the 68\% joint impulse responses and colour-sort them by the maximum sum of the responses of GDP and inflation for the euro area aggregate over the horizon of 36 months. \\
We perform this robustness analysis due to the following reasoning. First, the estimation of the joint credible sets for the country responses with respect to GDP and inflation captures the full uncertainty about the impulse responses as the conventional impulse responses depict marginal posterior distributions. In this regard, \citeA{ik22} also note that the posterior draws for the impulse response oftentimes do not contain the (exact) median impulse response. Therefore, the presentation of the joint posterior distribution is potentially less distorting than the median response and the (pointwise) error bands. Second, by ordering the posterior draws of the countries considered with respect to the maximum sum of responses for GDP and inflation of the euro area aggregate we show that the homogeneous transmission via the common cycle channels for GDP and inflation does not arise from an averaging out effect as the actual posterior draws exhibit a large degree of homogeneity (Figures \ref{fig_OA44} and \ref{fig_OA45}).

\vspace{-50pt}

\vspace{-50pt}

\section{Data Appendix}

\begin{table}[H]
	\centering
    \begin{threeparttable}
	\begin{tabular}{lll}
		\toprule
	    \textbf{Variable} &\textbf{Frequency} &\textbf{Source}  \\
		\midrule		
		Real Gross Domestic Product & Quarterly & FRED \\
        Industrial Production Index & Monthly & Eurostat \\
        Unemployment Rate & Monthly & ECB \\
		Harmonized Index of Consumer Prices & Monthly & Eurostat \\
        Eurostoxx50 & Monthly & ECB \\
        German 1 Year Government Bond Yield & Daily & Refinitiv \\
        Average Euro Area 1 Year Government Bond Yield & Daily & Refinitiv \\
        BBB Bond Spread & Daily  & FRED \\
        German 10 Year Government Bond Yield & Daily & Refinitiv \\
        Effective Real Exchange Rate & Monthly & ECB \\
        Loans to non-financial Institutions & Monthly & ECB \\
        1 Month OIS Rate Change around ECB Announcement & Daily & EA-MPD \\
        3 Month OIS Rate Change around ECB Announcement & Daily & EA-MPD \\
        6 Month OIS Rate Change around ECB Announcement & Daily & EA-MPD \\
        1 Year OIS Rate Change around ECB Announcement & Daily & EA-MPD \\
        Eurostoxx50 Change around ECB Announcement & Daily & EA-MPD \\
		\bottomrule
	\end{tabular}
    \begin{tablenotes}[flushleft]\footnotesize
    \linespread{.5}\small
    \item\hspace*{-\fontdimen2\font}\note This table summarizes the variables used in the different model specifications. The data on real gross domestic product, industrial production and unemployment are retrieved for the ten euro area countries considered and the euro area aggregate. The BBB bond spread is operationalized by the ICE BofA euro high yields index option-adjusted spread. The effective real exchange rate depicts the weighted and CPI-deflated average of the nominal exchange rates with the main trading partners. EA-MPD abbreviates the Euro Area Monetary Policy event study Database provided and regularly updated by \shortciteA{abgmr19}. Daily data are aggregated to monthly frequency. Quarterly data is interpolated to monthly frequency as described in the paper and the section on data transformation below.
    \end{tablenotes}
    \end{threeparttable}
    \label{tab_Data}
\end{table}

\clearpage

\newpage
\section{Data Transformation}

In this section, we describe the transformation applied to the data. As noted in the paper, we interpolate the seasonally adjusted real gross domestic products using seasonally adjusted industrial productions indices and unemployment rates following \citeA{cl71}. Before we perform the interpolation, we adjust for additive outliers as suggested by \citeA{eurostat20} and following method by \citeA{cl93}. The interpolated time series for GDP are then transformed to annual growth rates as described in the paper and in the online appendix section ``Alternative Annual Growth Rates''. Despite the interpolation and the outlier adjustment, the same transformation is applied to the harmonized index of consumer prices. \\
The German 1 year government bond yields, the average euro area 1 year government bond yields, the German 10 year government bond yields and the BBB bond spread enter the corresponding models as percentage values. The instruments for the monetary policy shocks enter the models as changes in the percentage values. The Eurostoxx50, the effective real exchange rate and the loans to non-financial institutions enter the models in logs.

\newpage
\section{Rotational Sign Restrictions}

As noted in section 3.2 on the structural identification, we follow \citeA{j22} and use rotational sign restrictions to untangle the \emph{pure} monetary policy shock, $m$ from the information shock, $cbi$. Here, we briefly outline the underlying approach. \\
First, we estimate the principal component, $i$, of the changes in the 1-, 3- and 6-month and the 1-year OIS rate around the ECB announcements.\footnote{As \citeA{j22}, we drop the coordinated announcement by the ECB and the FED on \nth{8} October 2008.} Then, we stack the principal component and the changes of the Eurostoxx50 around the announcement into a matrix, $U$ with dimension $T \times 2$. We apply the QR-decomposition of $U$ and obtain the orthonormal matrix $Q$ and the upper-triangular matrix R which diagonal elements are restricted to be positive. \\
Next, we establish a vector that equals $i$ if the principal component, $i$ and the change in Eurostoxx50 have different signs or 0 otherwise. Then, we estimate the variance of the non-zero elements in the vector divided by the total variance of the principal component, $i$ and define it as $\gamma$. We compute $\alpha =  \sqrt[2]{\gamma}$ and set:
\begin{eqnarray}
                P = \begin{bmatrix} 
            \cos(\alpha) & \sin(\alpha) \\
            -\sin(\alpha) & \cos(\alpha)           
            \end{bmatrix}. \nonumber
\end{eqnarray}
We employ $P$ to rotate $Q$. The resulting matrix consists of two vectors, $m{t}$ and $cbi_{t}$ that constitute the instruments of the monetary policy shock and the information shock. Both instruments are scaled such that the sum amounts to the principal component, $i$. Both vectors are aggregated to monthly frequency.\footnote{The results remain unchanged when the aggregation to monthly series is conducted before the untangling of the shocks.}

\newpage
\section{Data for the Correlation in 4.4}

\begin{table}[H]
	\centering
    \begin{threeparttable}
	\begin{tabular}{lll}
		\toprule
	    \textbf{Variable} &\textbf{Time Period} &\textbf{Source}  \\
		\midrule		
		GDP per capita & 2003-2023 & World Bank \\
        Public Debt to GDP ratio & 2003-2022 & World Bank \\
        Unemployment Rate & 2003-2023 & World Bank \\
		Ease of Doing Business & 2019 & World Bank \\
        Flexible Mortgage Rate & 2003-2023 & ECB \\
        Home Ownership & 2003-2023 & Eurostat \\
        Home Ownership w. Mortgage & 2003-2023 & Eurostat \\
        Home Ownership w/o Mortgage & 2003-2023  & Eurostat  \\
		\bottomrule
	\end{tabular}
    \begin{tablenotes}[flushleft]\footnotesize
    \linespread{.5}\small
    \item\hspace*{-\fontdimen2\font}\note In order to estimate the correlation coefficients and semi-partial correlation coefficients, we average the time series. The Ease of Doing Business Index is taken from year 2019. We use monthly data for the flexible mortgage rate. With respect to the Home Ownership, Home Ownership w. Mortgages and Home Ownership w/o Mortgage, the following observations are missing in the data set: AT:2003-2006, DE:2003, 2004, 2006-2009, GR: 2004-2006, ES: 2003-2006, FI: 2003, FR: 2003-2004, IT: 2003, NL: 2003-2004, PT:2003.
    \end{tablenotes}
     \end{threeparttable}\label{tab:Data:44}
\end{table}


\end{document}